\documentclass[11pt,a4paper]{article}
\usepackage{bbm}
\pdfoutput=1
\usepackage{jheppub}
\usepackage{amsmath}
\usepackage{epsfig}
\usepackage{amssymb}
\usepackage{graphics}
\usepackage[active]{srcltx}
\usepackage{epstopdf}

\newcommand{\bea}{\begin{eqnarray}}
\newcommand{\eea}{\end{eqnarray}}
\newcommand{\bean}{\begin{eqnarray*}}
\newcommand{\eean}{\end{eqnarray*}}

\def\O #1{\overline{#1}}

\def\det{\mathop{\rm det}}

\def\Label#1{\label{#1}%
  \smash{\hbox to0pt{\raise1ex\hbox{\tiny[#1]}\hss}}}

\def\tree{\tiny\mbox{tree}}

\def\inv{{\tiny \mbox{inv}}}

\title{Cross-ratio Identities and Higher-order Poles of CHY-integrand}
\author[a]{Carlos Cardona,}
\author[b,c]{Bo Feng,}
\author[d,e]{Humberto Gomez}
\author[b]{and Rijun Huang\footnote{The correspondence author}}

\affiliation[a]{Physics Division, National Center for Theoretical Sciences, National Tsing-Hua
University,\\ No.101, Section 2, Kuang-Fu Road, Hsinchu, Taiwan 30013, Republic of China.}
\affiliation[b]{Zhejiang Institute of Modern Physics, Department of Physics,
  Zhejiang University,\\ No.38, Zheda Road, Hangzhou, 310027, P.R. China.}
\affiliation[c]{Center of Mathematical Science,
  Zhejiang University,\\ No.38, Zheda Road, Hangzhou, 310027, P.R. China.}
\affiliation[d]{Instituto de Fisica -- Universidade de S\~ao Paulo,\\
Caixa Postal  66318, 05315-970 S\~ao Paulo, SP, Brazil.}
\affiliation[e]{Facultad de Ciencias Basicas,  Universidad Santiago de Cali,\\
Calle 5 $N^\circ$  62-00 Barrio Pampalinda, Cali, Valle, Colombia.}

\emailAdd{carlosandres@mx.nthu.edu.tw}
\emailAdd{fengbo@zju.edu.cn}
\emailAdd{humgomzu@gmail.com}
\emailAdd{huang@nbi.dk}

\date{\today}
\abstract{The evaluation of generic Cachazo-He-Yuan(CHY)-integrands is a big challenge and efficient
computational methods are in demand for practical evaluation. In this paper, we propose a systematic
decomposition algorithm by using cross-ratio identities, which provides an analytic and easy to implement
method for the evaluation of any CHY-integrand. This algorithm aims to decompose a given CHY-integrand
containing higher-order poles as a linear combination of CHY-integrands with only simple poles in a finite
number of steps, which ultimately can be trivially evaluated by integration rules of simple poles.
To make the method even more efficient for CHY-integrands with large number of particles and complicated
higher-order pole structures, we combine the $\Lambda$-algorithm and the cross-ratio identities,
and as a by-product it provides us a way to deal with CHY-integrands where the $\Lambda$-algorithm
was not applicable in its original formulation.}

\begin{document}

\begin{flushright}
\vspace{10pt} \hfill{NCTS-TH/1603} \vspace{20mm}
\end{flushright}
\maketitle \flushbottom

\section{Introduction}
\label{secIntroduction}

The $n$-particle scattering amplitude in arbitrary dimension can be
described by Cachazo-He-Yuan (CHY) formulation
\cite{Cachazo:2013gna,Cachazo:2013hca, Cachazo:2013iea,
Cachazo:2014nsa,Cachazo:2014xea} as
%
%
%
\bea \mathcal{A}_{n}^{\tree}&=&\int \Big( {\prod_{a=1}^nd z_a\over
\mbox{vol}~SL(2,\mathbb{C})}\Big)\Big({\prod}'\delta(\mathcal{E}_a)\Big)~\mathcal{I}(z)~~~~\label{CHYformula0}\\
&=&\int
\Big(z_{rs}z_{st}z_{tr}\prod_{a\in\{1,2,\ldots,n\}\setminus\{r,s,t\}}
dz_a\Big)\Big({z_{ij}z_{jk}z_{ki}}\prod_{a\in\{1,2,\ldots,n\}\setminus\{i,j,k\}}\delta(\mathcal{E}_a)\Big)~\mathcal{I}(z)~,~~~\label{CHYfoumula}\eea
respecting M\"obius $SL(2,\mathbb{C})$ invariance. The scattering
equations $\mathcal{E}_a$'s form an algebraic system of $n$ rational
functions in $n$ complex variables $z_a,a=1,2,\ldots, n$ as
\bea
0=\mathcal{E}_a=\sum_{b\in\{1,2,\ldots,n\}\setminus\{a\}}{s_{ab}\over
z_a-z_b}~~~\mbox{for}~~~a=1,2,\ldots,n~,~~~\label{seqn}\eea
where $\{1,2,\ldots,n\}\setminus\{a\}$ denotes a set of
$\{1,2,\ldots, n\}$ extracting the element $a$. Instead of
(\ref{seqn}), it has been shown in \cite{Dolan:2014ega} that an
equivalent polynomial form of scattering equations exists, with its
geometric structure investigated in \cite{He:2014wua}, makes the
evaluation of (\ref{CHYfoumula}) well-suited in the algebraic
geometry context. The M\"obius invariance implies that among the $n$
distinct punctured points $z_a, a=1,2,\ldots, n$ in Riemann sphere,
any three of them, say $z_a, a\in \{r,s,t\}$ , can be fixed to
particular locations, conventionally chosen as $z_r=\infty, z_s=1,
z_t=0$, such that the $n$-dimensional integration module the volume
of $SL(2,\mathbb{C})$ in (\ref{CHYformula0}) can be written as the
one in the first parenthesis of (\ref{CHYfoumula}), which is a
$(n-3)$-dimensional integration.
%
%
The M\"obius invariance also implies that only $(n-3)$ scattering
equations are linearly independent, and it forces us to write the
${\prod}'\delta(\mathcal{E}_a)$ in (\ref{CHYformula0}) as the one in
the second parenthesis of (\ref{CHYfoumula}),
%
%
to ensure that ${\prod}'\delta(\mathcal{E}_a)$ is independent of
removing any three scattering equations $\mathcal{E}_a, a\in
\{i,j,k\}$, leaving only $(n-3)$ linearly independent delta
functions. Hence formula (\ref{CHYfoumula}) is in fact a
$(n-3)$-dimensional integration constrained by $(n-3)$ delta
functions, allowing a representation
\bea \mathcal{A}_n^{\tree}=\sum_{z\in~{\small\mbox{solutions}}}
{z_{ij}z_{jk}z_{ki}z_{rs}z_{st}z_{tr}\over \mbox{Jacobian}}
\mathcal{I}(z)~,~~~\label{CHYrep1}\eea
on the $(n-3)!$ solutions of scattering equations, where
\bea \mbox{Jacobian}=\det\Big[{\partial \mathcal{E}_a\over
\partial z_b}\Big]_{(n-3)\times (n-3)}~~,~~\mbox{for}~~a\in\{1,2,\ldots,n\}\setminus\{i,j,k\}~~,~~b\in\{1,2,\ldots,n\}\setminus\{r,s,t\}~,~~~\label{jacobian}\eea
coming from the evaluation of $(n-3)$ delta functions. The so called
{\sl CHY-integrand} $\mathcal{I}(z)$ is a rational function of
$z_{ij}\equiv z_i-z_j$, external momenta $k_i$'s as well as the
polarization vectors $\epsilon_i$'s, whose explicit definition
varies with the field theories under consideration, while systematic
and compact construction of $\mathcal{I}(z)$ exists for bi-adjacent
cubic-scalar, pure Yang-Mills, Gravity theories, NLSM, DBI as well
as mixing among them
\cite{Cachazo:2014xea,Cachazo:2014nsa,Cachazo:2013iea}.

Although conceptually simple and elegant, it is in fact impossible
to analytically evaluate (\ref{CHYfoumula}) by (\ref{CHYrep1}), due
to the well-known {\sl Abel-Ruffini theorem} that there is no
algebraic solution to the general polynomial equations of degree
five or higher with arbitrary coefficients. There are a few studies
on the solutions of scattering equations in four-dimension and at
special kinematics
\cite{Kalousios:2013eca,Weinzierl:2014vwa,Lam:2014tga,Du:2016blz},
but not generic. Even in five-point case where analytic solution of
scattering equations is available, the $(5-3)!=2$ solutions are
radical functions of Mandelstam variables. Only after summing over
two solutions we get rational functions as the final simple result.
This infers that there must be better evaluation techniques, and it
motives various approaches towards the evaluation of
(\ref{CHYfoumula}) avoiding the explicit solutions of scattering
equations, based on algebraic geometry techniques. In
\cite{Kalousios:2015fya}, a method is proposed and applied to
analytically evaluating all five-point amplitudes, by the well-known
Vieta formulae that relates the sum of solutions of a polynomial
equations to the coefficient of polynomials. The elimination theory
elaborated therein for rewriting multivariate polynomials as an
univariate polynomial is further developed in
\cite{Cardona:2015eba,Cardona:2015ouc,Dolan:2015iln}, to deal with
more generic $n$-particle scattering system. In
\cite{Huang:2015yka}, companion matrix method is introduced, which
rephrases the computation of summing over $(n-3)!$ solutions as
computing the trace of certain $(n-3)!\times (n-3)!$ matrix, and
provides an intuitive interpretation that the final analytic result
is indeed rational functions. This is later proven to be equivalent
to the elimination method \cite{Cardona:2015eba,Dolan:2015iln}. In
\cite{Sogaard:2015dba}, Bezoutian matrix method is introduced to
evaluate the total sum of residues of (\ref{CHYfoumula})
algebraically, without working out individual residues (or solutions
of scattering equations). In \cite{Bosma:2016ttj,Zlotnikov:2016wtk},
polynomial reduction techniques are investigated based on the
polynomial form \cite{Dolan:2014ega} of scattering equation. All
above algebraic geometry based methods are in principle
generalizable to the evaluation of any $n$-particle scattering,
however the elimination algorithm as well as the polynomial
reduction rewrite the polynomials in a form such that the
coefficients are rather involved. Furthermore, a complete
computation also depends on the explicit expression of Jacobian
(\ref{jacobian}) which could be very complicated for large $n$.
These make the analytic computation in practical difficult even for
lower-point amplitude. A resulting trouble is that the amplitude
computed by these methods is not in the form with manifest physical
poles, and simplification of the result takes quite a long time. For
example, using the companion matrix method \cite{Huang:2015yka}, it
would be possible to spend hours to produce a six-point amplitude of
scalar theory.

In the demand of computational efficiency, many other methods come
to rescue. In \cite{Cachazo:2015nwa}, graph theory knowledge is
introduced in the contour integration of (\ref{CHYfoumula}), to
expand a generic CHY-integrand into basis of known simple
CHY-integrands, named after {\sl building blocks}. This idea is
further developed in \cite{Gomez:2016bmv}, resulting into the so
called {\sl $\Lambda$-algorithm}, to recursively cut a generic
CHY-integrand into lower-point sub-CHY-integrands until to certain
 basic building blocks. Although this method is able to deal with
off-shell configurations, which have allowed to generalize it to
loop-level recently in \cite{Cardona:2016bpi}, and despite it can
compute CHY-integrands of higher-order poles, it still has some
limitations. The cutting procedure for CHY-integrands is general,
but when the so called {\sl singular configurations} appear in
sub-CHY-integrands, evaluation becomes difficult. However for many
CHY-integrands, the singular configuration is un-avoidable. Another
approach is inspired by the string amplitude computation, where
combinatorial rules for integration (integration rules) of
(\ref{CHYfoumula}) is derived
\cite{Baadsgaard:2015voa,Baadsgaard:2015ifa}. In
\cite{Lam:2015sqb,Lam:2016tlk}, auxiliary Feynman-like diagrams are
introduced to compute the global residue, quite similar to that of
\cite{Baadsgaard:2015voa,Baadsgaard:2015ifa,Baadsgaard:2015hia},
while in \cite{Mafra:2016ltu}, Berends-Giele recursion relations are
applied to the computation of CHY-integrands which are products of
two Parke-Taylor type factors that containing only simple poles. On
the other hand, the integration rule is a rather simple and
efficient technique. In fact, the actual computation can be carried
out without any information of the solutions of scattering equations
as well as the detailed expression of Jacobian in CHY-formulation,
but only the information of CHY-integrand as a rational function of
$z_{ij}$. However, it suffers from a disadvantage that only
CHY-integrand with simple poles can be perfectly evaluated, while a
generic CHY-integrand from Yang-Mills or gravity theories could have
many higher-order poles. There are two intuitive ways of bypassing
this disadvantage. The first is to create rules for higher-order
poles, which is investigated in \cite{Huang:2016zzb}, but yet a
complete set of rules for any types of higher-order poles are
required therein before it could be a complete method. The other is
to decompose a CHY-integrand of higher-order poles into several
CHY-integrands of simple poles by non-trivial identities relating
different rational terms of $z_{ij}$. The idea of decomposition is
already discussed in \cite{Baadsgaard:2015voa}, where Pfaffian
identities are introduced to take on the task. But the power of
Pfaffian identities is limited to certain examples with lower-point
scattering, and no systematic implementation can be elaborated with
them\footnote{For even-integer $n$, a Pfaffian identity relates
$2(n-3)!!$ terms, while for odd-integer $n$, it relates $(n-2)!!$
terms, so the number of terms involved in an identity grows
factorially. But a more severe problem is that the Mandelstam
variables involved in Pfaffian identity are only $s_{ij}$, and a
pole of ${1\over s_{i_1i_2\cdot i_m}}$ with $m>2$ is not obviously
exist.}. In a recent paper \cite{Bjerrum-Bohr:2016juj}, identities
originated from monodromy relations are proposed to systematically
decompose a CHY-integrand of higher-order poles into those of simple
poles.

In this paper, we propose another kind of identities, namely the
{\sl cross-ratio identities}, to overcome the difficulties towards a
systematic and complete evaluation of (\ref{CHYfoumula}) in
practice. These identities are applied in two levels. For a
CHY-integrand with reasonable number of scattering particles and not
so complicated higher-order pole structures\footnote{From the
practice, this can be detailed as around $n=10$ with about five
higher-order poles (double or triple pole), or any number of
particles with around two higher-order poles. They can be computed
in a reasonably short time.}, the cross-ratio identities can be
applied directly to decompose a CHY-integrand into terms with only
simple poles. While in the situation that a naive decomposition of a
CHY-integrand would result in far too much terms that slow the
computation, the $\Lambda$-algorithm developed in
\cite{Gomez:2016bmv} would then be introduced to split a
CHY-integrand into products of several lower-point CHY-integrands
which can not be computed by $\Lambda$-algorithm, followed by the
application of cross-ratio identities to reformulate the lower-point
ones as those that can be computed by integration rules of simple
poles or $\Lambda$-algorithm.

This paper is structured as follows. In \S \ref{secCrossratio}, we
present the construction of cross-ratio identities and its generic
formulation for an arbitrary pole $1/s$, and demonstrate the
decomposition of CHY-integrands of higher-order poles by cross-ratio
identities with two examples. in \S\ref{secAlgorithm}, we propose a
systematic algorithm aims to completely decompose any CHY-integrand
of higher-order poles within finite steps by cross-ratio identities,
and illustrate the algorithm by a highly non-trivial eight-point
example. In \S\ref{secLalgorithm}, we illustrate how the
$\Lambda$-algorithm and decomposition algorithm with cross-ratio
identities can work together to make an even more efficient
computational method. Recurrence relations for a particular type of
CHY-integrands are presented as an example to show that for many
complicated CHY-integrands with large $n$, iterative cutting
procedure can be applied to rewrite them as lower-point
sub-CHY-integrands which are easy to compute. \S\ref{secConclusion}
comes the conclusion, while in Appendix \ref{secSignRule}, a
practical algorithm is proposed for a complete implementation of
integration rules of simple poles. In Appendix \ref{secMonodromy},
comparison is provided between the identities from monodromy
relations \cite{Bjerrum-Bohr:2016juj} and cross-ratio identities,
while in Appendix \ref{secPT2}, \ref{sec2PT2}, quantities needed for
the recurrence relations by $\Lambda$-algorithm are provided. Very
brief introduction on integration rules of simple poles can be found
in \S \ref{secAlgorithm} and Appendix \ref{secSignRule}, but we
suggest \cite{Huang:2016zzb} for more detailed explanation. For
detailed description of $\Lambda$-algorithm please refer to
\cite{Gomez:2016bmv,Cardona:2016bpi}.

\section{The Cross-Ratio Identities}
\label{secCrossratio}
\begin{figure}[h]
  \centering
  \includegraphics[width=4in]{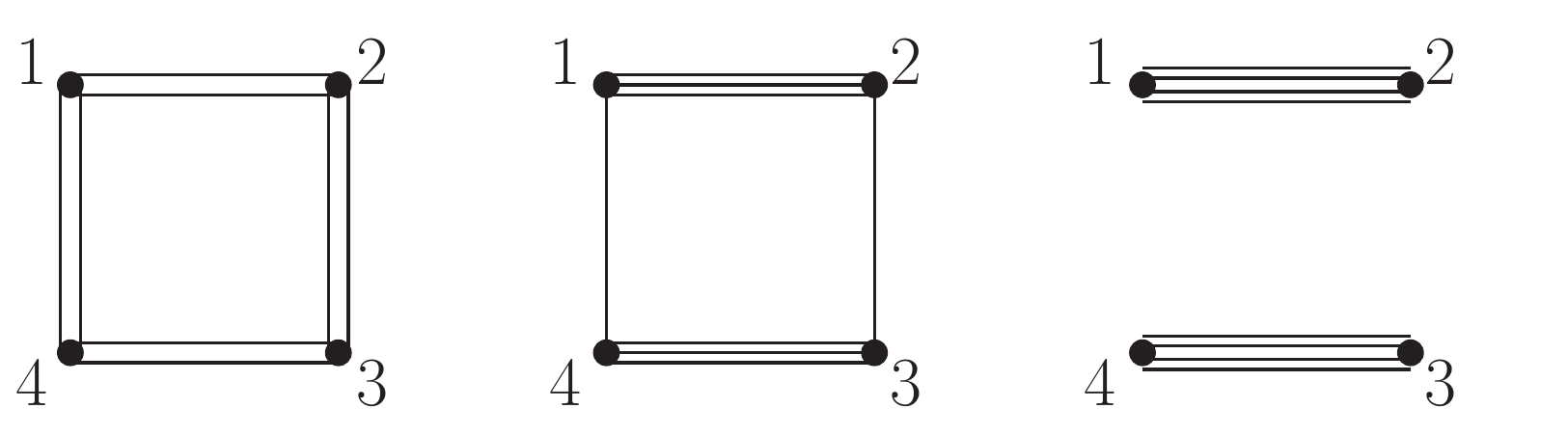}\\
  \caption{The {\sl 4-regular} graphs of four-point CHY-integrands with simple, double and triple pole respectively.}\label{Fig4ptAll}
\end{figure}

Following the decomposition idea, the primary problem is to find a
better identity rather than Pfaffian identities, which can be
applied to CHY-integrands with any pole structures. Although it is
extensively discussed in the literature, let us take, for conceptual
inspiration, some four-point CHY-integrands as starting point. Three
typical CHY-integrands are given as
\bea \mathcal{I}_4^{{\tiny\mbox{S}}}={1\over
z_{12}^2z_{23}^2z_{34}^2z_{41}^2}~~~,~~~\mathcal{I}_4^{{\tiny\mbox{D}}}={1\over
z_{12}^3z_{23}z_{34}^3z_{41}}~~~,~~~\mathcal{I}_4^{{\tiny\mbox{T}}}={1\over
z_{12}^4z_{34}^4}~,~~~\eea
with their {\sl 4-regular} graphs shown in Figure \ref{Fig4ptAll}.

\begin{figure}
  \centering
  \includegraphics[width=5in]{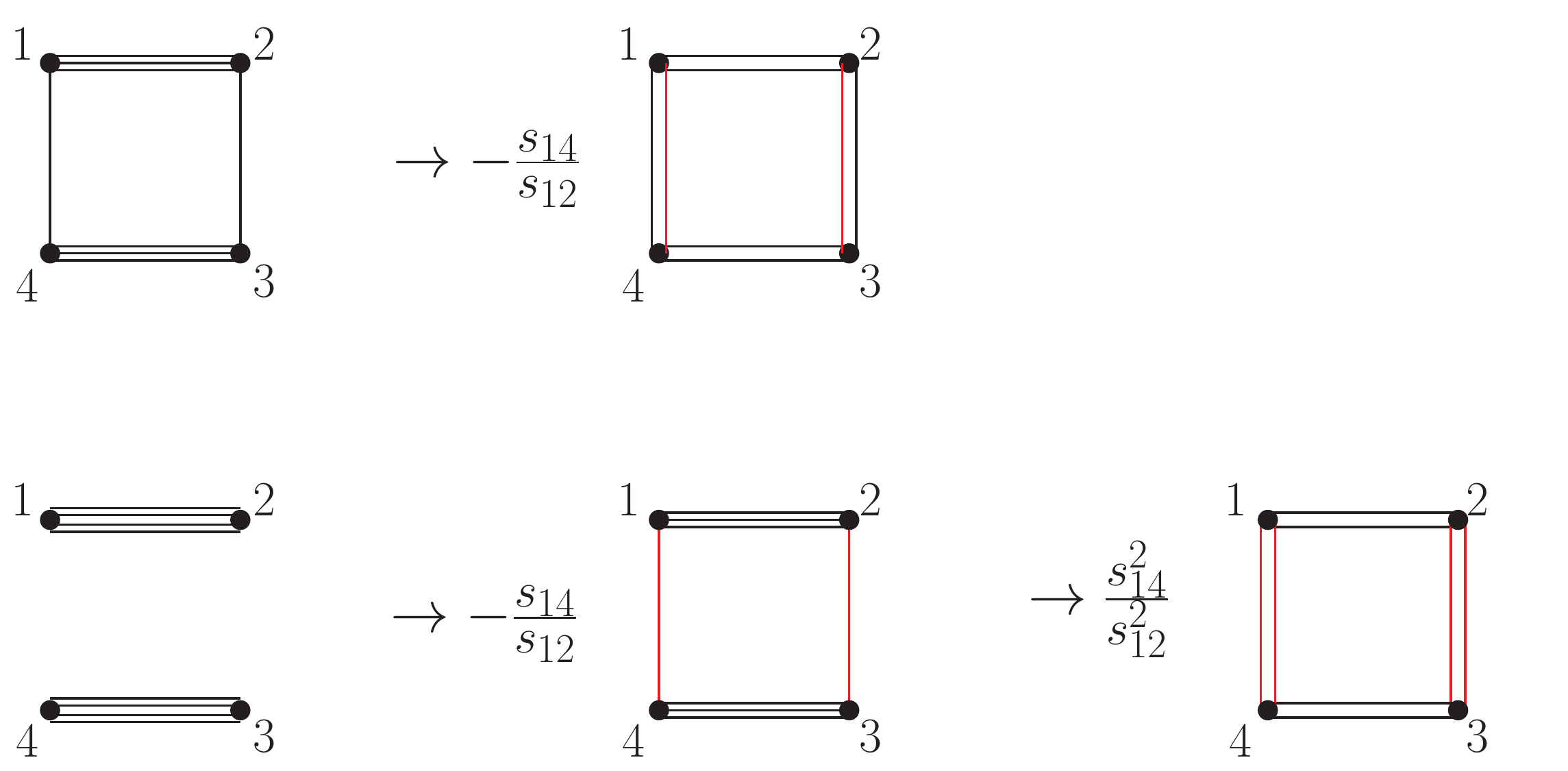}\\
  \caption{The diagrammatic presentation of how CHY-integrands with different pole structures can be
  related by cross-ratio identities.}\label{Fig4ptId}
\end{figure}

The explicit results for them are given as
\bea \mathcal{A}(\mathcal{I}_4^{{\tiny\mbox{S}}})=-{1\over
s_{12}}-{1\over
s_{14}}~~~,~~~\mathcal{A}(\mathcal{I}_4^{{\tiny\mbox{D}}})=-{s_{13}\over
s_{12}^2}~~~,~~~\mathcal{A}(\mathcal{I}_4^{{\tiny\mbox{T}}})={s_{13}s_{14}\over
s_{12}^3}~.~~~\eea
The CHY-integrand $\mathcal{I}_4^{{\tiny\mbox{S}}}$ contains only
simple poles, and can be readily evaluated by integration rules,
while $\mathcal{I}_4^{{\tiny\mbox{D}}}$ and
$\mathcal{I}_4^{{\tiny\mbox{T}}}$ contain double pole and triple
pole respectively, which makes it impossible to apply integration
rules of simple poles. However, a simple factorization of these
results
\bea \mathcal{A}(\mathcal{I}_4^{{\tiny\mbox{D}}})=-{1\over
s_{12}}\left({s_{13}\over
s_{12}}\right)~~~,~~~\mathcal{A}(\mathcal{I}_4^{{\tiny\mbox{T}}})={1\over
s_{12}^2}\left({s_{13}s_{14}\over s_{12}}\right) \eea
indicates that, if explicit ${1\over s}$ factor can be introduced in
the original CHY-integrand to compensate the extra degrees of $s$
from the higher-order poles, then the expressions in the parenthesis
are likely to be produced by CHY-integrands of simple poles (dressed
with appropriate $s$). To explain the above statement, let us start
from a scattering equation
\bea \mathcal{E}_1={s_{12}\over z_{12}}+{s_{13}\over
z_{13}}+{s_{14}\over z_{14}}=0~,~~~\eea
and modify it as
\bea 0={z_{12}\over s_{12}}\mathcal{E}_1&=&1+{s_{13}\over
s_{12}}{z_{12}\over z_{13}}+{s_{14}\over s_{12}}{z_{12}\over z_{14}}
=1-{s_{12}+s_{14}\over s_{12}}{z_{12}\over z_{13}}+{s_{14}\over
s_{12}}{z_{12}\over z_{14}}\nonumber\\
&&~~~~~~~~~=\left(1-{z_{12}\over z_{13}}\right)+{s_{14}\over
s_{12}}\left({z_{12}\over z_{14}}-{z_{12}\over
z_{13}}\right)={z_{23}\over z_{13}}+{s_{14}\over
s_{12}}{z_{12}z_{43}\over z_{14}z_{13}}~.~~~\eea
From the last expression we end up with an identity
\bea 1=-{s_{14}\over s_{12}}{z_{12}z_{43}\over
z_{14}z_{23}}~.~~~\eea
Since the $z_{ij}$'s in the identity are arranged as cross-ratios,
which are invariant under M\"obius transformation, we call it the
{\sl cross-ratio identity}. Note that a pole ${1\over s_{12}}$ is
apparent in the identity. Multiplying it with
$\mathcal{I}_4^{{\tiny\mbox{D}}}$ leads to
\bea \mathcal{I}_4^{{\tiny\mbox{D}}}=\left({1\over
z_{12}^3z_{23}z_{34}^3z_{41}}\right)\times\left(-{s_{14}\over
s_{12}}{z_{12}z_{43}\over z_{14}z_{23}}\right)=-{s_{14}\over
s_{12}}~\mathcal{I}_4^{{\tiny\mbox{S}}}~,~~~\eea
while multiplying it two times with
$\mathcal{I}_4^{{\tiny\mbox{T}}}$ leads to
\bea \mathcal{I}_4^{{\tiny\mbox{T}}}=\left({1\over
z_{12}^4z_{34}^4}\right)\times\left(-{s_{14}\over
s_{12}}{z_{12}z_{43}\over z_{14}z_{23}}\right)^2={s_{14}^2\over
s_{12}^2}~\mathcal{I}_4^{{\tiny\mbox{S}}}~,~~~\eea
which are diagrammatically shown in Figure \ref{Fig4ptId}.
Hence for four-point case, an evaluation of CHY-integrands with
simple poles is sufficient to produce results of all other
CHY-integrands with higher-order poles.

The idea encoded in above decomposition procedure is readily
generalized to any CHY-integrands of higher-order poles. For a given
$n$-point CHY-integrand, the poles as well as the order of poles can
be determined by {\sl Pole Condition}. Whenever there is any
higher-order pole ${1\over s_A^{\alpha}}$, we can reduce the order
of pole by multiplying identities containing that pole $s_{A}$. The
decomposition proceeds by iteratively reducing the higher-order
poles, until every resulting CHY-integrand is composed of simple
poles. In practical computation, this strategy can be carried out
only when there are enough identities of any possible poles, and as
mentioned, the Pfaffian identity is obviously not a good candidate.
However, the {\sl cross-ratio identity} that appeared in the
four-point examples can be generalized to satisfy the computational
demand, which we will describe below.

\subsection*{The cross-ratio identities}

For a $n$-point scattering system, let us pick up an arbitrary
scattering equation $\mathcal{E}_a$. For $p,q\neq a$, we can modify
the scattering equation as
\bea 0  =  {z_{aq}\over s_{aq}}\mathcal{E}_a = 1+ \sum_{b\neq a,q}
{z_{aq}\over s_{aq}} {s_{ab}\over z_{ab}}&=&1+ \sum_{b\neq a,q,p}
{z_{aq}\over s_{aq}} {s_{ab}\over z_{ab}}+ {-\sum_{t\neq a,p}
s_{at}\over z_{ap}} {z_{aq}\over s_{aq}}\nonumber\\
&=&\left(1-{z_{aq}\over z_{ap}}\right)+\sum_{b\neq
a,q,p}{s_{ab}\over s_{aq}}\left({z_{aq}\over z_{ab}}-{z_{aq}\over
z_{ap}}\right)~,~~~\eea
where in the last step of line one we have rewritten $s_{ap}$ by
momentum conservation. Using $z_{ij}-z_{ik}=z_{kj}$, we get
\bean 0 & =&  {z_{qp}\over z_{ap}}+\sum_{b\neq a,q,p} {s_{ab}\over
s_{aq}} { z_{aq}z_{bp}\over z_{ab} z_{ap}}~.\eean
Hence the cross-ratio identity for $s_{ij}$-type pole can be
formulated as
\bea \boxed{1=- \sum_{b\neq a,q,p} {s_{ab}\over s_{aq}}{
z_{aq}z_{bp}\over z_{ab} z_{qp}} ~.~~~}\label{CR-Ide-1}\eea
Again we remark that in each term of the identity, $z_{ij}$'s are
arranged as cross-ratios, which are invariant under M\"obius
transformation. The identities (\ref{CR-Ide-1}) will be called {\sl
fundamental cross-ratio identities} since all other identities of
given pole $s_{i_1i_2\cdots i_m}$ can be derived from them.

Before presenting the cross-ratio identity for a generic pole
$s_{i_1i_2\cdots i_m}$, let us show the construction by taking the
pole $s_{ijkl}$ as a warm up. The fundamental cross-ratio identity
(\ref{CR-Ide-1}) can be rewritten as
\bea -s_{ij} &= & \sum_{b\neq i,j,p} {s_{ib}}{ z_{bp}\over z_{ib} }{
z_{ij} \over z_{jp}}~.~~~\eea
Let us take an arbitrary $p\neq i,j,k,l$, and consider the following
three fundamental cross-ratio identities
\bea -s_{ij} &= & {s_{ik}}{ z_{kp}z_{ij} \over z_{ik}z_{jp}}
+{s_{il}}{ z_{lp}z_{ij} \over z_{il} z_{jp}}+\sum_{b\neq i,j,k,l,p}
{s_{ib}}{ z_{bp}z_{ij} \over z_{ib} z_{jp}}~,~~~\\
-s_{kj} &= &{s_{ki}}{ z_{ip}z_{kj} \over z_{ki} z_{jp}} +{s_{kl}}{
z_{lp}z_{kj}\over z_{kl} z_{jp}}+\sum_{b\neq i,j,k,l,p} {s_{kb}}{
z_{bp}z_{kj}\over z_{kb} z_{jp}}~,~~~\\
-s_{lj} & = & {s_{li}}{ z_{ip}z_{lj}\over z_{li} z_{jp}}+{s_{lk}}{
z_{kp}z_{lj}\over z_{lk} z_{jp}}+\sum_{b\neq i,j,k,l,p} {s_{lb}}{
z_{bp}z_{lj}\over z_{lb} z_{jp}}~.~~~\eea
It is easy to see that
\bea {s_{ik}}{ z_{kp}z_{ij} \over z_{ik}z_{jp}}+{s_{ki}}{
z_{ip}z_{kj} \over z_{ki} z_{jp}}=s_{ik}~~,~~{s_{il}}{ z_{lp}z_{ij}
\over z_{il} z_{jp}}+{s_{li}}{ z_{ip}z_{lj}\over z_{li}
z_{jp}}=s_{il}~~,~~{s_{kl}}{ z_{lp}z_{kj}\over z_{kl}
z_{jp}}+{s_{lk}}{ z_{kp}z_{lj}\over z_{lk} z_{jp}}=s_{lk}~,~~~\eea
so summing over the three identities, we obtain
\bea -s_{ijkl} =   \sum_{b\neq i,j,k,l,p} \left({s_{ib}}{
z_{bp}z_{ij} \over z_{ib} z_{jp}} +{s_{kb}}{ z_{bp}z_{kj}\over
z_{kb} z_{jp}}+{s_{lb}}{ z_{bp}z_{lj}\over z_{lb}
z_{jp}}\right)~.~~~\eea
This immediately gives the cross-ratio identity of $s_{ijkl}$-type
pole as
\bea \boxed{ 1=  -{1\over s_{ijkl}} \sum_{b\neq i,j,k,l,p}
\left({s_{ib}}{ z_{bp}z_{ij} \over z_{ib} z_{jp}} +{s_{kb}}{
z_{bp}z_{kj}\over z_{kb} z_{jp}}+{s_{lb}}{ z_{bp}z_{lj}\over z_{lb}
z_{jp}}\right)~.~~~}\label{s[ijkl]-construction}\eea

Some remarks are in order for identity (\ref{s[ijkl]-construction}).
The first is about the factors $z_{ij}, z_{kj}, z_{lj}$ in the
numerator of each term respectively. They are crucial for reducing
the number of lines that connecting nodes $\{i,j,k,l\}$, hence
consequently reducing the degree of pole $s_{ijkl}$ by one. This is
the key point of our algorithm. The second is about the $p$ index.
It could be any one in $\{1,2,\ldots, n\}$ except $i,j,k,l$.
Otherwise for instance $p=k$, there would always be a factor
$z_{jp}\equiv z_{jk}$ in the denominator of
(\ref{s[ijkl]-construction}), such that the number of lines
connecting nodes $\{i,j,k,l\}$ will increase by one. So the order of
poles will not be reduced. The third is about the $j$ index. In the
derivation, we have chosen to fix the index $j$ and consider three
fundamental cross-ratio identities for poles $s_{ij}, s_{kj},
s_{lj}$. Such choice breaks the symmetry among indices
$\{i,j,k,l\}$, leaving only $S_3$ permutation symmetry on
$\{i,k,l\}$ manifest. Similarly, the choice of $p$ also breaks the
symmetry among remaining indices. As a consequence, different choice
of $j,p$ leads to different cross-ratio identity for the same pole
$s_{ijkl}$, and during computation we can choose an appropriate one
to simplify the decomposition.

The construction of cross-ratio identity for generic pole ${1\over
s_A}$ follows exactly the same derivation. Let $A$ be a subset of
$\{1,2,...,n\}$, and assume $\O A$ to be its complement. Because of
momentum conservation, $s_A=s_{\O A}$. Then the cross-ratio identity
for pole $s_A$ with selected index $j\in A$ and $p\in \O A$ is
\bea \boxed{1=-\sum_{i\in A\setminus\{j\}}\sum_{b\in \O
A\setminus\{p\}} {s_{ib}\over s_A}{ z_{bp}z_{i j} \over z_{i b}
z_{jp}}\equiv\mathbb{I}_n[A,j,p]~,~~~}\label{generalCR} \eea
where $A\setminus\{j\}$ denotes the set $A$ extracting the element $j$.
In the notation $\mathbb{I}_{n}[A,j,p]$, $A$ is the subset
associated to the pole $s_A$ of identity, and $j,p$ are explicitly
written down to emphasize the special choice. We remark again that,
the cross-ratio identity is invariant under permutation on
$A\setminus\{j\}$ as well as permutation on $\O A\setminus\{p\}$, so
there are in total $k(n-k)$ different identities for the pole $s_A$
if $A$ is a length-$k$ subset.

With the general construction of cross-ratio identities, we can now
implement an algorithm for the decomposition of any CHY-integrands
with higher-order poles. In the next section, we will present a
systematic decomposition algorithm, but now let us follow two
examples to explore some details during the decomposition.

\subsection*{Two examples}
\begin{figure}[h]
  \centering
  \includegraphics[width=2in]{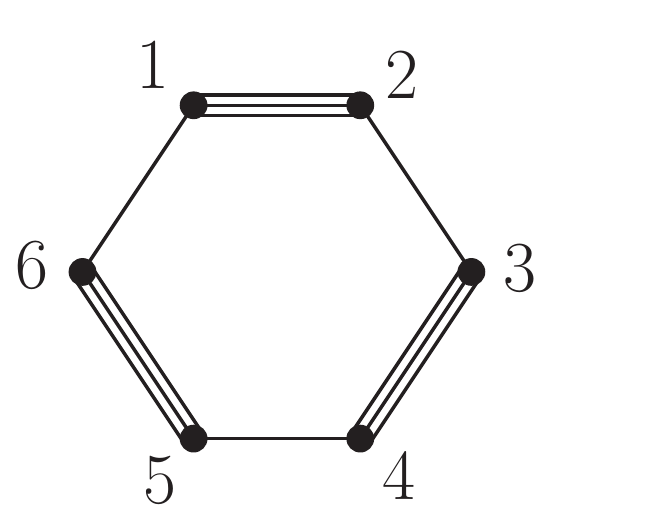}\\
  \caption{The {\sl 4-regular} graph of a six-point CHY-integrand with three double poles.}\label{FigA61}
\end{figure}

The first example considers the CHY-integrand
\bea \mathcal{I}_6^{[1]}={1\over
z_{12}^3z_{23}z_{34}^3z_{45}z_{56}^3z_{61}}~,~~~\label{intA61}\eea
with its {\sl 4-regular} graph shown in Figure \ref{FigA61}.
From the counting of lines among subsets of nodes, we know the
result for this CHY-integrand should be
\bea {N(s_{ij})\over
s_{12}^2s_{34}^2s_{56}^2s_{123}s_{234}s_{345}}~,~~~\eea
where $N(s_{ij})$ is some polynomial function of $s_{ij}$ which is
irrelevant for the purpose of the current example. So in order to
perform the decomposition, we need to multiply it with a cross-ratio
identity of the pole ${1\over s_{12}}$, an identity of the pole
${1\over s_{34}}$ and an identity of the pole ${1\over s_{56}}$.
Since for each $s_{ij}$, there are $2(6-2)=8$ cross-ratio
identities, so naively we have $8^3=512$ possibilities of
multiplying three identities. However, not all of them can
successfully decompose $\mathcal{I}_6^{[1]}$ into terms with only
simple poles, since with the multiplication of $z_{ij}$, new
higher-order poles would appear in some terms (as long as there are
more than two terms for a new higher-order pole, such that after
summing over these terms the new higher-order pole is still
canceled.). Of course, the best expectation is that we can find at
least one multiplication such that the original CHY-integrand can be
decomposed into terms with only simple poles. Fortunately in this
example, if taking the following three fundamental cross-ratio
identities
\bea &&\mathbb{I}_6[\{1,2\},2,5]=-\left(\frac{s_{13} z_{12}
z_{35}}{s_{12} z_{13} z_{25}}+\frac{s_{14} z_{12} z_{45}}{s_{12}
z_{14} z_{25}}+\frac{s_{16} z_{12} z_{65}}{s_{12} z_{16}
z_{25}}\right)~,~~~\nonumber\\
&&\mathbb{I}_6[\{3,4\},3,5]=-\left(\frac{s_{46} z_{65}
z_{43}}{s_{34} z_{35} z_{46}}+\frac{s_{14} z_{15} z_{43}}{s_{34}
z_{35} z_{41}}+\frac{s_{24} z_{25} z_{43}}{s_{34} z_{35}
z_{42}}\right)~,~~~\nonumber\\
&&\mathbb{I}_6[\{5,6\},5,1]=-\left(\frac{s_{26} z_{21}
z_{65}}{s_{56} z_{51} z_{62}}+\frac{s_{36} z_{31} z_{65}}{s_{56}
z_{51} z_{63}}+\frac{s_{46} z_{41} z_{65}}{s_{56} z_{51}
z_{64}}\right)~,~~~\eea
i.e.,
\bea
1&=&\mathbb{I}_6[\{1,2\},2,5]\mathbb{I}_6[\{3,4\},3,5]\mathbb{I}_6[\{5,6\},5,1]~,~~~
\eea
then all the resulting $3^3=27$ terms after expanding
$\mathcal{I}_6^{[1]}\times
\mathbb{I}_6[\{1,2\},2,5]\times\mathbb{I}_6[\{3,4\},3,5]\times\mathbb{I}_6[\{5,6\},5,1]$
are CHY-integrands with only simple poles. By using integration
rules for simple poles, it is confirmed that summing over these 27
terms indeed produces correct answer.

\begin{figure}
  \centering
  \includegraphics[width=2in]{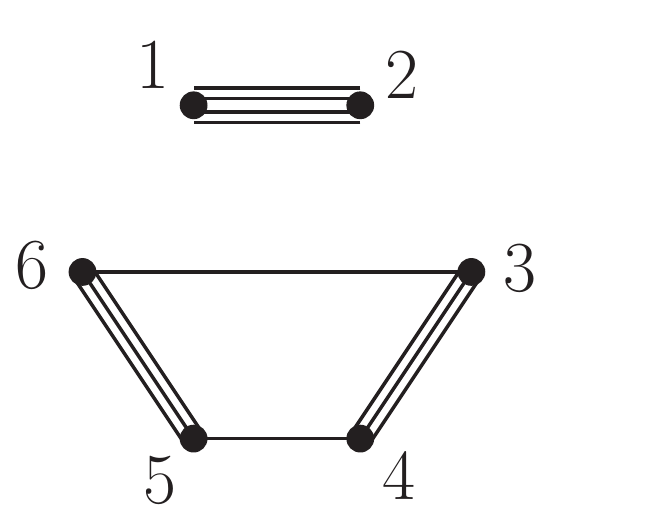}\\
  \caption{The {\sl 4-regular} graph of a six-point CHY-integrand with one triple pole and two double poles.}\label{FigA63}
\end{figure}

The second example considers the CHY-integrand
\bea \mathcal{I}_6^{[2]}={1\over
z_{12}^4z_{34}^3z_{45}z_{56}^3z_{63}}~,~~~ \label{intA63}\eea
with its {\sl 4-regular} graph shown in Figure \ref{FigA63}.
Again the result can be inferred as
\bea {N(s_{ij})\over
s_{12}^3s_{34}^2s_{56}^2s_{123}s_{124}s_{125}s_{126}}~,~~~\eea
where $N(s_{ij})$ is another polynomial function of $s_{ij}$. So we
need to multiply it with four fundamental cross-ratio identities,
two of the pole ${1\over s_{12}}$ (counting multiplicity), one of
the pole ${1\over s_{34}}$ and one of the pole ${1\over s_{56}}$.
Since each identity contains 3 terms, after multiplying four
identities the original CHY-integrand would be decomposed into
$3^4=81$ terms.

For each $s_{ij}$, there are 8 cross-ratio identities. We have gone
through all ${8(8+1)\over 2}\times 8\times 8=2304$ possible
multiplications of four fundamental cross-ratio identities which
have the poles ${1\over s_{12}^2s_{34}s_{56}}$, and find none of
them can decompose CHY-integrand (\ref{intA63}) into 81 terms with
only simple poles. The best situation is that, 4 out of 81 terms are
CHY-integrands with a double pole while the remaining 77 terms with
simple poles. Such a multiplication can be constructed from the
following four cross-ratio identities
\bea &&\mathbb{I}_6[\{1,2\},2,5]=-\left(\frac{s_{13} z_{12}
z_{35}}{s_{12} z_{13} z_{25}}+\frac{s_{14} z_{12} z_{45}}{s_{12}
z_{14} z_{25}}+\frac{s_{16} z_{12} z_{65}}{s_{12} z_{16}
z_{25}}\right)~,~~~\nonumber\\
&&\mathbb{I}_6[\{1,2\},1,5]=-\left(\frac{s_{23} z_{21}
z_{35}}{s_{12} z_{15} z_{23}}+\frac{s_{24} z_{21} z_{45}}{s_{12}
z_{15} z_{24}}+\frac{s_{26} z_{21} z_{65}}{s_{12} z_{15}
z_{26}}\right)~,~~~\nonumber\\
&&\mathbb{I}_6[\{3,4\},4,1]=-\left(\frac{s_{35} z_{51}
z_{34}}{s_{34} z_{35} z_{41}}+\frac{s_{36} z_{61} z_{34}}{s_{34}
z_{36} z_{41}}+\frac{s_{23} z_{21} z_{34}}{s_{34} z_{32}
z_{41}}\right)~,~~~\nonumber\\
&&\mathbb{I}_6[\{5,6\},6,1]=-\left(\frac{s_{25} z_{21}
z_{56}}{s_{56} z_{52} z_{61}}+\frac{s_{35} z_{31} z_{56}}{s_{56}
z_{53} z_{61}}+\frac{s_{45} z_{41} z_{56}}{s_{56} z_{54}
z_{61}}\right)~.~~~\eea
After multiplying
$\mathbb{I}_6[\{1,2\},2,5]\times\mathbb{I}_6[\{1,2\},1,5]\times\mathbb{I}_6[\{3,4\},4,1]\times\mathbb{I}_6[\{5,6\},6,1]$
to $\mathcal{I}_6^{[2]}$, we get 81 terms, while the following four
terms contain double pole, explicitly as
\bea -{s_{14} s_{24} s_{35} s_{36}\over s_{12}^2 s_{34}
s_{56}}\frac{ z_{13} z_{45}}{ z_{12}^2 z_{14}^2 z_{15} z_{24} z_{25}
z_{34}^2 z_{35} z_{36}^2 z_{56}^2}~~~,~~~-{s_{14} s_{24}
s_{35}^2\over s_{12}^2 s_{34} s_{56}}\frac{ z_{13} z_{45}}{ z_{12}^2
z_{14}^2 z_{16} z_{24} z_{25} z_{34}^2 z_{35}^2 z_{36}
z_{56}^2}~~~~\label{I62hpole1}\eea
with double pole $s_{124}$, and
\bea -{s_{16} s_{26} s_{35}^2\over s_{12}^2 s_{34} s_{56}}\frac{
z_{13}}{ z_{12}^2 z_{14} z_{16}^2 z_{25} z_{26} z_{34}^2 z_{35}^2
z_{36} z_{45}}~~~,~~~-{s_{16} s_{26} s_{35} s_{45}\over s_{12}^2
s_{34} s_{56} }\frac{1}{z_{12}^2 z_{16}^2 z_{25} z_{26} z_{34}^2
z_{35} z_{36} z_{45}^2}~~~~\label{I62hpole2}\eea
with double pole $s_{126}$. Note that $s_{124}$ and $s_{126}$ are
not double poles of original CHY-integrand, so it should not present
in the final answer. The two terms of each double pole guarantee the
cancelation. For a complete decomposition, we can further multiply
the cross-ratio identity
\bea \mathbb{I}_6[\{1,2,4\},2,3]=-\left(\frac{s_{15} z_{12}
z_{53}}{s_{124} z_{15} z_{23}}+\frac{s_{45} z_{42} z_{53}}{s_{124}
z_{23} z_{45}}+\frac{s_{16} z_{12} z_{63}}{s_{124} z_{16}
z_{23}}+\frac{s_{46} z_{42} z_{63}}{s_{124} z_{23}
z_{46}}\right)~,~~~\eea
to the two terms in (\ref{I62hpole1}), and the identity
\bea \mathbb{I}_6[\{1,2,6\},2,3]=-\left(\frac{s_{14} z_{12}
z_{43}}{s_{126} z_{14} z_{23}}+\frac{s_{46} z_{62} z_{43}}{s_{126}
z_{23} z_{64}}+\frac{s_{15} z_{12} z_{53}}{s_{126} z_{15}
z_{23}}+\frac{s_{56} z_{53} z_{62}}{s_{126} z_{23}
z_{65}}\right)~~~~\eea
to the two terms in (\ref{I62hpole2}). Then the four terms with
double poles can be further decomposed into terms with simple poles.
Hence the CHY-integrand (\ref{intA63}) is decomposed into
$77+4\times 4=93$ CHY-integrands of simple poles, and summing over
these 93 terms indeed produces the correct answer.

The second example clearly shows that, for generic CHY-integrand of
complicated higher-order poles, it is usually not possible to
completely decompose it within one step. Even after trying all
possible multiplication of cross-ratio identities, there would be
some resulting CHY-integrands which need a second and even more
steps on cross-ratio identity decomposition. Another thing is about
the various cross-ratio identities for the same pole $s_A$. To
compensate the higher-order pole $s_A$, different cross-ratio
identities provide different decomposition. Some cross-ratio
identities will reduce the degree of higher-order poles in each
resulting terms while some will introduce other higher-order poles.
These suggest us to implement a decomposition algorithm step by
step, and a complete decomposition could be guaranteed only if there
are appropriate cross-ratio identities to reduce, or at least not
increase, the degree of higher-order poles in each step. Since there
are pretty much cross-ratio identities for a pole $s_A$, and also
various possibilities of decomposition steps, it seems to give us
enough information such as the decomposition procedure can always
continue until we get a decomposition with CHY-integrands of simple
poles.

With above preparations, we are ready to propose a systematic
decomposition algorithm in the following section.

\section{A Systematic Decomposition Algorithm}
\label{secAlgorithm}

\subsection{The algorithm}

In order to decompose a CHY-integrand $\mathcal{I}(z_{ij})$ of
higher-order poles into terms with only simple poles by cross-ratio
identities, we can start from an arbitrary higher-order pole $s_A$
and multiply an appropriate cross-ratio identity of pole $s_A$ to
the original CHY-integrand. This leads to several CHY-integrands
with the order of higher-order poles reduced\footnote{ There are
subtleties that require a careful treatment and we will discuss them
along this section}. For each resulting CHY-integrand, we again
reduce the order of poles by multiplying an appropriate cross-ratio
identity, and iteratively perform this procedure until all resulting
terms contain simple poles. Here we present a systematic
decomposition algorithm aims to decompose any CHY-integrand of
higher-order poles into terms with only simple poles in finite
steps.

Let us start from a generic $n$-point CHY-integrand
$\mathcal{I}(z_{ij})$, as a rational function of $z_{ij}$, as
\bea \mathcal{I}={1\over \prod_{1\leq i<j\leq n
}z_{ij}^{\beta_{ij}}}~,~~~\label{genericCHY}\eea
In the {\sl 4-regular} graph representation, the $\beta_{ij}$ is
represented by lines connecting nodes $z_i,z_j$. A positive integer
$\beta_{ij}$ is represented by the corresponding number of solid
lines, a negative integer $\beta_{ij}$ (which stands for a
non-trivial numerator) is represented by the corresponding number of
dashed lines. In order to respect the M\"obius invariance, for each
node, the number of connected solid lines minus the number of
connected dashed lines is four.

For the length-$n$ set $\{1,2,\ldots,n\}$, since a subset is
considered to be equivalent to its complement due to momentum
conservation, we have the following independent subsets
$\widetilde{A}$,
\begin{itemize}
  \item If $n$ is odd, $\widetilde{A}_\alpha=\{i_1,i_2,\ldots,i_k\}$, ~~$i_1,i_2,\ldots,i_k\in
  \{1,2,\ldots,n\}$,~~$2\leq k\leq \lfloor {n\over 2}\rfloor$,
  \item If $n$ is even, $\widetilde{A}_\alpha=\{i_1,i_2,\ldots,i_k\}$, ~~$i_1,i_2,\ldots,i_k\in
  \{1,2,\ldots,n\}$, for $2\leq k\leq \lfloor {n\over
  2}\rfloor-1$, and $i_1=1$, $i_2,\ldots,i_k\in \{2,\ldots,n\}$
  for $k=\lfloor {n\over 2}\rfloor$.
\end{itemize}

For a subset $\widetilde{A}_{\alpha}=\{i_1,i_2,\ldots,i_k\}$, the
number of lines $\mathbb{L}[\widetilde{A}_{\alpha}]$ connecting
nodes $z_{i_1},z_{i_2},\ldots,z_{i_k}$ is given by
\bea \mathbb{L}[\widetilde{A}_{\alpha}]=\sum_{i',j'\in
\widetilde{A}_{\alpha}} \beta_{i'j'}~,~~~\eea
so the {\sl pole index} $\chi_{\widetilde{A}_{\alpha}}$ for this
subset is
\bea\label{pole_index}
\chi_{\widetilde{A}_{\alpha}}=\mathbb{L}[\widetilde{A}_{\alpha}]-2(|\widetilde{A}_{\alpha}|-1)=\Big(\sum_{i',j'\in
\widetilde{A}_{\alpha}} \beta_{i'j'}\Big)-2(k-1)~.~~~\eea
From the CHY-integrand (\ref{genericCHY}), we can directly read out
the pole index for every subset. Each subset
$\widetilde{A}_{\alpha}$ corresponds to a pole ${1\over
s_{\alpha}}$. From the {\sl Pole Condition} follows that, if
$\chi_{\widetilde{A}_{\alpha}}<0$, the pole will not present in the
final result, while if $\chi_{\widetilde{A}_{\alpha}}= 0$, a simple
pole ${1\over s_{\alpha}}$ will present, and if
$\chi_{\widetilde{A}_{\alpha}}>0$, a pole of order ${1\over
s_{\alpha}^{\chi+1}}$ will appear in the final result.

Assuming that for a given CHY-integrand $\mathcal{I}$, there are $m$
independent subsets $A_{\alpha_1},A_{\alpha_2},\ldots, A_{\alpha_m}$
with $\chi_{A_{\alpha_i}}\geq 0$, we define the order of poles of
the CHY-integrand as
\bea
\Upsilon[\mathcal{I}]=\sum_{i=1}^{m}\chi_{A_{\alpha_i}}~,~~~\eea
which can be readily computed from $\beta_{ij}$ by using
(\ref{pole_index}). The $\Upsilon$ is the number of poles to be
compensated by cross-ratio identities, i.e., in order to completely
decompose a CHY-integrand, we need to multiply at least $\Upsilon$
cross-ratio identities. $\Upsilon[\mathcal{I}]=0$ means the
corresponding CHY-integrand $\mathcal{I}$ contains only simple
poles. We will use it as a criteria in the decomposition algorithm.

Before stating the algorithm, let us have a look at the cross-ratio
identities. For a generic pole $s_{i_1i_2\cdots i_k}$, from
definition (\ref{generalCR}) we know that there are $k(n-k)$
identities
\bea
\mathbb{I}_n[\{i_1,i_2,\ldots,i_k\},j,p]~~~&,&~~~j\in\{i_1,i_2,\ldots,i_{k}\}~,~~~\nonumber\\
~~~&,&~~~p\in
\{1,2,\ldots,n\}\setminus\{i_1,i_2,\ldots,i_{k}\}~.~~~\eea
Each identity gives a different decomposition of CHY-integrand with
higher-order pole of $s_{i_1i_2\cdots i_k}$, and we need to choose
an appropriate one in the algorithm.

Now let us state the decomposition algorithm. For a generic
CHY-integrand $\mathcal{I}$, we
\begin{enumerate}
  \item Compute the order of poles $\Upsilon[\mathcal{I}]$. If
  $\Upsilon[\mathcal{I}]=0$, return $\mathcal{I}$ itself. If
  $\Upsilon[\mathcal{I}]>0$, list all independent subsets with
  $\chi>0$(assuming there are $m'$)
  \bea
  A'_{\alpha_1}~~,~~A'_{\alpha_2}~~,~~\ldots~~,~~A'_{\alpha_{m'}}~.~~~\eea
  \item Take the first $A'_{\alpha_1}$, and list all $|A'_{\alpha_1}|(n-|A'_{\alpha_1}|)$ cross-ratio
  identities of $s_{\alpha_1}$,
  \bea \mathbb{I}_n[A'_{\alpha_1},j,p]~~~~\mbox{where}~~~~j\in
  A'_{\alpha_1}~~,~~p\in \{1,2,\ldots,n\}\setminus
  A'_{\alpha_1}~.~~~\eea
  \item Decompose the CHY-integrand $\mathcal{I}$ with the first
  cross-ratio identity in step 2,
  \bea \mathcal{I}=\mathcal{I}\times
  \mathbb{I}_n[A'_{\alpha_1},j,p]=\sum_{\ell}
  c_{\ell}\mathcal{I}'_{\ell}~,~~~\eea
  where $\mathcal{I'}$'s are resulting new CHY-integrands, and
  $c_{\ell}$'s are rational functions of Mandelstam variables.
  \item Compute all $\Upsilon[\mathcal{I}']$,
  \begin{itemize}
  \item If all $\Upsilon[\mathcal{I}']<\Upsilon[\mathcal{I}]$,
  return $\sum_{\ell} c_{\ell}\mathcal{I}'_{\ell}$,
  \item If any $\Upsilon[\mathcal{I}']\geq\Upsilon[\mathcal{I}]$, try
  the second cross-ratio identity in step 2 and so on, until
  find a cross-ratio identity satisfying all
  $\Upsilon[\mathcal{I}']<\Upsilon[\mathcal{I}]$. By this
  way, the order of poles of CHY-integrand is at least
  reduced by one.
  \item If we can always find a cross-ratio identity such that all
  $\Upsilon[\mathcal{I}']<\Upsilon[\mathcal{I}]$, then after
  at most $\Upsilon[\mathcal{I}]$ steps, the CHY-integrand
  can be decomposed into terms with only simple poles. This
  happens for some CHY-integrands but not for all. If after
  running over all cross-ratio identities of the pole
  $s_{\alpha_1}$, we still can not find an identity such
  that all $\Upsilon[\mathcal{I}']<\Upsilon[\mathcal{I}]$,
  then start from the first cross-ratio identity in step 2
  again, but now stop at an identity such that all
  $\Upsilon[\mathcal{I}']\leq \Upsilon[\mathcal{I}]$. By
  this way, some of $\mathcal{I}'$ would have the same order
  of poles as $\mathcal{I}$ but different rational functions
  of $z_{ij}$. Anyway we return $\sum_{\ell}
  c_{\ell}\mathcal{I}'_{\ell}$.


  \end{itemize}

\end{enumerate}
After above procedure, we get
$\mathcal{I}=\sum_{\ell}c_{\ell}\mathcal{I}'_{\ell}$. Then let each
$\mathcal{I}'_{\ell}$ go through the procedure recursively, until
all resulting CHY-integrands contain only simple poles. If after
some steps (larger than $\Upsilon[\mathcal{I}]$), the terms of
higher-order poles in resulting CHY-integrands keep growing, then we
shall restart the algorithm again, and choose $A'_{\alpha_2}$ to
start the decomposition, etc. The whole algorithm will end in finite
steps, with the judgement that all $\Upsilon[\mathcal{I}']=0$. Above
algorithm can be easily implemented in {\sc Mathematica}.

\subsection{An illustrative example}

\begin{figure}
  \centering
  \includegraphics[width=2in]{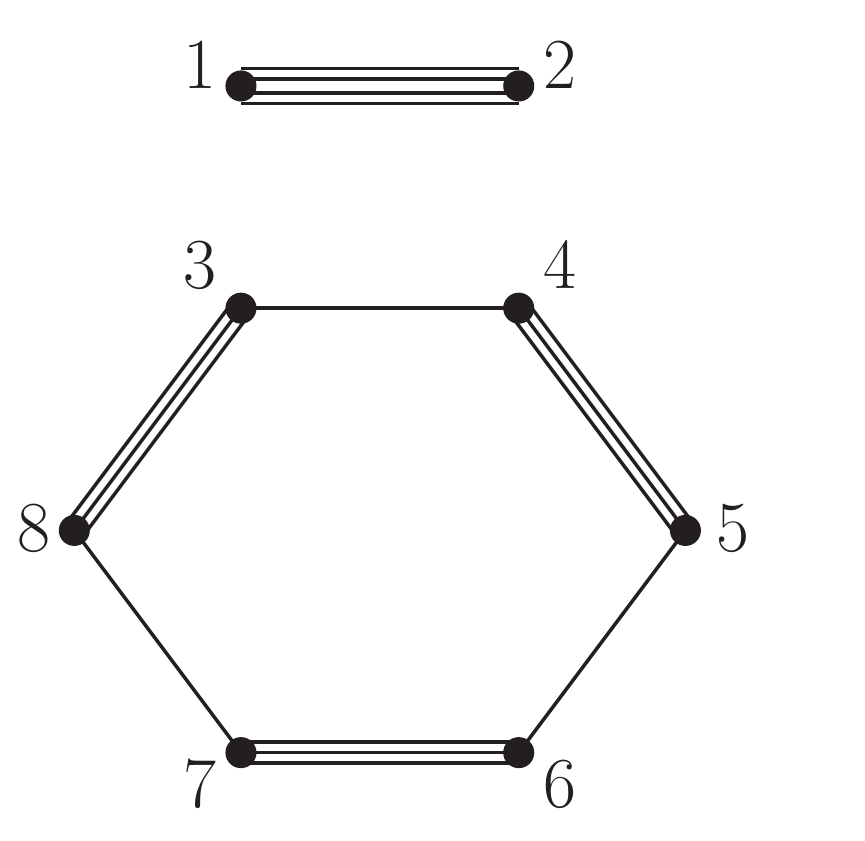}\\
  \caption{The {\sl 4-regular} graph of an eight-point CHY-integrand.}\label{FigA86}
\end{figure}

As a highly non-trivial example to illustrate the above mentioned
algorithm, let us consider the CHY-integrand
\bea \mathcal{I}_8=\frac{1}{z_{12}^4 z_{34} z_{45}^3 z_{56} z_{67}^3
z_{78} z_{38}^3}~,~~~\label{intA86}\eea
with its {\sl 4-regular} graph shown in Figure \ref{FigA86}.
It can be computed that
\bea \Upsilon[\mathcal{I}_8]=8~,~~~\eea
and all subsets with $\chi>0$ are list below,
\bea
\{1,2\}~~,~~\{3,8\}~~,~~\{4,5\}~~,~~\{6,7\}~~,~~\{1,2,3,8\}~~,~~\{1,2,4,5\}~~,~~\{1,2,6,7\}~,~~~\eea
where $\chi_{\{1,2\}}=2$ and $\chi=1$ for the remaining six subsets.
Let us start from subset $\{3,8\}$ and multiply $\mathcal{I}_8$ with
a cross-ratio identity of $s_{38}$\footnote{According to the
algorithm, we have taken the first subset $\{1,2\}$ to start the
decomposition. However after 10 rounds of decompositions, we still
can not get a result with CHY-integrands of only simple poles. Then
we restart the algorithm with the second subset $\{3,8\}$. }. There
are in total $2(8-2)=12$ cross-ratio identities for pole $s_{38}$,
list as follows,
\bea
&&\mathbb{I}_8[\{3,8\},3,1]~~,~~\mathbb{I}_8[\{3,8\},3,2]~~,~~\mathbb{I}_8[\{3,8\},3,4]~~,~~\mathbb{I}_8[\{3,8\},3,5]~~,~~\mathbb{I}_8[\{3,8\},3,6]~~,~~\mathbb{I}_8[\{3,8\},3,7]~,~~~\nonumber\\
&&\mathbb{I}_8[\{3,8\},8,1]~~,~~\mathbb{I}_8[\{3,8\},8,2]~~,~~\mathbb{I}_8[\{3,8\},8,4]~~,~~\mathbb{I}_8[\{3,8\},8,5]~~,~~\mathbb{I}_8[\{3,8\},8,6]~~,~~\mathbb{I}_8[\{3,8\},8,7]~.~~~\nonumber\eea
Each identity contains five terms, so after decomposition we get
five terms
\bea
\mathcal{I}_8=c_1\mathcal{I}'^{[1]}_8+c_2\mathcal{I}'^{[2]}_8+c_3\mathcal{I}'^{[3]}_8+c_4\mathcal{I}'^{[4]}_8+c_5\mathcal{I}'^{[5]}_8~.~~~\eea
Let us take for example the first identity for decomposition, and
compute the order of poles of resulting five terms, as
\bea
\Upsilon[\mathcal{I}'^{[1]}_8]=4~~,~~\Upsilon[\mathcal{I}'^{[2]}_8]=7~~,~~\Upsilon[\mathcal{I}'^{[3]}_8]=7~~,~~\Upsilon[\mathcal{I}'^{[4]}_8]=8~~,~~\Upsilon[\mathcal{I}'^{[5]}_8]=8~.~~~\eea
It dose not satisfy the condition that all
$\Upsilon[\mathcal{I}'^{[i]}_8]<8$, so we look for the next
identity. It can be found that the first identity satisfying this
condition is $\mathbb{I}_8[\{3,8\},3,6]$\footnote{In fact, among the
12 identities, there are four satisfying the request. But since we
only need to find one identity, we do not need to check the
remaining ones when a required one is obtained.}, with which we have
\bea
\Upsilon[\mathcal{I}'^{[1]}_8]=7~~,~~\Upsilon[\mathcal{I}'^{[2]}_8]=7~~,~~\Upsilon[\mathcal{I}'^{[3]}_8]=6~~,~~\Upsilon[\mathcal{I}'^{[4]}_8]=6~~,~~\Upsilon[\mathcal{I}'^{[5]}_8]=4~.~~~\eea
This finishes the {\sl Round 1} decomposition, and we should go
through {\sl Round 2} decomposition with each $\mathcal{I}'^{[i]}_8$
going through the strategy, until all resulting terms satisfying
$\Upsilon[\mathcal{I}'^{[i]}_8]=0$. Below is a table showing the
number $\#[\mbox{ALL}]$ of resulting terms and the number
 $\#[\mbox{H}]$ of terms of higher-order poles in each {\sl Round} decomposition,

\begin{center}
\begin{tabular}{|c|c|c|c|c|c|c|c|c|}
  \hline
  ~ & Round 1 & Round 2 & Round 3 & Round 4 & Round 5 & Round 6 & Round 7 & Round 8 \\
  \hline
  $\#[\mbox{ALL}]$ & 5 & 25 & 121 & 613 & 2779 & 7543 & 9914 & 9922 \\
  \hline
  $\#[\mbox{H}]$ & 5 & 25 & 121 & 464 & 615 & 301 & 2 & 0 \\
  \hline
\end{tabular}
\end{center}

It can be seen from the table that, after four rounds of
decomposition, some resulting terms are already those of simple
poles. After five rounds of decomposition, terms of higher-order
poles start decreasing, until to the {\sl round 8} decomposition,
where all terms of higher-order poles are reduced, leaving 9922
CHY-integrands of simple poles. Computing these 9922 terms via
integration rules of simple poles takes a few minutes by {\sc
Mathematica} in a laptop, and the result is confirmed numerically.

This algorithm can be applied to higher-$n$ scattering process
without difficulty. The efficiency of decomposition mostly depends
on the number of terms of higher-order poles in each round
decomposition but not the number $n$ of scattering points. If
$\Upsilon$ of original CHY-integrand is not very large, the
algorithm can be easily finished in a short time. However, if
$\Upsilon$ is large (for instance $\Upsilon>10$), the decomposition
can still proceed, but might take some time.

\section{The $\Lambda$-Algorithm and the Cross-Ratio Identities}\label{secLalgorithm}

In the previous section, we apply the cross-ratio identities to the
systematic decomposition of CHY-integrand with higher-order poles.
For CHY-integrands with large $n$ and $\Upsilon$, the resulting
terms of simple poles can easily reach a number of millions, hence
slow the computation. For those CHY-integrands where decomposition
algorithm is significantly slow, we can nevertheless combine the
cross-ratio identities with $\Lambda$-algorithm, to pursue a more
efficient realization. In this section, we will describe such a
combination.

The $\Lambda$-algorithm has been recently developed by one of the
authors to compute CHY-integrands. It has some interesting features
as it can support up to three off-shell particles as well as it
factorizes the original graph representing the CHY-integrand into
sub-graphs with less number of vertices by means of an iterative
algorithm. Nevertheless, it depends on the gauge-fixing and it does
not work on singular configurations. At some point on the iterative
process one usually reach some sub-graphs containing those singular
configurations and then the algorithm cease to work. Here we show
how the $\Lambda$-algorithm is improved by using the cross-ratio
identities on graphs containing singular configurations. In
addition, we find some recurrence relations for particular types of
CHY-integrands. Ultimately, the cross-ratio identities in
conjunction with the $\Lambda$-algorithm provide a more efficient
and systematic way to deal with amplitudes with a large number of
particles.

\subsection{Some notations}\label{secNotations}

For reader's convenience, let us briefly introduce some notations
here, which will be useful in the computation of some non-trivial
examples soon after. We define the {\sl stripped} Mandelstam
variables $\widetilde{s}_{a_1\ldots a_m}$ as
\begin{align}
\widetilde{s}_{a_1\ldots a_m}:=\sum_{a_i<a_j}^m k_{a_i}\cdot k_{a_j}~,~~~
\end{align}
which equals to the standard Mandelstam variables $s_{a_1\ldots
a_m}=(k_{a_1}+\cdots +k_{a_m})^2/2$ when all $k_{a_i}$'s are
massless. We also follow the conventional definition
\bea k_{a_1\ldots a_m}:=k_{a_1}+k_{a_2}+\cdots + k_{a_m}~,~~~ \eea
and whenever $[a_1,\ldots,a_m]$ is present in $\mathcal{I}(z)$ or
$\widetilde{s}$, it stands for the punctured point merged from
points $\{z_{a_1},\ldots, z_{a_m}\}$ in  graph
associated to the massive momentum $k_{a_1\ldots a_m}$. The colored
notation in \cite{Gomez:2016bmv} which is needed to apply the
$\Lambda$-algorithm is adopted,
\begin{center}
\includegraphics[width=5.5in]{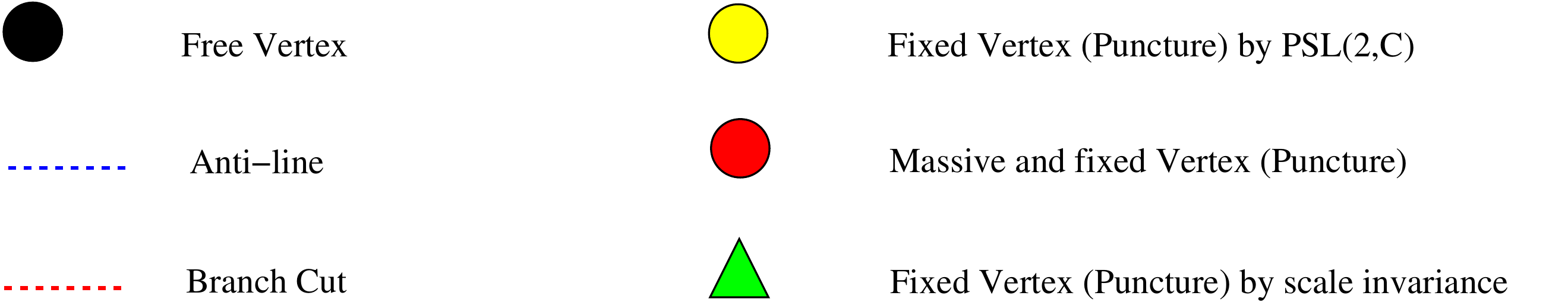}.
\end{center}
and for details please refer to \cite{Gomez:2016bmv}.

All the results obtained from the $\Lambda$-algorithm can be written
as a linear combination of the following fundamental diagrams and
its powers
\begin{equation}\label{4_Fundamental}
\begin{aligned}
B(a,b|c,d)~~~~:=\raisebox{-15mm}{\includegraphics[keepaspectratio = true, scale = 0.35] {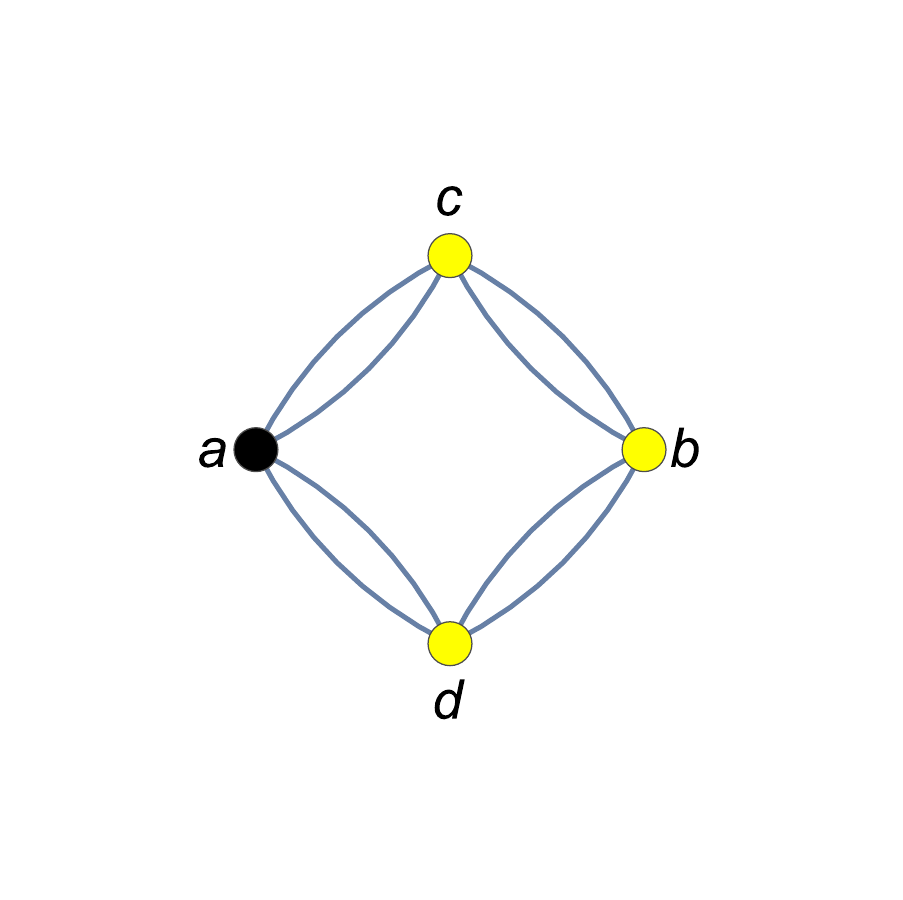}}
=~~~ \frac{1}{\widetilde{s}_{ac}}+ \frac{1}{\widetilde{s}_{ad}}~,~~~
\end{aligned}
\end{equation}
where $\{k_b, k_c,k_d\}$  can be off-shell particles. Clearly,
$B(a,b|c,d)=B(a,b|d,c)$.

It is simple to check, using the  ${\cal E}_a$ scattering equation,
that
\begin{equation}\label{4_points_one}
\begin{aligned}
&\raisebox{-15mm}{\includegraphics[keepaspectratio = true, scale = 0.35] {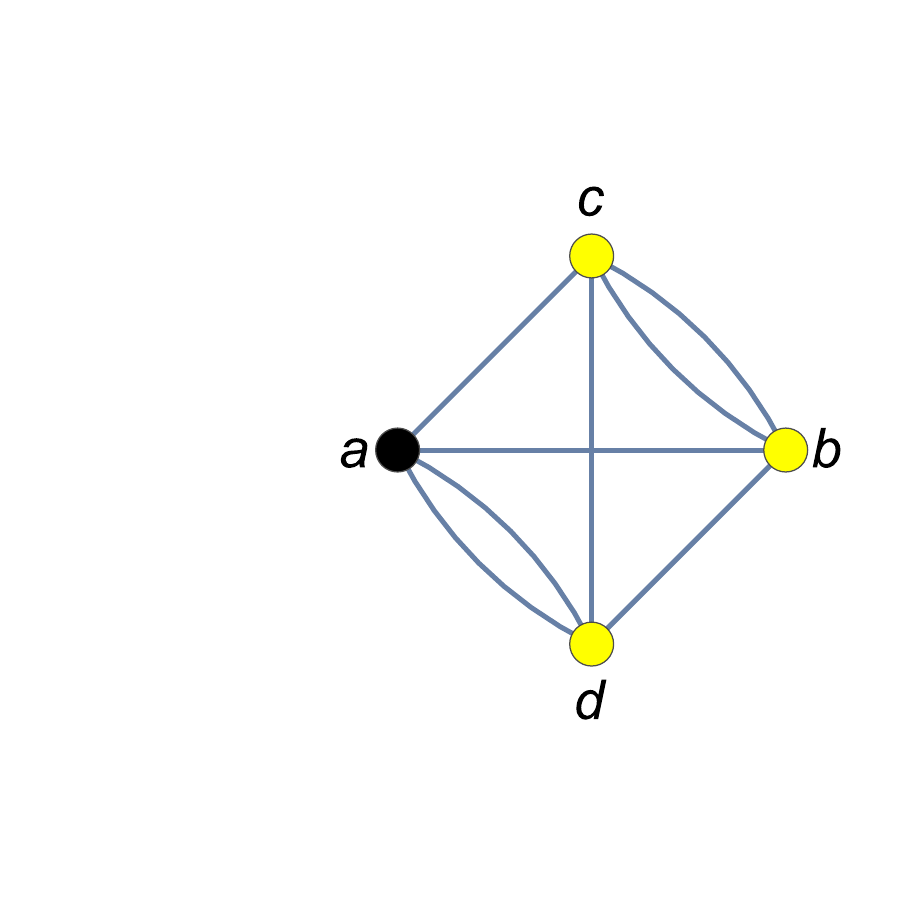}}
=~-\frac{\widetilde{s}_{ac}}{\widetilde{s}_{ab}} \,B(a,b|c,d)\, ,
\raisebox{-15mm}{\includegraphics[keepaspectratio = true, scale = 0.35] {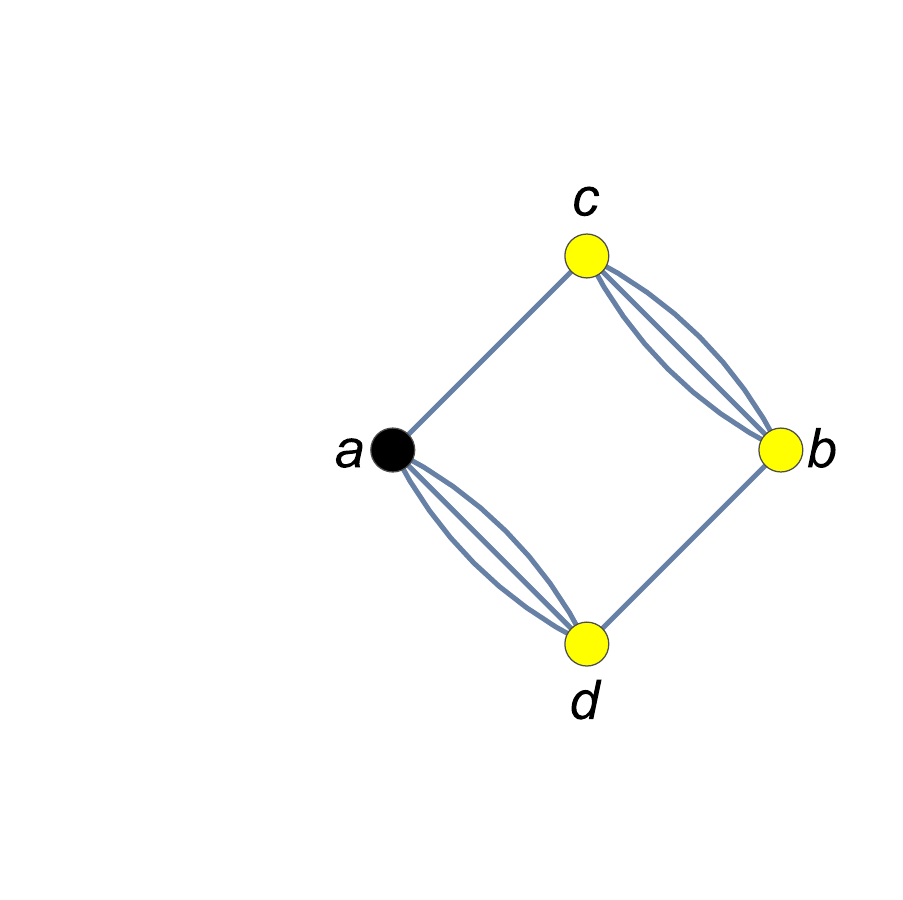}}
=~-\frac{\widetilde{s}_{ac}}{\widetilde{s}_{ad}} \,B(a,b|c,d)\, ,\\
&\raisebox{-15mm}{\includegraphics[keepaspectratio = true, scale = 0.35] {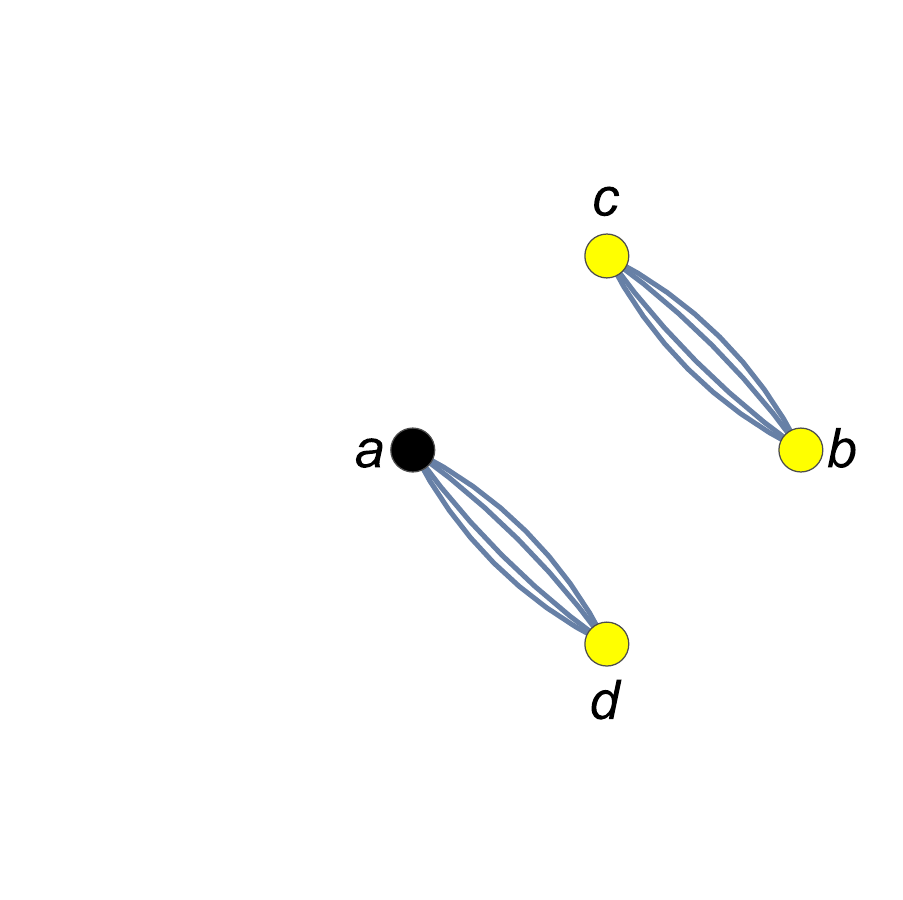}}
=~\left(\frac{\widetilde{s}_{ac}}{\widetilde{s}_{ad}}\right)^2 \,B(a,b|c,d)\,\, ,
\end{aligned}
\end{equation}
where $\{k_b, k_c,k_d\}$  could be off-shell particles.

\subsection{A simple example}\label{secSimpleexample}

Before a general discussion, let us start from a simple but
non-trivial six-point example with the following geometry $({\rm
Parke-Taylor})^2\oplus ({\rm Parke-Taylor})^2$ as
\begin{equation}\label{example6pts_3}
\begin{aligned}
{\cal I}_6(1,2,3|4,5,6)~~~=
\raisebox{-17mm}{\includegraphics[keepaspectratio = true, scale = 0.43] {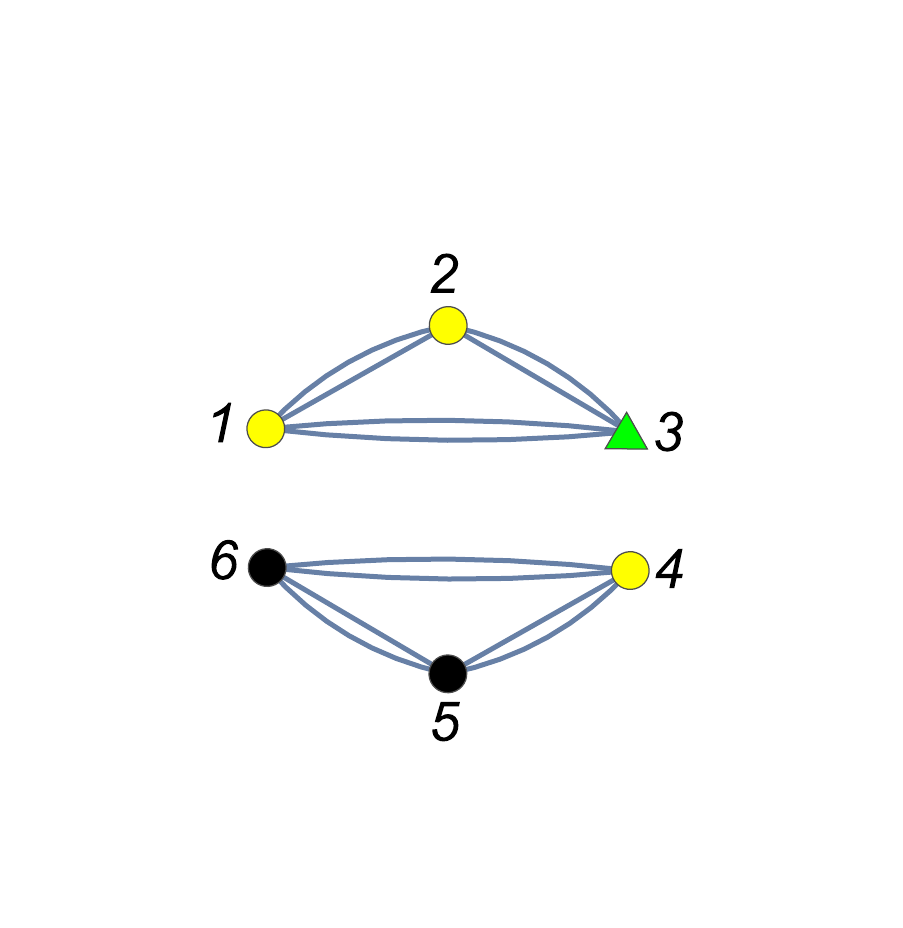}} ,
\end{aligned}
\end{equation}
where we have set the gauge-fixing so as to avoid singular
configurations \cite{Gomez:2016bmv}. All  non-zero allowable
configurations for this diagram are given by \cite{Gomez:2016bmv}
\begin{equation}\label{example6pts_3_C}
\begin{aligned}
&\raisebox{-18mm}{\includegraphics[keepaspectratio = true, scale = 0.4] {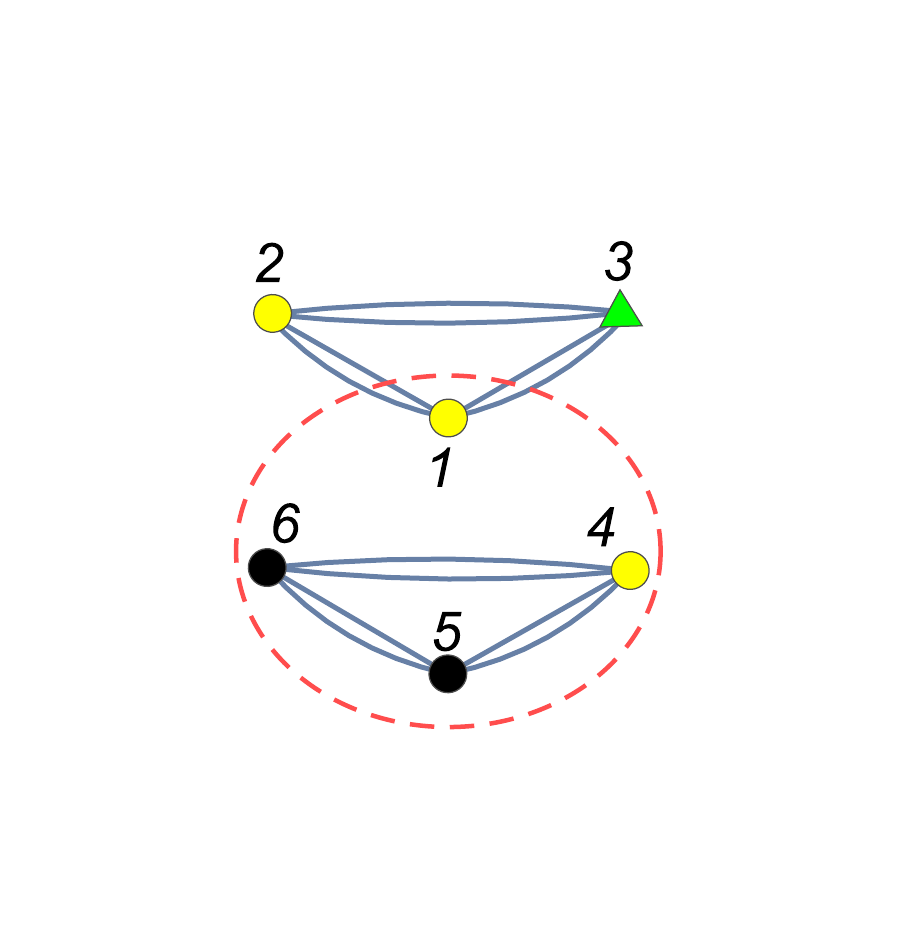}}
= ~~\frac{1}{\widetilde{s}_{23}}
\raisebox{-17mm}{\includegraphics[keepaspectratio = true, scale = 0.4] {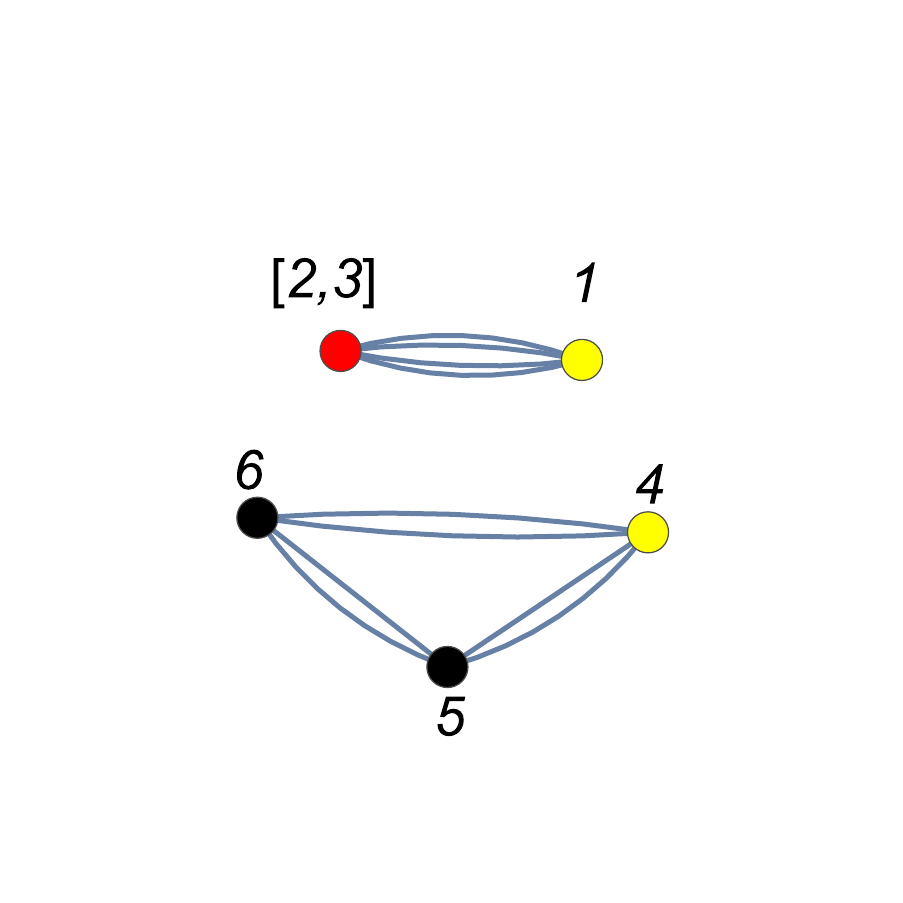}},
\raisebox{-18mm}{\includegraphics[keepaspectratio = true, scale = 0.4] {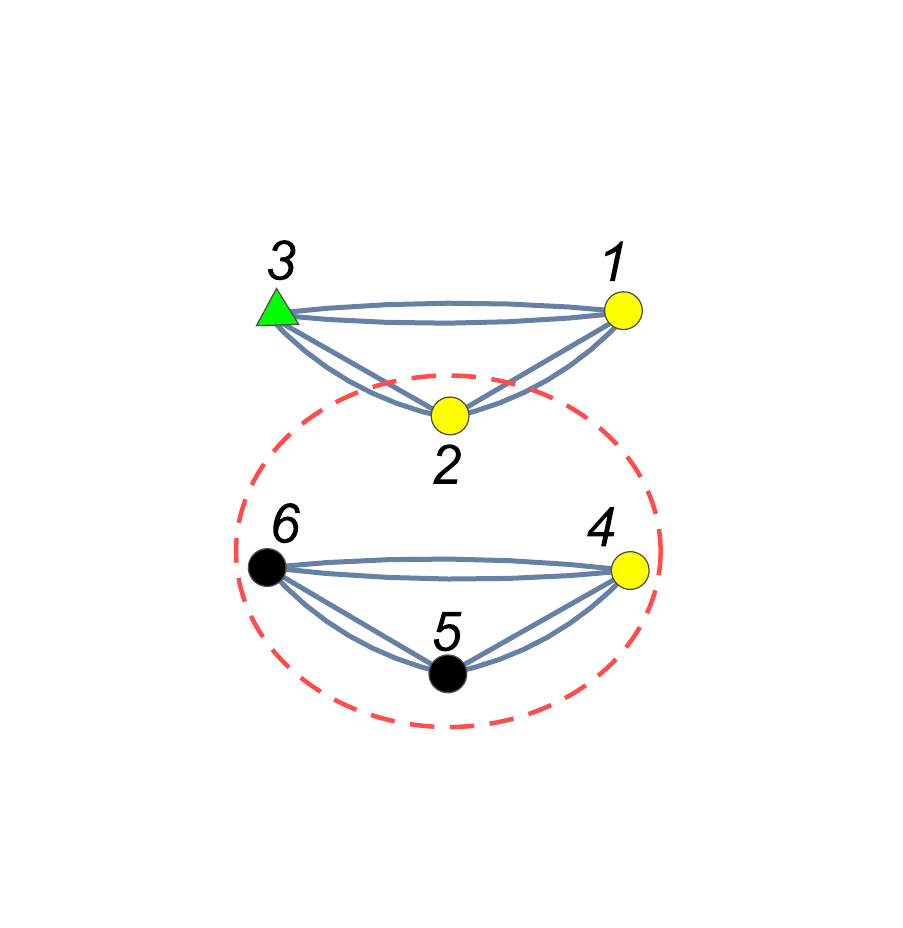}}
= ~~\frac{1}{\widetilde{s}_{13}}
\raisebox{-17mm}{\includegraphics[keepaspectratio = true, scale = 0.4] {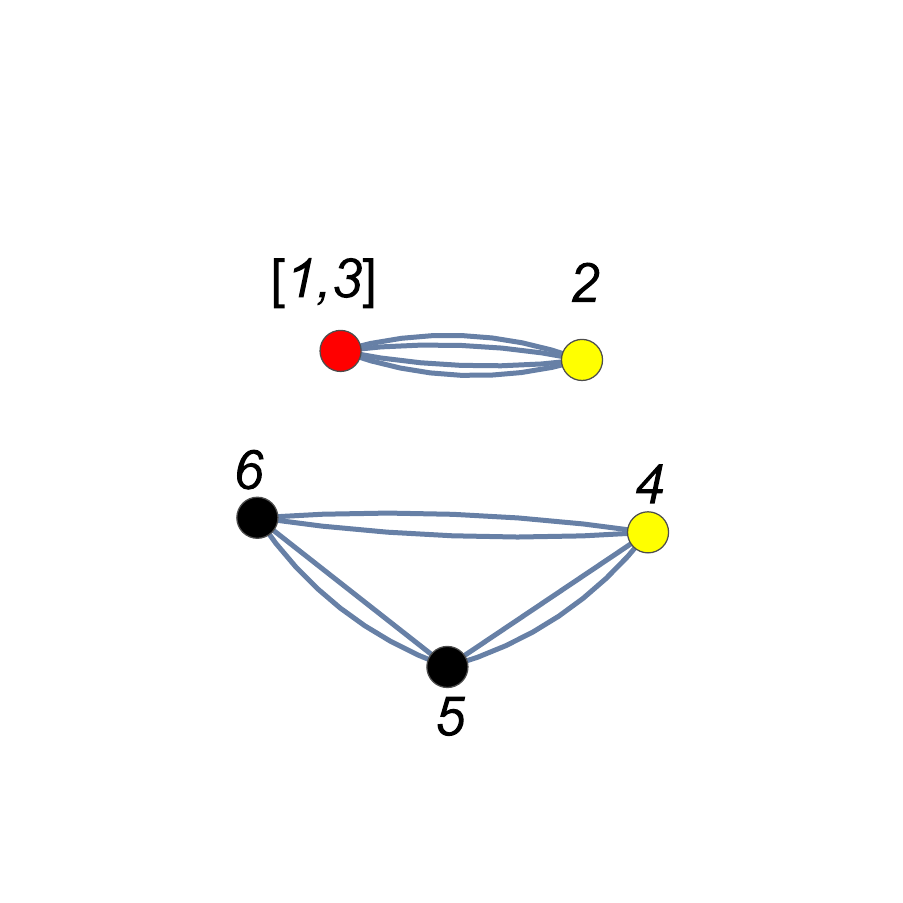}},\nonumber\\
&\raisebox{-18mm}{\includegraphics[keepaspectratio = true, scale = 0.4] {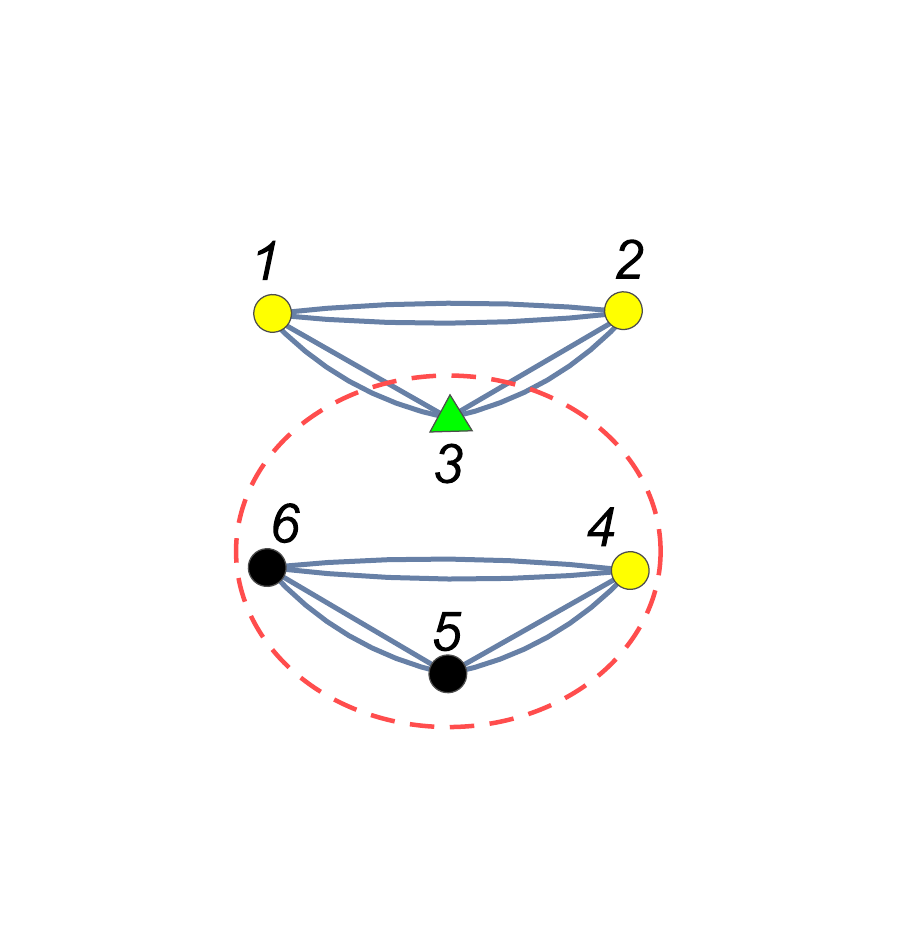}}
= ~~\frac{1}{\widetilde{s}_{3456}}
\raisebox{-17mm}{\includegraphics[keepaspectratio = true, scale = 0.4] {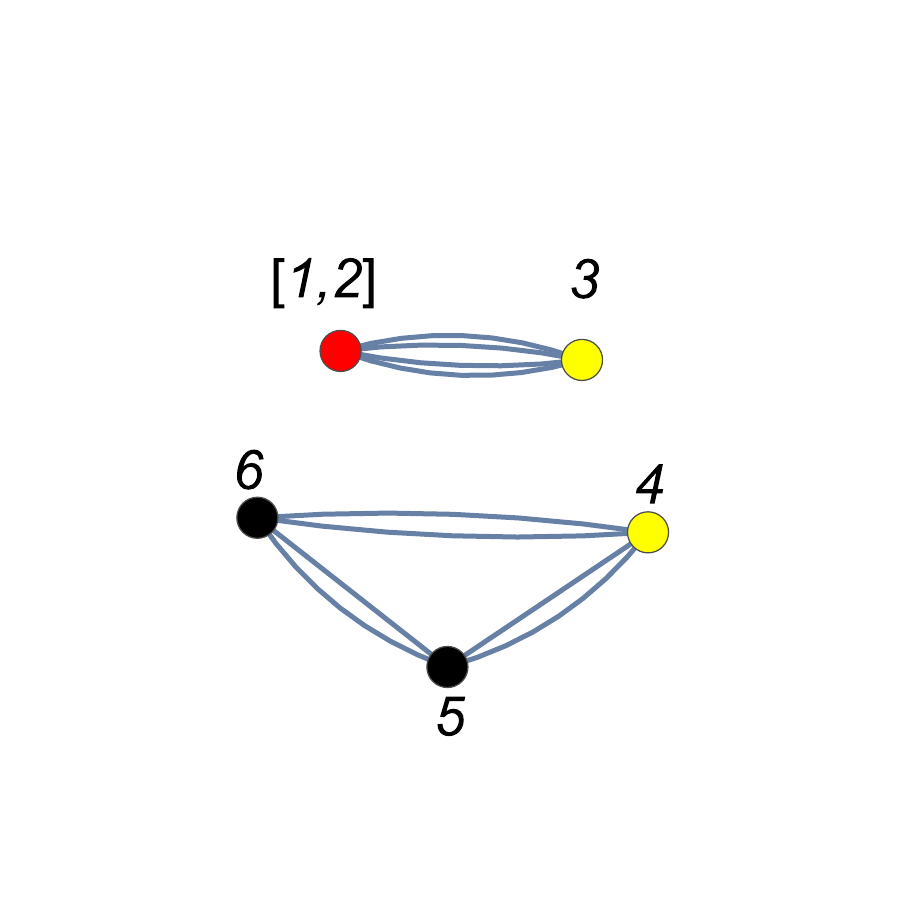}}.
\end{aligned}
\end{equation}
The $\Lambda$-algorithm stops at this step, since we have reached
sub-diagrams containing singular configurations
\begin{equation}\label{example5pts_3}
\begin{aligned}
{\cal I}_5(a,b|c,d,e)~~~=\raisebox{-15mm}{\includegraphics[keepaspectratio = true, scale = 0.4] {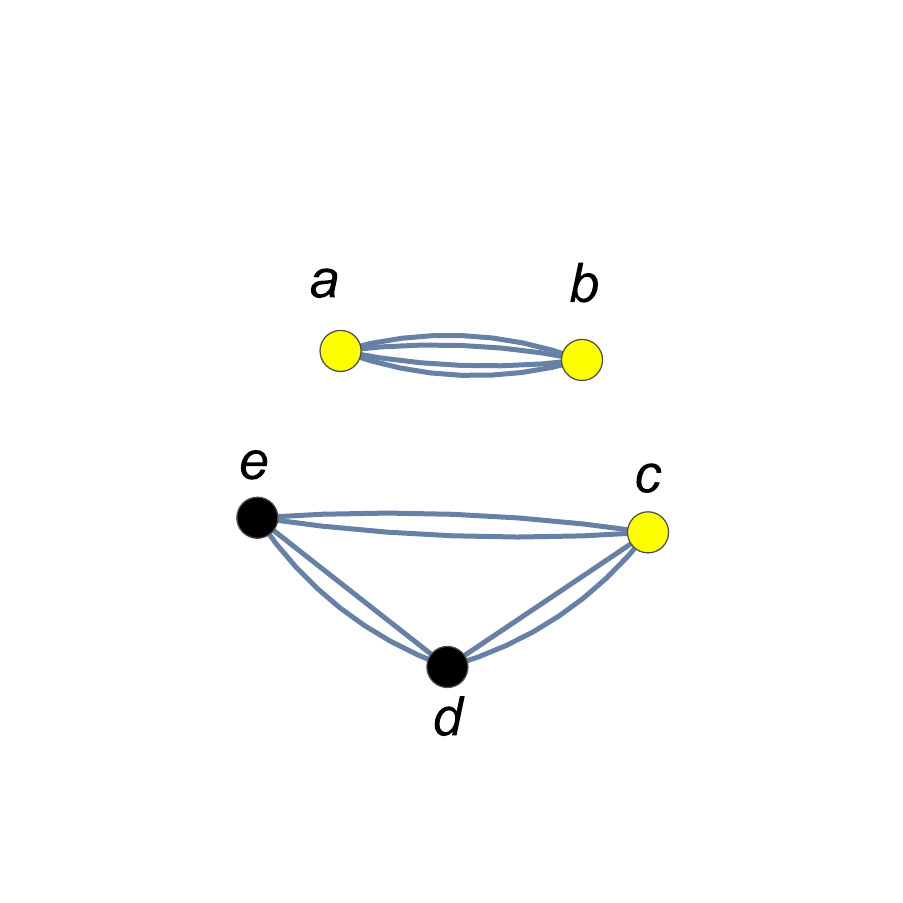}},
\end{aligned}
\end{equation}
which needs to be rewritten in terms of non-singular sub-diagrams by
using the cross-ratio identities, as we shall explain soon after.
Let us remember that in this graph, $\{k_a,k_b,k_c \}$ can be
off-shell.

First of all, notice that only the ${\cal E}_d$ and ${\cal E}_e$
scattering equations can be used, since the remaining points are
already fixed.  Moreover, clearly this graph has a triple pole,
$1/\widetilde{s}_{cde}^{~3}$. So, it is simple to show that using
the ${\cal E}_d$ and ${\cal E}_e$ scattering equations one obtains
the following cross-ratio identity
\begin{equation}
\widetilde{s}_{cde}=\widetilde{s}_{ad}\left( \frac{z_{ab} z_{cd} }{z_{bc} z_{ad} }   \right) + \widetilde{s}_{ae}\left( \frac{z_{ab} z_{ce} }{z_{bc} z_{ae} }   \right)~,~~~
\end{equation}
which agrees with \eqref{generalCR}. We can represent the identity
above by the  graph
\begin{equation}\label{CRidentity5}
\begin{aligned}
\widetilde{s}_{cde}~~=~~\widetilde{s}_{ad}
\raisebox{-17mm}{\includegraphics[keepaspectratio = true, scale = 0.38] {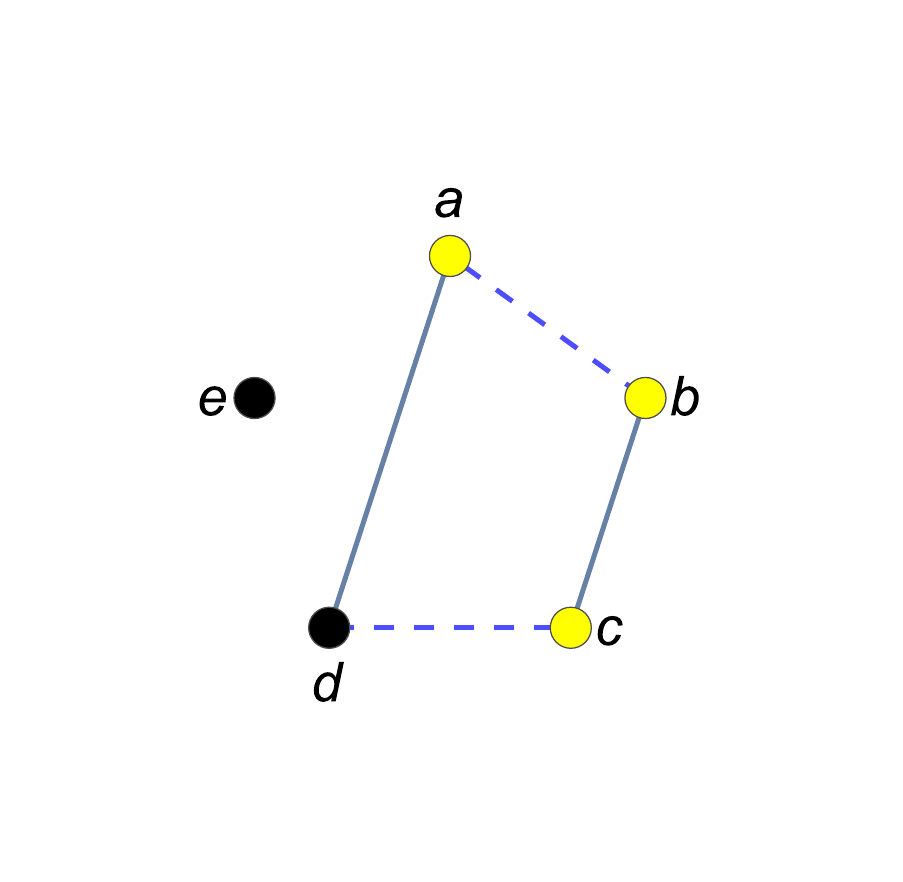}}
+~~\widetilde{s}_{ae}
\raisebox{-17mm}{\includegraphics[keepaspectratio = true, scale = 0.38] {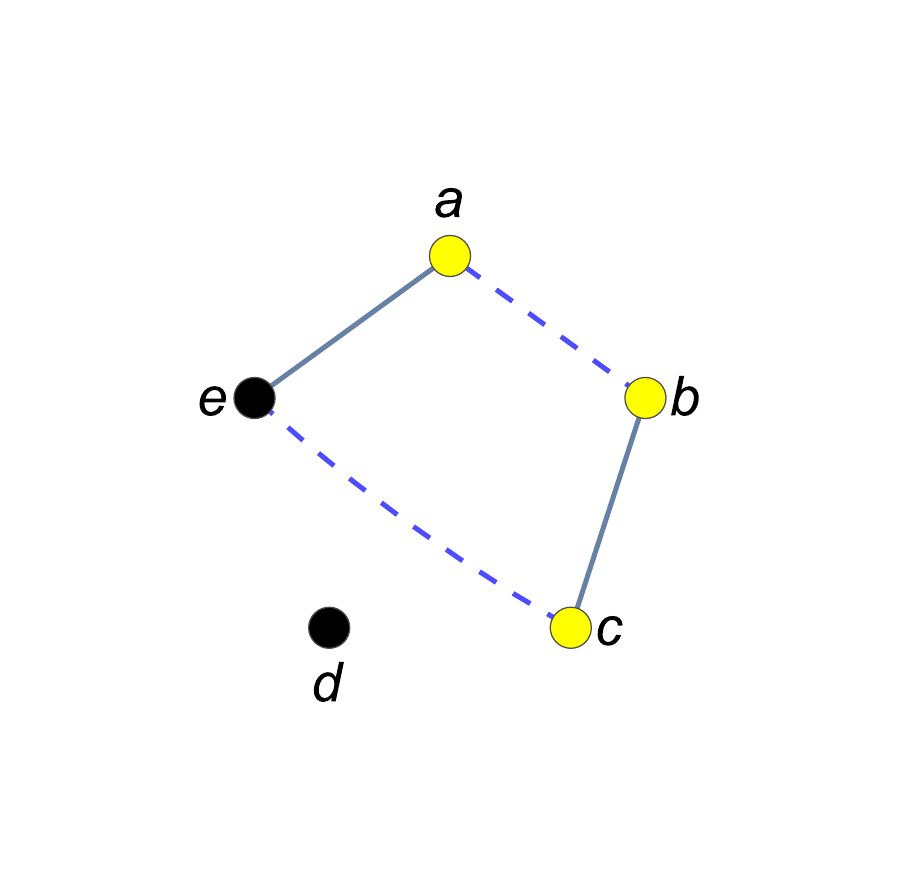}}.
\end{aligned}
\end{equation}
In order to eliminate the triple pole we take the square of
\eqref{CRidentity5}
\begin{equation}\label{CR2identity5}
\begin{aligned}
\widetilde{s}_{cde}^2~~=~~\widetilde{s}_{ad}^2
\raisebox{-17mm}{\includegraphics[keepaspectratio = true, scale = 0.38] {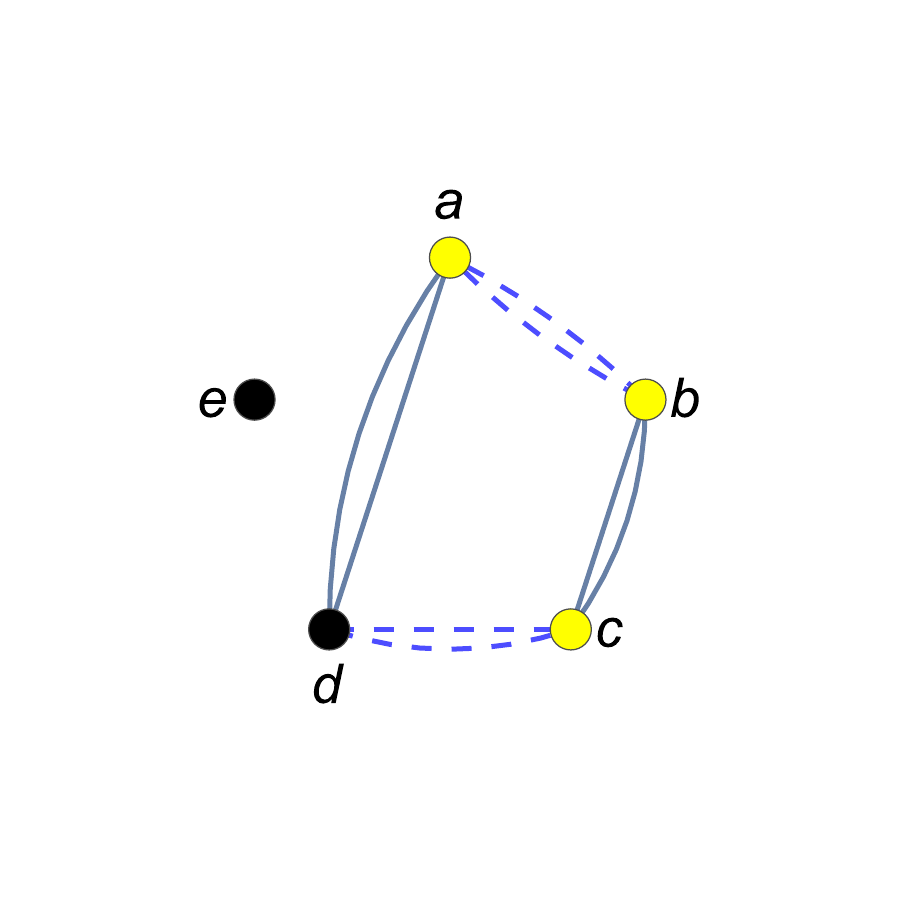}}
+~~\widetilde{s}_{ae}^2
\raisebox{-17mm}{\includegraphics[keepaspectratio = true, scale = 0.38] {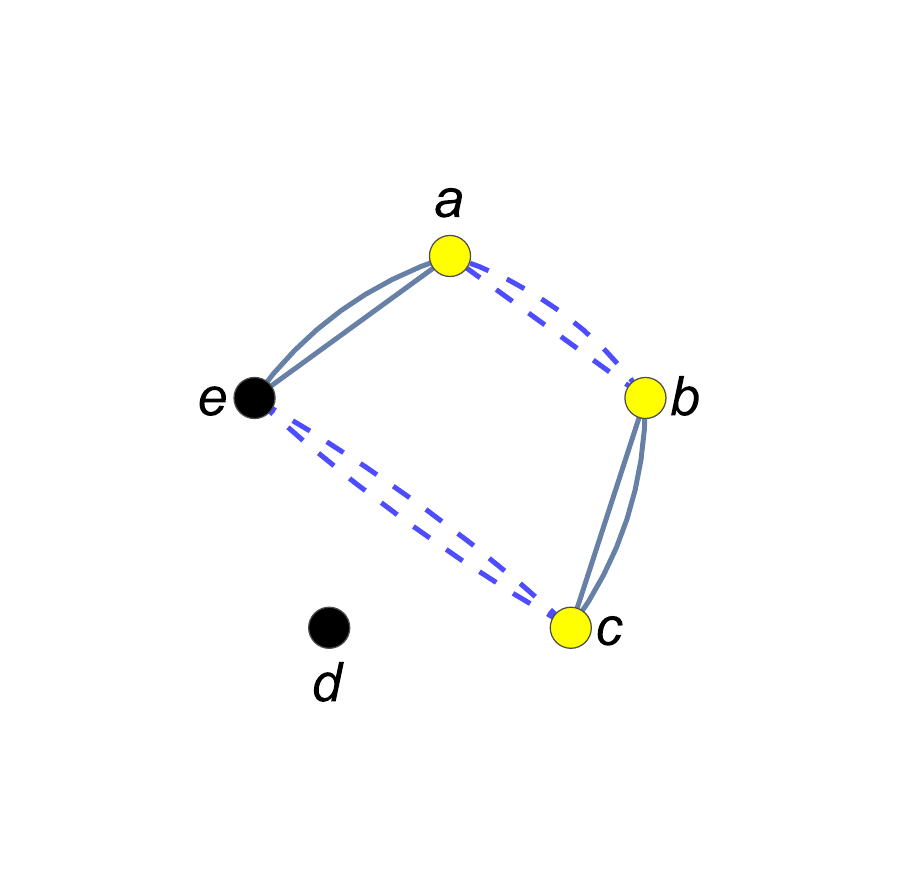}}
+~~2\widetilde{s}_{ad}\widetilde{s}_{ae}
\raisebox{-17mm}{\includegraphics[keepaspectratio = true, scale = 0.38] {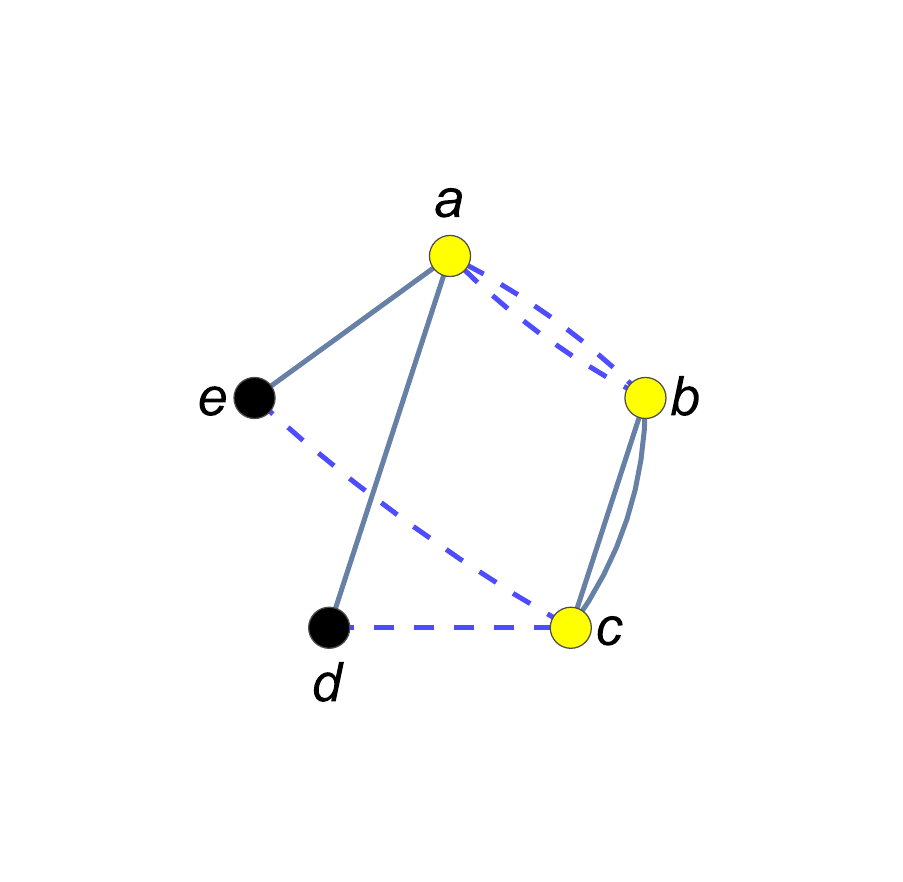}}.
\end{aligned}
\end{equation}
Therefore, using this identity over the CHY diagram in
\eqref{example5pts_3}, we obtain the expansion
\begin{equation*}\label{example5pts_4}
\begin{aligned}
\raisebox{-15mm}{\includegraphics[keepaspectratio = true, scale = 0.38] {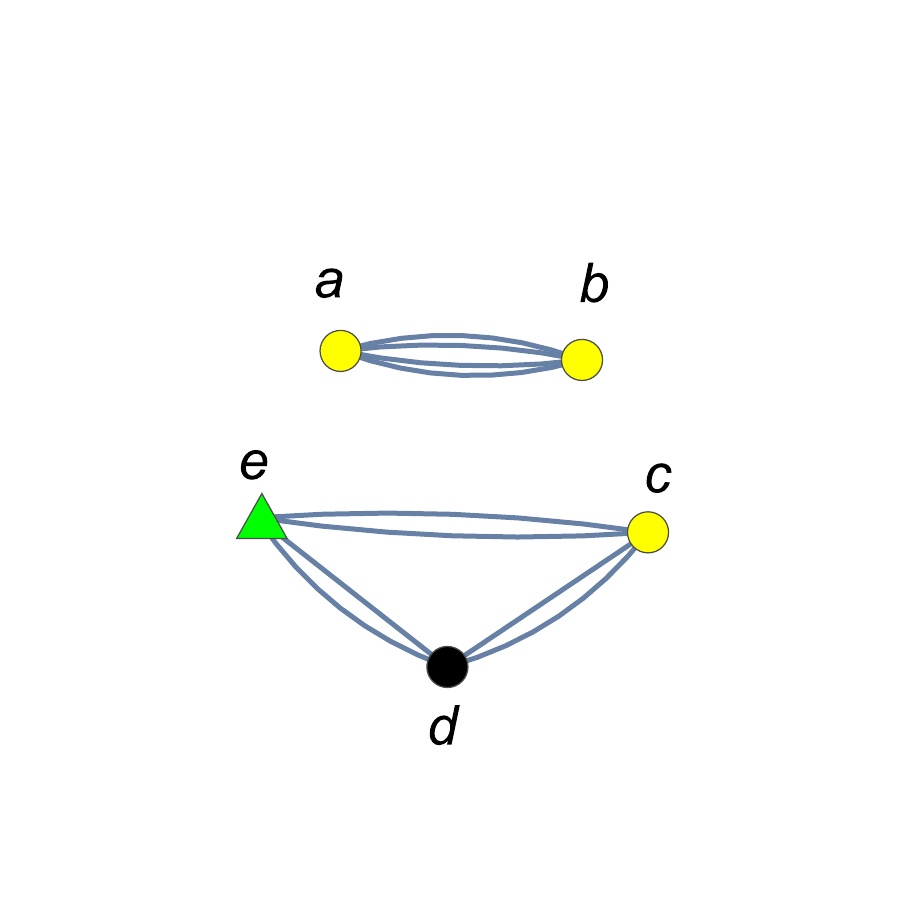}}
=\frac{\widetilde{s}_{ad}^2}{\widetilde{s}_{cde}^2}
\raisebox{-15mm}{\includegraphics[keepaspectratio = true, scale = 0.38] {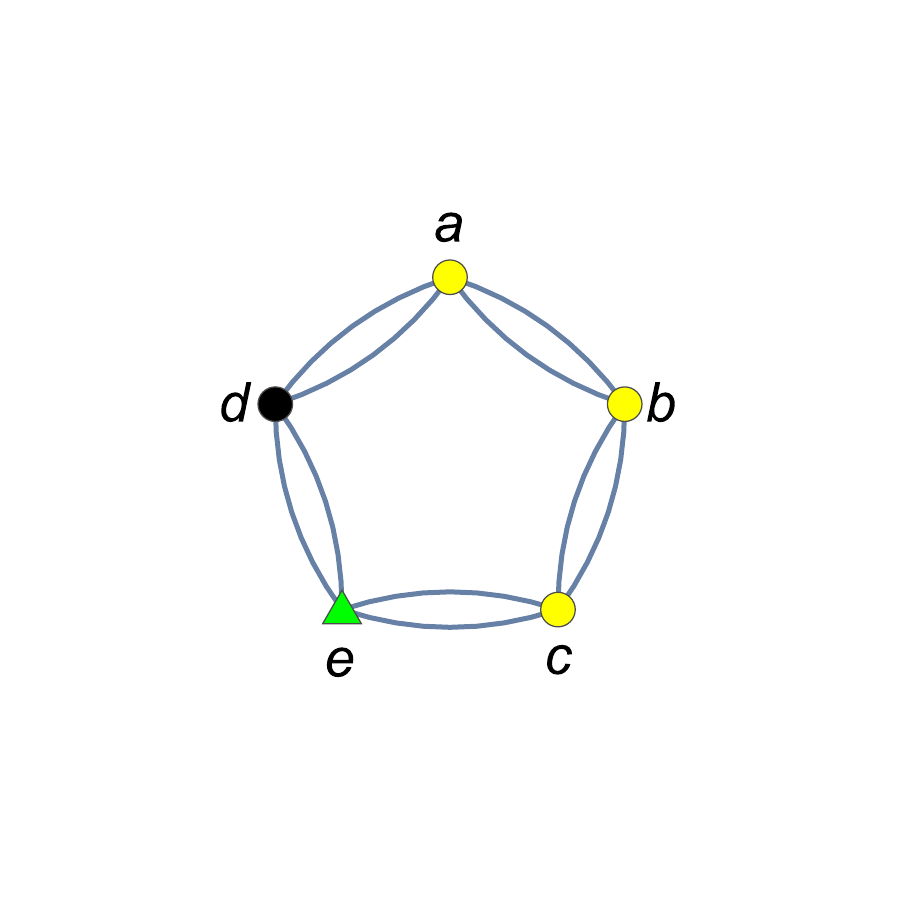}}+
\frac{\widetilde{s}_{ae}^2}{\widetilde{s}_{cde}^2}
\raisebox{-15mm}{\includegraphics[keepaspectratio = true, scale = 0.38] {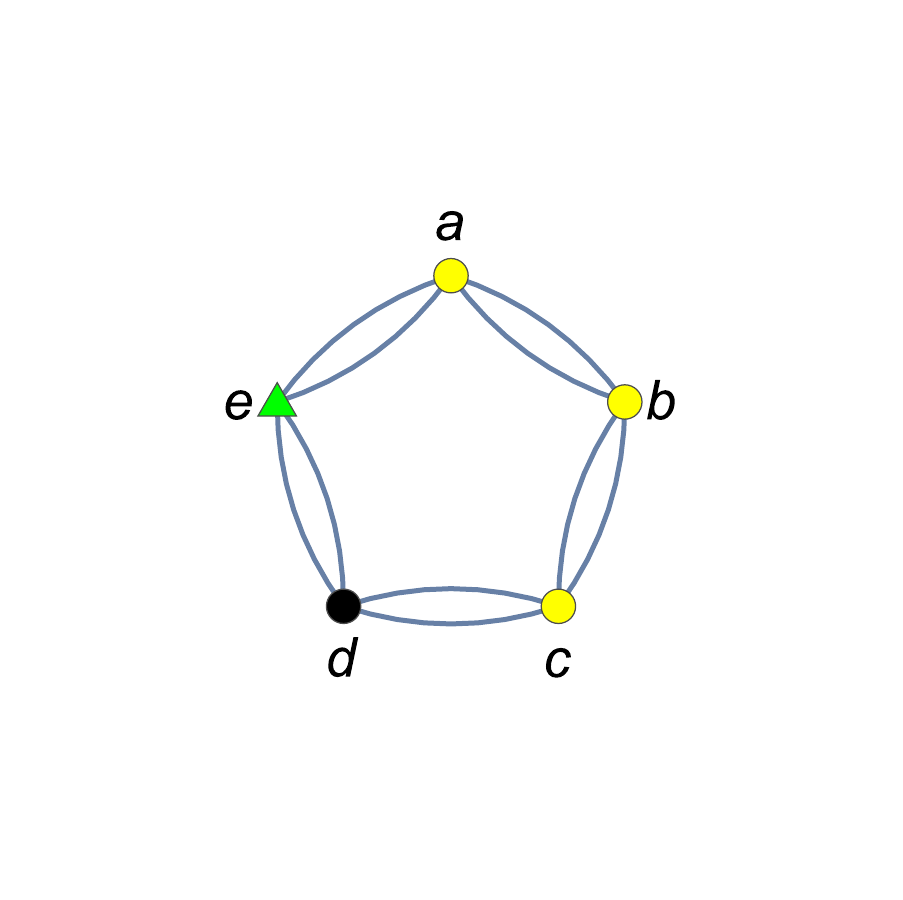}}+
\frac{2\,\widetilde{s}_{ad}\, \widetilde{s}_{ae}}{\widetilde{s}_{cde}^2}
\raisebox{-15mm}{\includegraphics[keepaspectratio = true, scale = 0.38] {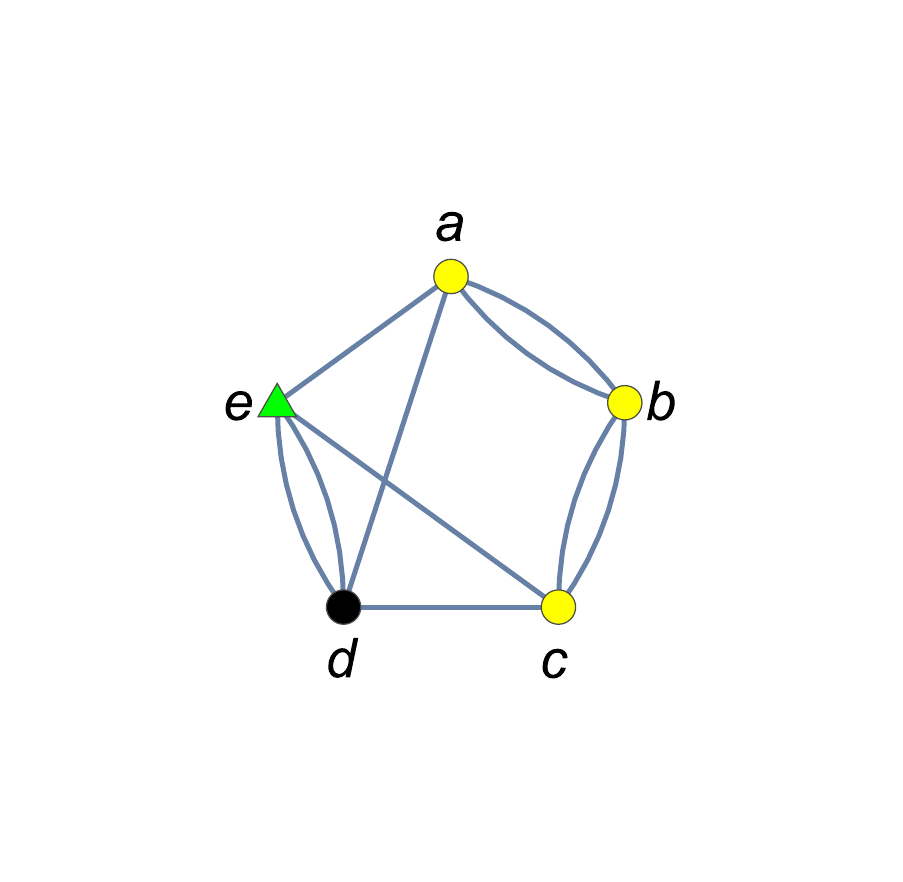}}
\end{aligned}
\end{equation*}
These three graphs are  computed straightforwardly from the
$\Lambda$-algorithm resulting in,
\begin{align}\label{I_2_3}
{\cal I}_5 (a,b|c,d,e) &= \frac{\widetilde{s}_{ad}^2}{\widetilde{s}_{cde}^2}\left( \frac{B(d,c|[a,b],e)}{\widetilde{s}_{cde}}+ \frac{B(d,[b,c]|a,e)}{\widetilde{s}_{dea}}   +  \frac{B(d,b|a,[c,e])}{\widetilde{s}_{ce}}  \right)\\
& +
\frac{\widetilde{s}_{ae}^2}{\widetilde{s}_{cde}^2}\left( \frac{B(d,[a,b]|c,,e)}{\widetilde{s}_{cde}}+ \frac{B(d,a|[b,c],e)}{\widetilde{s}_{dea}} + \frac{B(d,b|c,[a,e])}{\widetilde{s}_{ae}}  \right) \nonumber \\
&+
\frac{2\,\widetilde{s}_{ae}\, \widetilde{s}_{ad}}{\widetilde{s}_{cde}^2}\left( -\frac{ \widetilde{s}_{dc}  }{\widetilde{s}_{cde}\, \widetilde{s}_{d[a,b]}}B(d,[a,b]|c,e)-\frac{ \widetilde{s}_{d[b,c]}  }{\widetilde{s}_{dea}\, \widetilde{s}_{da}}B(d,a|[b,c],e)  \right)\nonumber~.~~~
\end{align}
Hence, the final answer for the CHY-integrand in
(\ref{example6pts_3}) is given in terms of ${\cal I}_5 (a,b|c,d,e) $
by the simple expression
\begin{align}
{\cal I}_6(1,2,3|4,5,6)=\frac{{\cal I}_{5}([2,3],1|4,5,6)}{\widetilde{s}_{23}}+\frac{{\cal I}_{5}([1,3],2|4,5,6)}{\widetilde{s}_{13}}+\frac{{\cal I}_{5}([1,2],3|4,5,6)}{\widetilde{s}_{3456}}~,~~~
\end{align}
which was checked numerically.

We remark that, for this particular example where we have combined
the $\Lambda$-algorithm with the cross-ratio identities, we solved a
total of three CHY-integrands.  On the other hand, applying directly
the cross-ratio identities over the CHY-integrand in
(\ref{example6pts_3}), one must compute ten CHY graphs, So, although
the decomposition technique with the cross-ratio identities is a
good method by itself, in combination with the $\Lambda$-algorithm
it produces an even more efficient approach. The simple example
above is enough to show how both methods work out together, and
instead of going into harder examples we choose to present a new
recurrence relation that can be obtained from this combination.

\subsection{$({\rm Parke}-{\rm Taylor)^2}\oplus({\rm Parke}-{\rm Taylor)^2}$ geometry and recurrence relations}

In this subsection, we would like to generalize the discussion in
previous subsection to a particular geometry given by two ${\rm
PT^2}$ (i.e., ${\rm PT^2}\oplus {\rm PT^2}$) CHY-integrands, e.g.,
\begin{equation}\label{examplePTPT}
\begin{aligned}
\raisebox{-1mm}{\includegraphics[keepaspectratio = true, scale = 0.5] {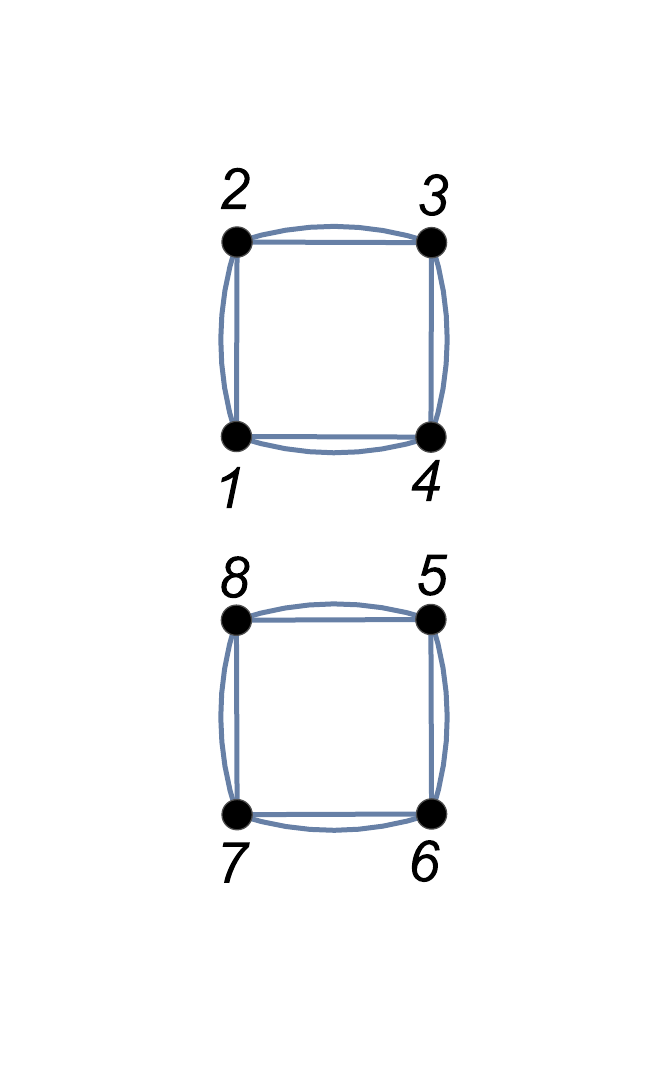}}
\raisebox{-1mm}{\includegraphics[keepaspectratio = true, scale = 0.5] {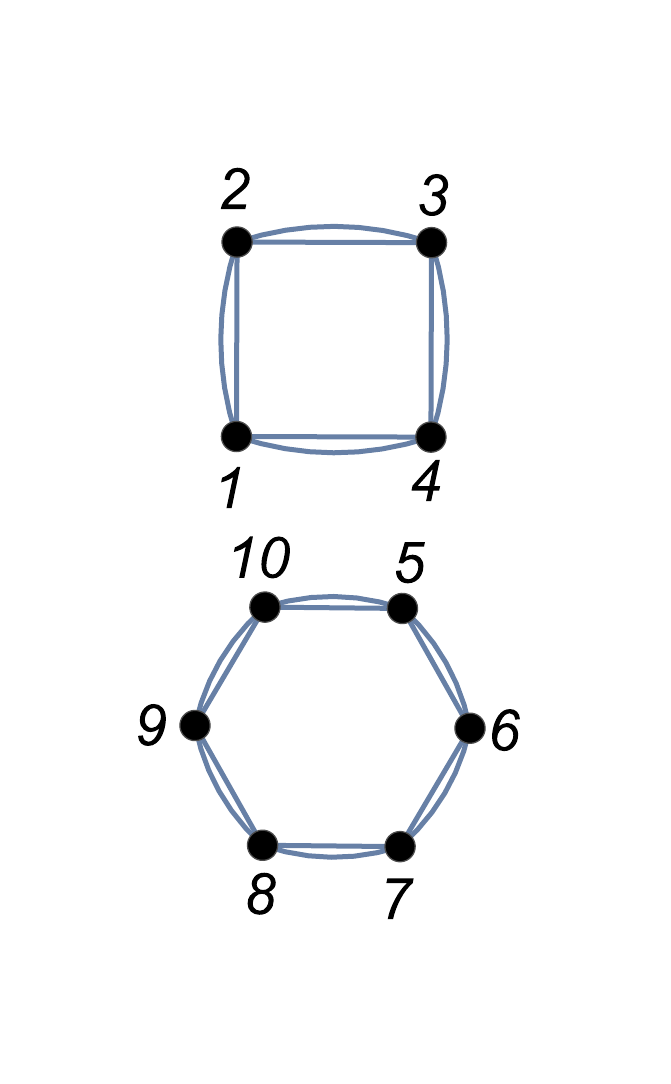}}
\raisebox{-4mm}{\includegraphics[keepaspectratio = true, scale = 0.5] {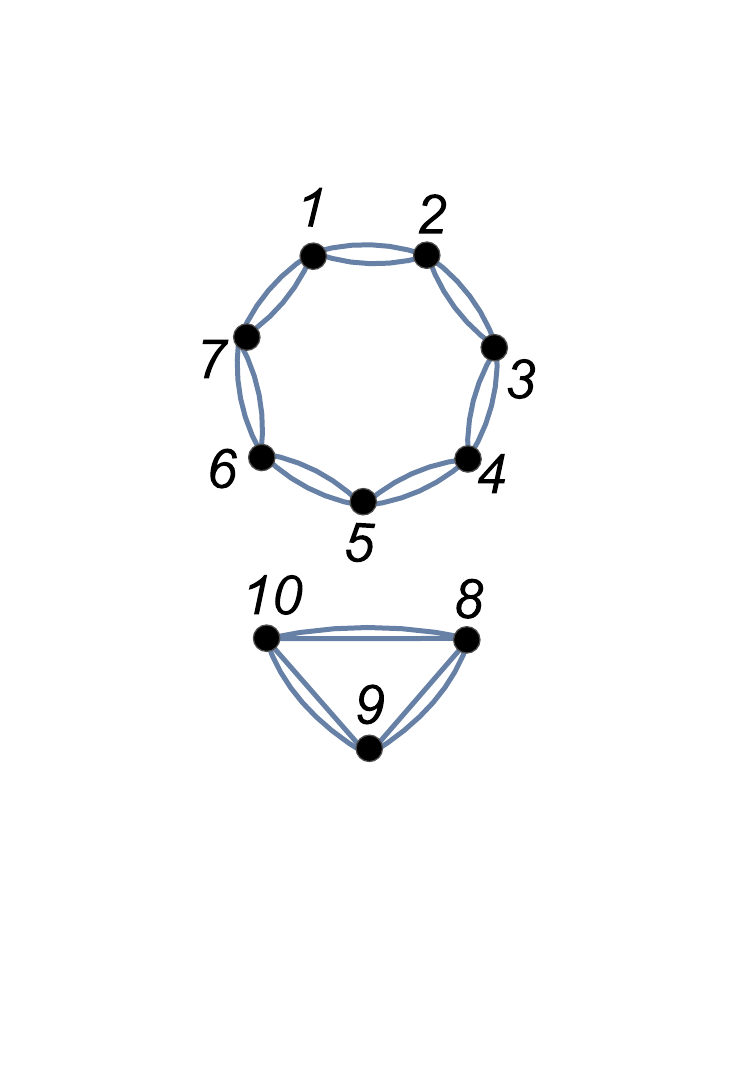}}
\end{aligned}
\end{equation}
We denote this type of graphs by ${\cal I}^{{\rm PT^2}\oplus {\rm
PT^2}}_{m,n-m}(1,\ldots,m\,|\,m+1,\ldots,n)$, where $m$  is the
number of vertices for the first ${\rm PT^2}$ sub-graph and $n-m$ is
the number of  vertices for the second one. For instance, the graphs
given in \eqref{examplePTPT} are denoted as ${\cal I}^{{\rm
PT^2}\oplus {\rm PT^2}}_{4,4}(1,2,3,4\,|\,5,6,7,8)$, ${\cal I}^{{\rm
PT^2}\oplus {\rm PT^2}}_{4,6}(1,2,3,4\,|\,5,6,7,8,9,10)$ and  ${\cal
I}^{{\rm PT^2}\oplus {\rm PT^2}}_{7,3}(1,2,3,4,5,6,7\,|\,8,9,10)$,
respectively. Note also that for the graphs in \eqref{example6pts_3}
and \eqref{example5pts_3}  one has ${\cal I}_{6}(1,2,3|4,5,6)={\cal
I}^{{\rm PT^2}\oplus {\rm PT^2}}_{3,3}(1,2,3\,|\,4,5,6)$ and ${\cal
I}_{5}(a,b |c,d,e)={\cal I}^{{\rm PT^2}\oplus {\rm
PT^2}}_{2,3}(a,b\,|\,c,d,e)$, respectively.

It is very well-known that this type of CHY-integrands is highly
non-trivial and one can solve them from the cross-ratio
identities\footnote{In fact, if there is only one higher-order
poles, either double or triple pole, the decomposition algorithm
with cross-ratio identities can produce the result instantly for any
points.}. However, the price to pay is that the number of
CHY-integrands of simple poles to be computed is very large. So, in
order to partially simplify the decomposition procedure, we seek
help from the $\Lambda$-algorithm. Nevertheless, the
$\Lambda$-algorithm alone is not enough to obtain the final answer,
so we use the cross-ratio identities but now over smaller
sub-graphs. Note that for above particular type of diagrams with
${\rm PT^2}\oplus {\rm PT^2}$ geometry, after applying the
$\Lambda$-algorithm, the resulting sub-graphs are of the same type,
i.e. ${\rm PT^2}\oplus {\rm PT^2}$, or just ${\rm PT^2}$. This
feature suggests a recurrence relation.

So as to find this recurrence relation, let us consider the
following ten-point example
\begin{equation}\label{figure11}
\begin{aligned}
\raisebox{-25mm}{\includegraphics[keepaspectratio = true, scale = 0.55] {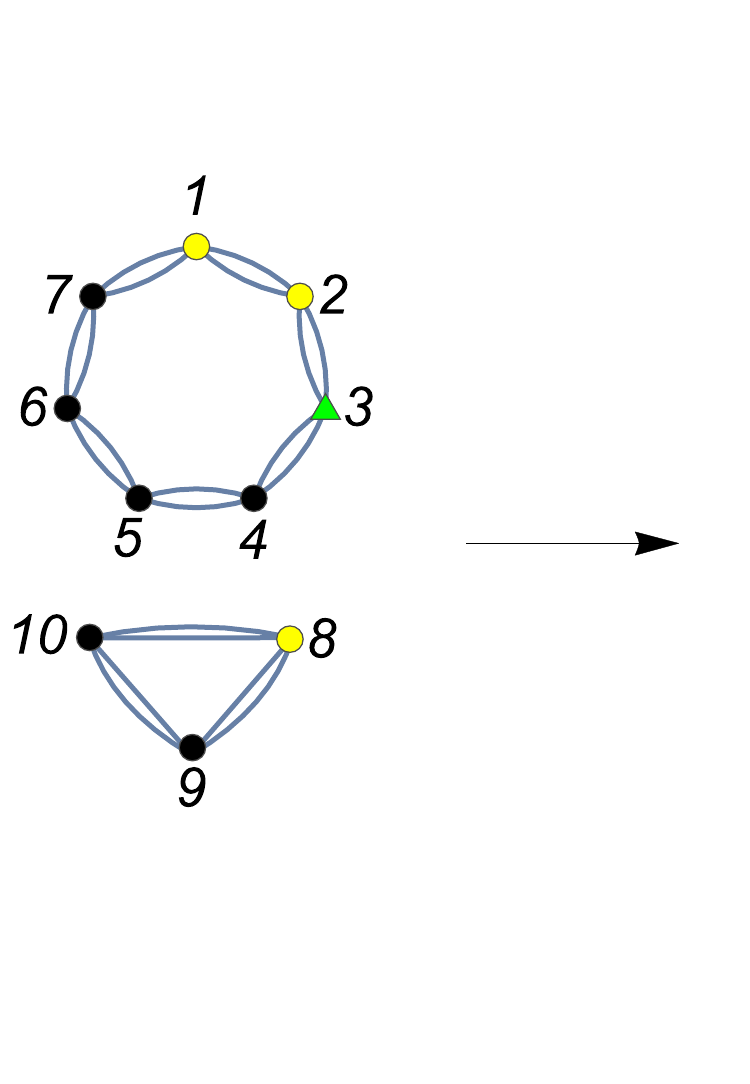}}
\raisebox{-25mm}{\includegraphics[keepaspectratio = true, scale = 0.55] {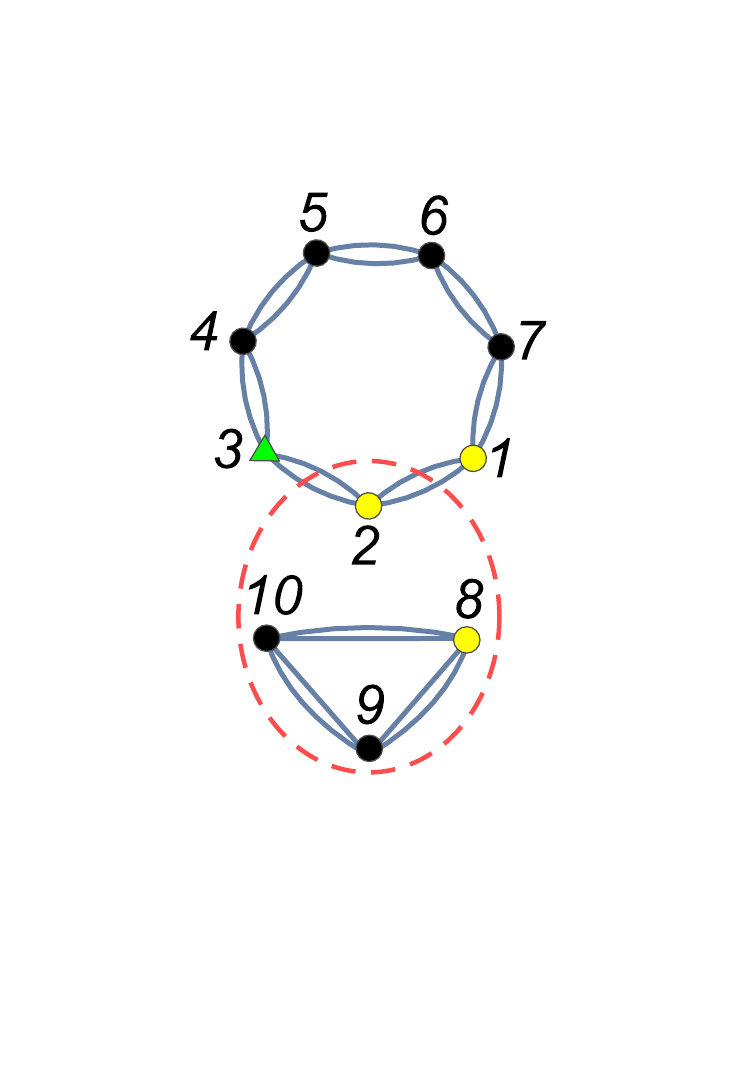}}
\raisebox{-25mm}{\includegraphics[keepaspectratio = true, scale = 0.55] {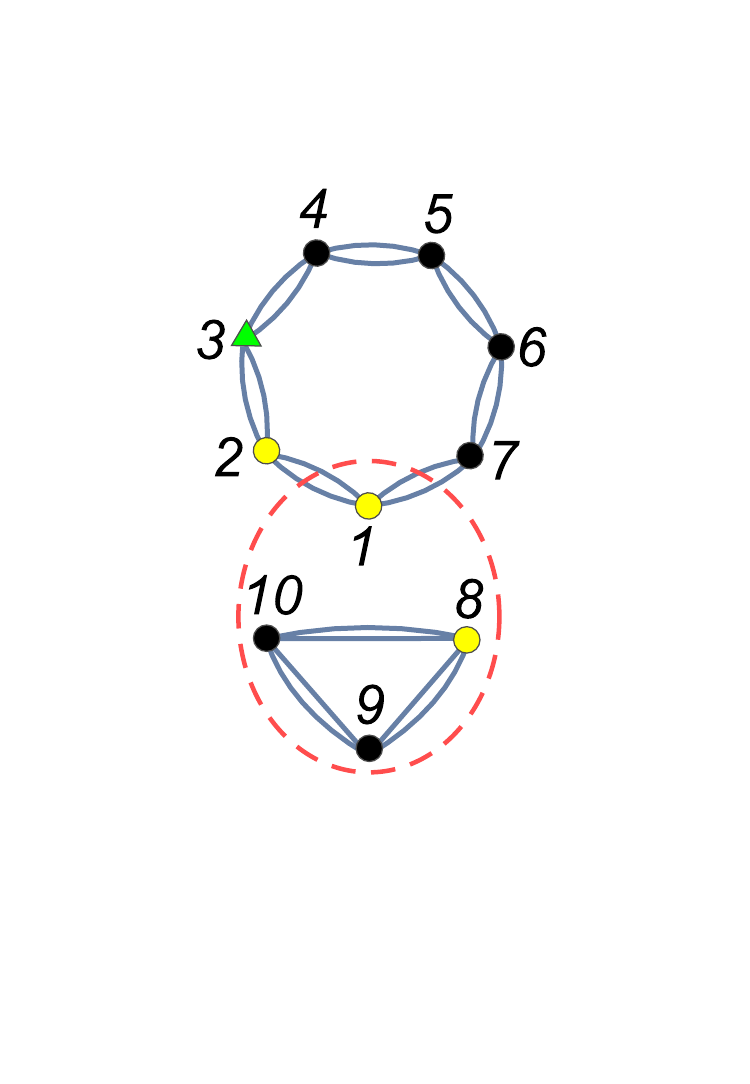}}
\raisebox{-25mm}{\includegraphics[keepaspectratio = true, scale = 0.55] {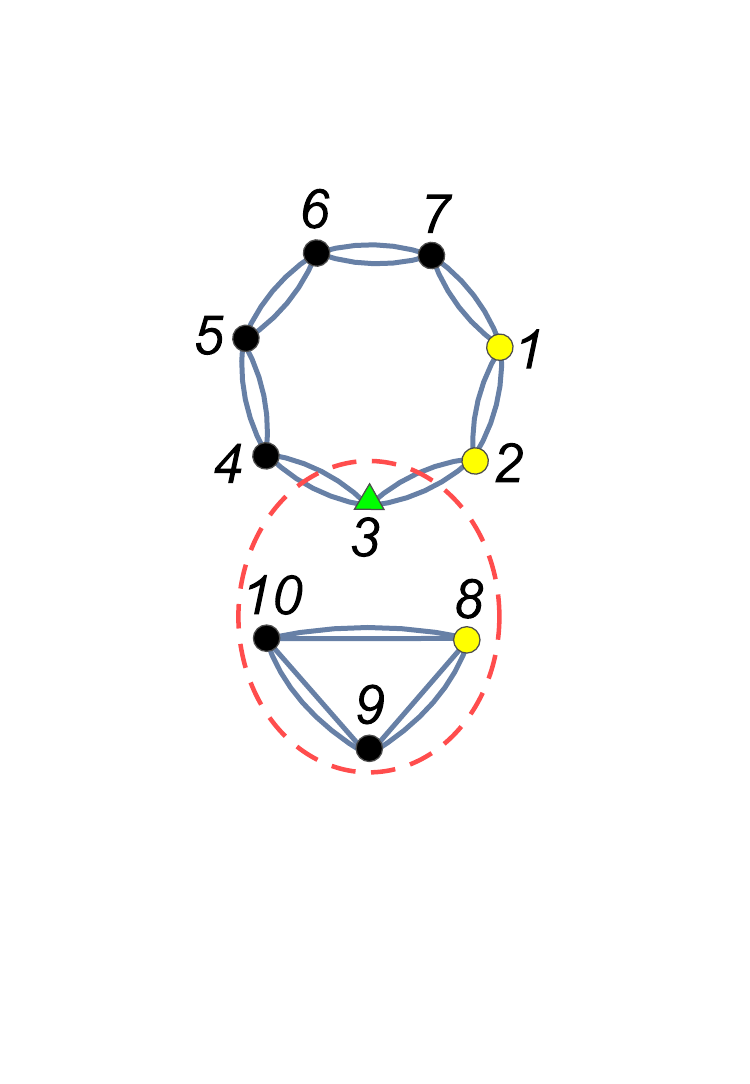}}
\end{aligned}
\end{equation}
\vspace{-1in}
\begin{equation*}
\begin{aligned}
\raisebox{-25mm}{\includegraphics[keepaspectratio = true, scale = 0.55] {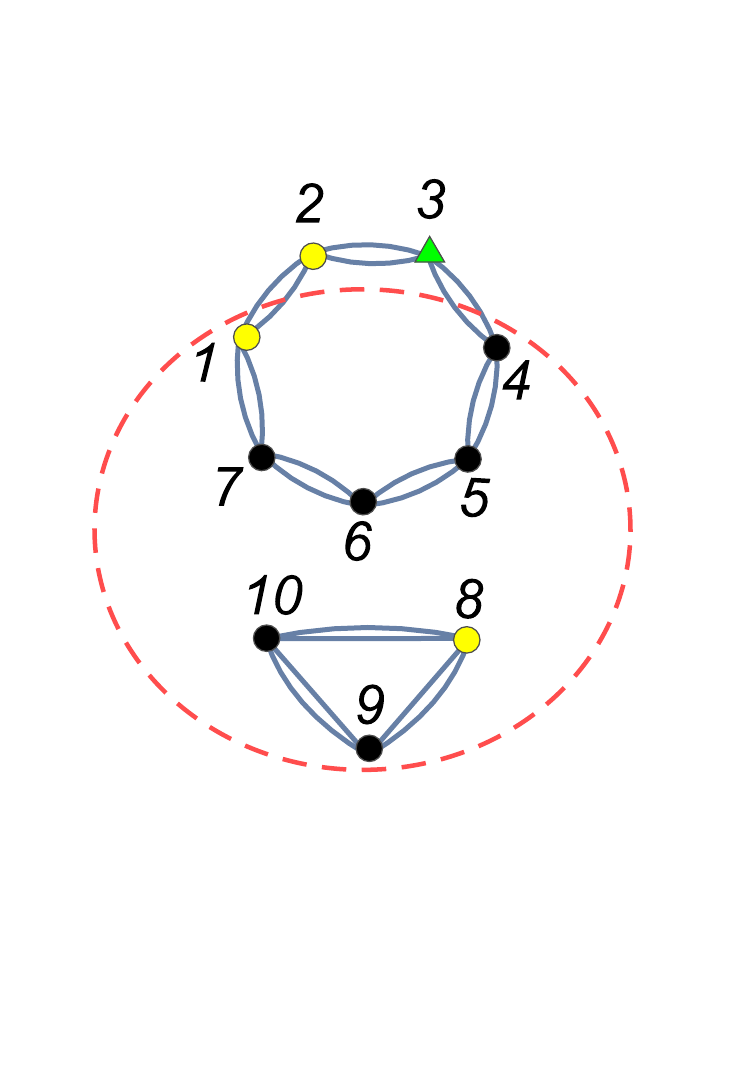}}
\raisebox{-25mm}{\includegraphics[keepaspectratio = true, scale = 0.55] {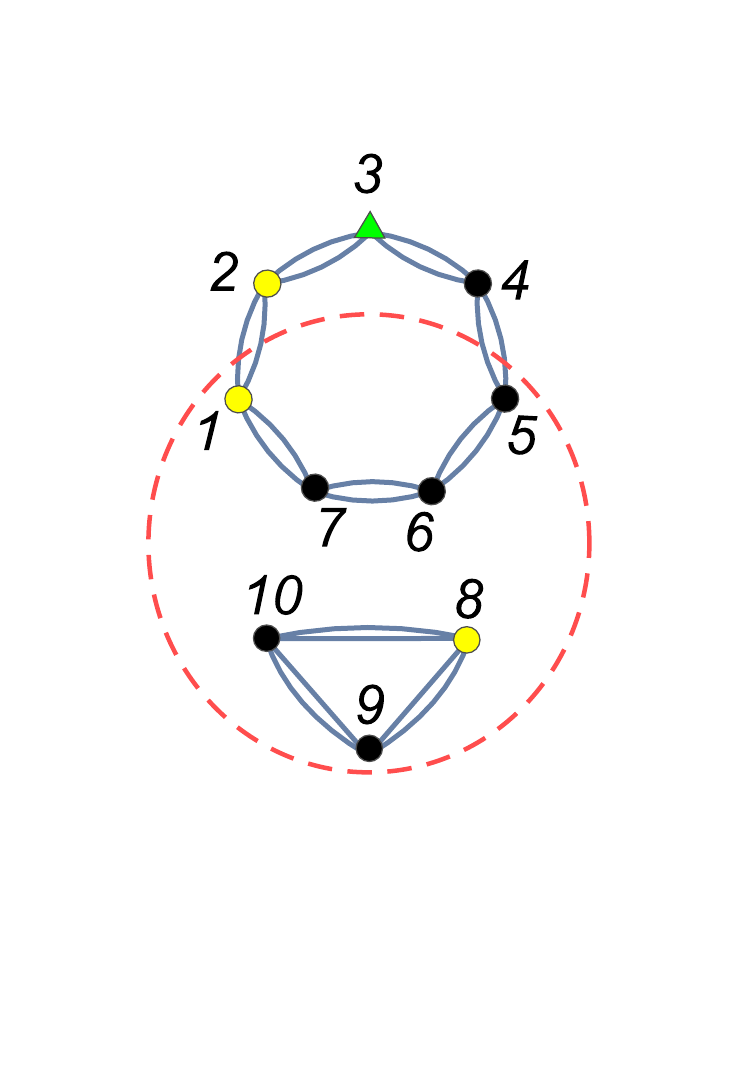}}
\raisebox{-25mm}{\includegraphics[keepaspectratio = true, scale = 0.55] {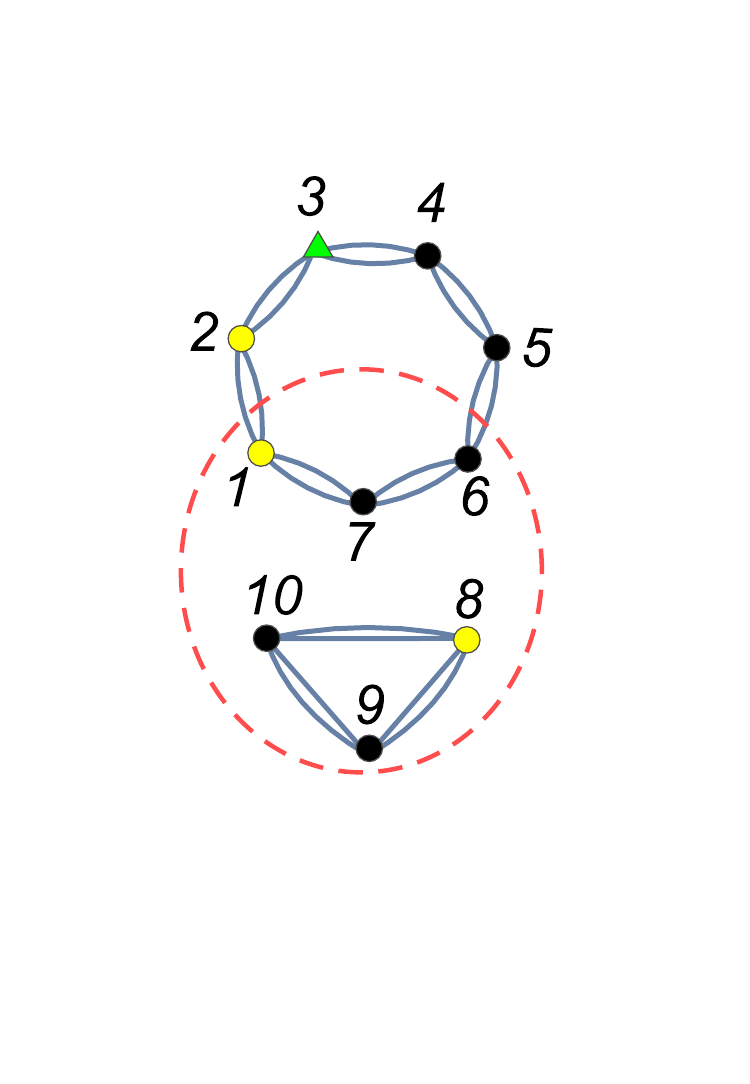}}
\raisebox{-25mm}{\includegraphics[keepaspectratio = true, scale = 0.55] {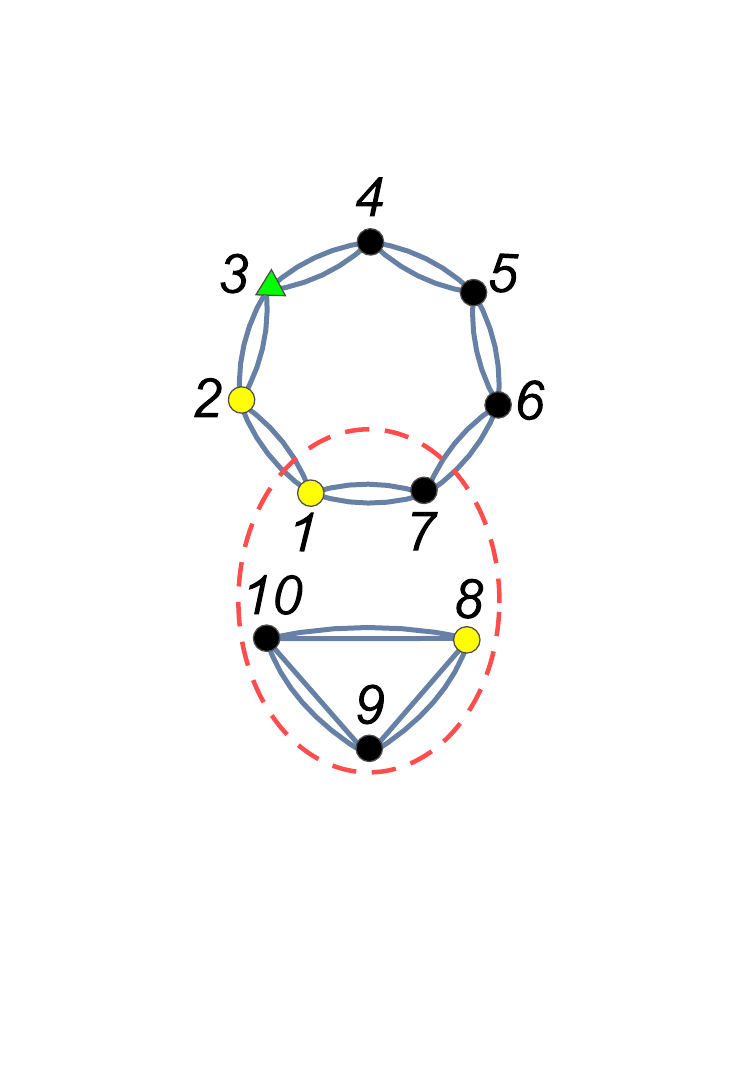}}
\end{aligned}
\end{equation*}
\vspace{-0.8in}
\begin{equation*}
\begin{aligned}
\raisebox{-25mm}{\includegraphics[keepaspectratio = true, scale = 0.55] {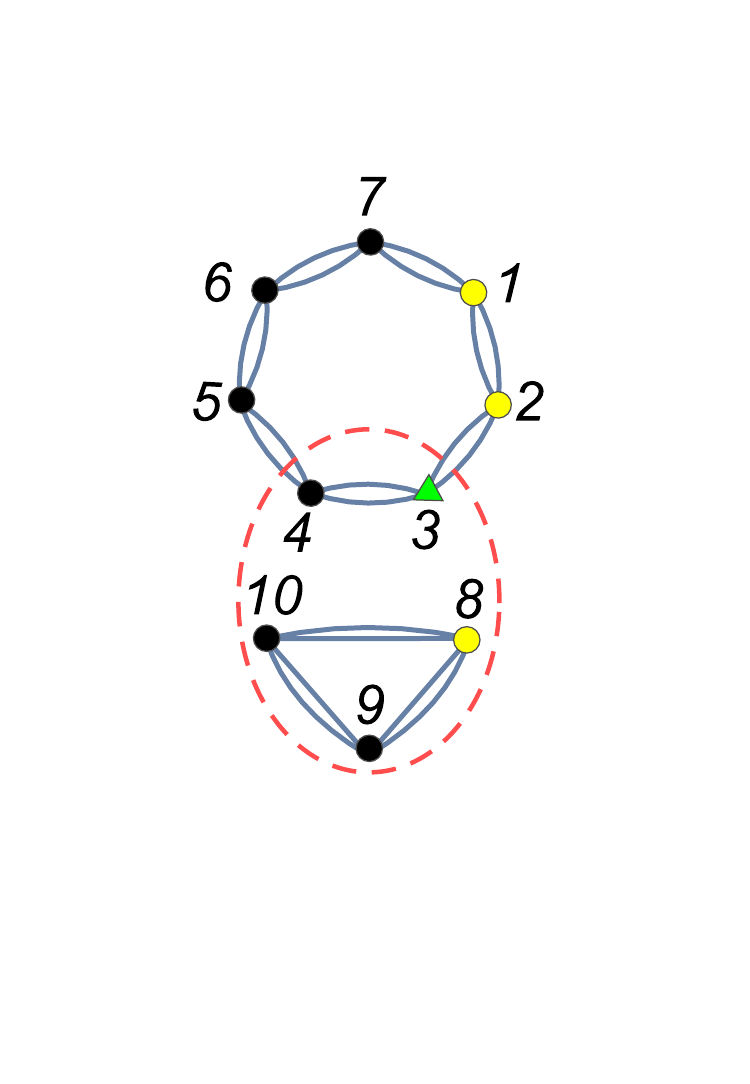}}
\raisebox{-25mm}{\includegraphics[keepaspectratio = true, scale = 0.55] {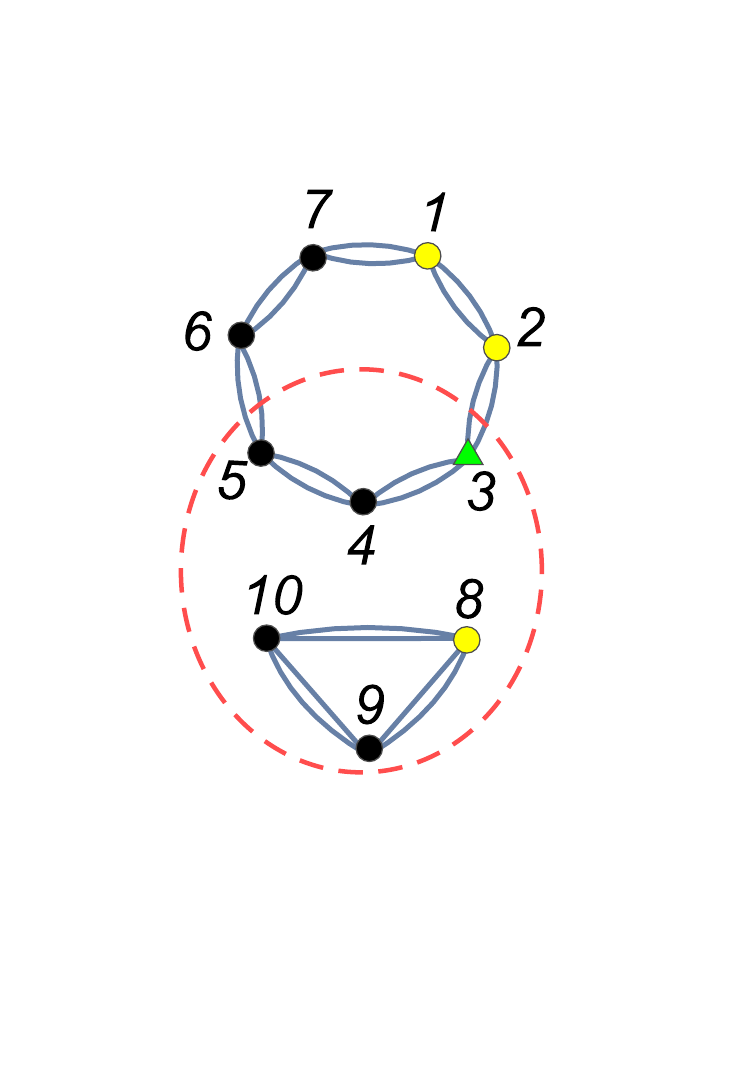}}
\raisebox{-25mm}{\includegraphics[keepaspectratio = true, scale = 0.55] {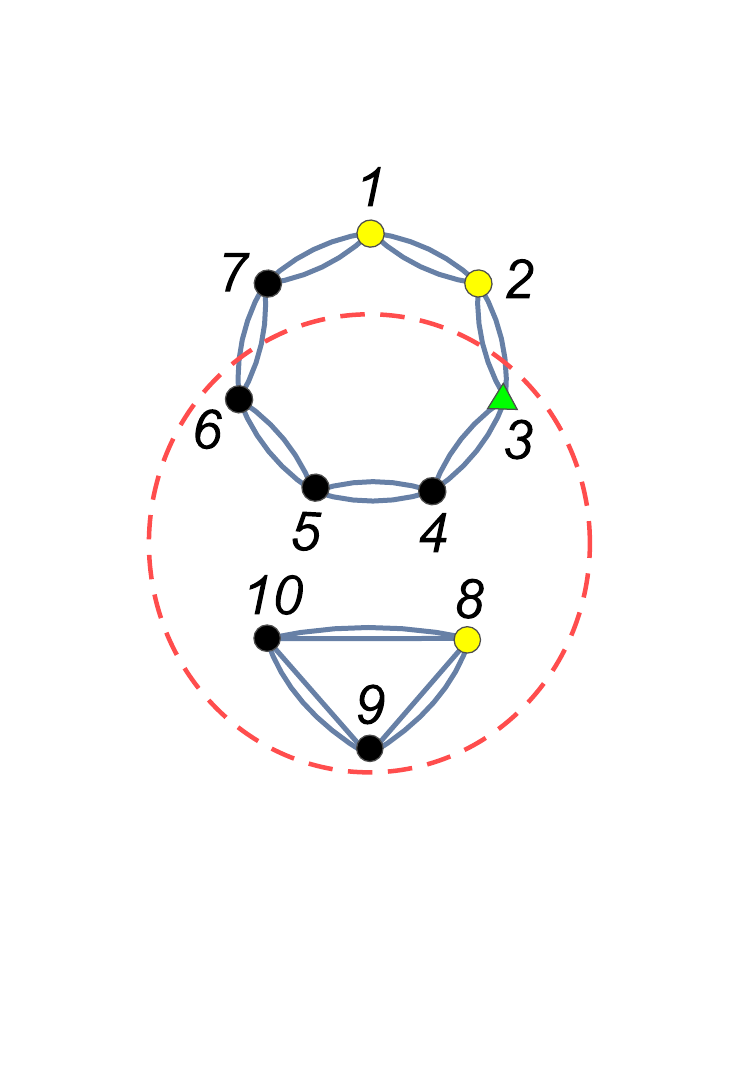}}
\raisebox{-25mm}{\includegraphics[keepaspectratio = true, scale = 0.55] {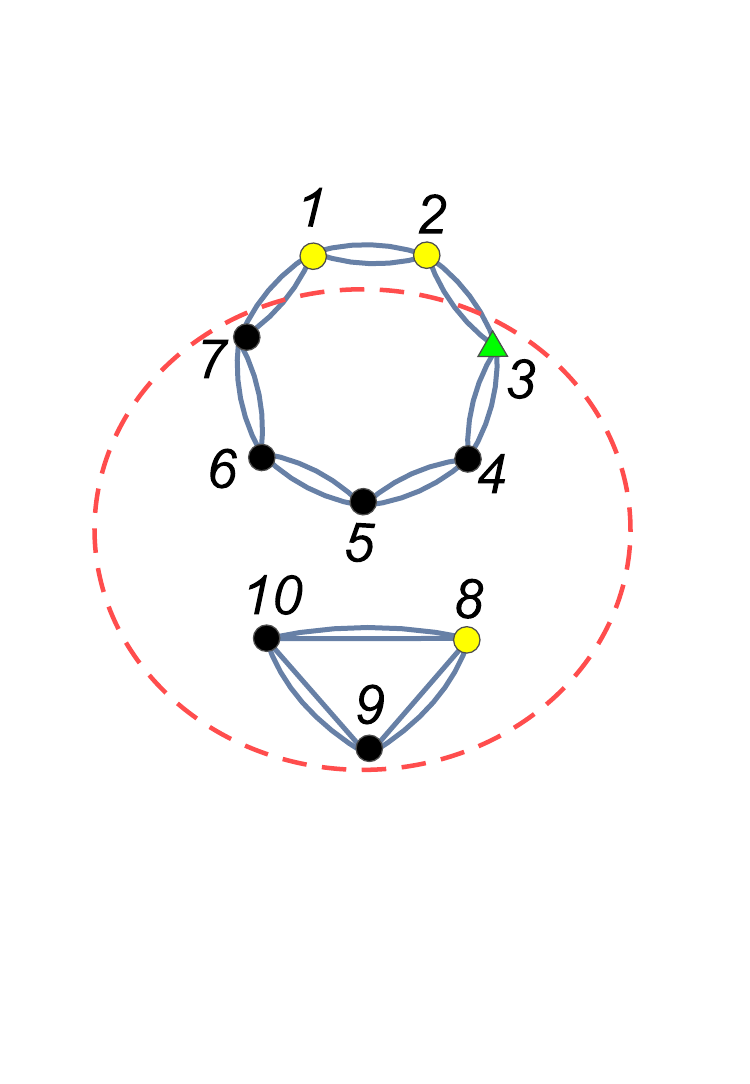}}
\end{aligned}
\end{equation*}
where we have chosen a proper gauge in order to avoid singular
configurations and we have drawn all non-zero allowable
configurations. Following this example, we can intermediately deduce
the new recurrence relation.

By using the notation in Appendix \ref{secPT2}, we make the
following definitions. Let $\mathbbm{o}_i$ be the set of ordered
elements given by
\begin{equation}
\mathbbm{o}_i:= \{4,5,6,\ldots i+3 \}~~~,~~~ {\rm where}~~  i\in\{1,\ldots, m-3\}~,~~~
\end{equation}
and we define $\mathbbm{o}_0=\emptyset$. Clearly, $
\mathbbm{o}_1=\{4\},\,\mathbbm{o}_2=\{4,5 \},\,
\mathbbm{o}_3=\{4,5,6\} $ and so on.

We also denote $\O {\mathbbm{o}}_i$ as the ordered complement of
$\mathbbm{o}_i$, as
\begin{equation}
\O {\mathbbm{o}}_i:=\{4,5,6,\ldots, m\}  \setminus {\mathbbm{o}}_i~,~~~
\end{equation}
for example
\begin{equation}
\O {\mathbbm{o}}_0=\{4,\ldots , m  \}~~, ~~ \O {\mathbbm{o}}_1=\{5,\ldots , m  \}~~,~~\ldots~~,~~\O {\mathbbm{o}}_{m-3}=\emptyset~.~~~
\end{equation}
With these definitions in mind, the recurrence relation has the form
\bea
&&{\cal I}^{{\rm PT^2}\oplus {\rm PT^2}}_{m,n-m}(1,2,\ldots ,m\,|\, m+1,\ldots, n )\label{recurrenceR2}\\
&&=~~~{{\cal I}^{{\rm PT^2}\oplus {\rm PT^2}}_{2,n-m}([3,4,\ldots
,m,1],2\,|\,m+1,\ldots,n) \times {\cal I}^{{\rm
PT^2}}_{m}(1,[2,m+1,\ldots, n ],3\ldots ,m)
\over \widetilde{s}_{3,4,\cdots,m,1}} \nonumber  \\
&&~~~~+\sum_{i=0}^{m-3}
{{\cal I}^{{\rm PT^2}\oplus {\rm PT^2}}_{m-1-i,n-m}(1,[2,3, {\mathbbm{o}}_i],\O {\mathbbm{o}}_i\,|\,m+1,\ldots,n)\times{\cal I}^{\rm PT^2}_{3+i}([1, \O {\mathbbm{o}}_i,m+1,\ldots n],2,3, {\mathbbm{o}}_i)\over \widetilde{s}_{2,3,{\mathbbm{o}}_i}}\,  \nonumber\\
&&~~~~~~+\sum_{i=0}^{m-3} {{\cal I}^{{\rm PT^2}\oplus {\rm
PT^2}}_{2+i,n-m }([1,2,\O {\mathbbm{o}}_i],3,
{\mathbbm{o}}_i\,|\,m+1,\ldots,n)\times{\cal I}^{\rm
PT^2}_{m-i}(1,2,[3, {\mathbbm{o}}_i,m+1,\ldots n],\O
{\mathbbm{o}}_i)\over \widetilde{s}_{3,{\mathbbm{o}}_i,
m+1,\cdots,n}}~,~~~\nonumber \eea
where remind again that $[a_1,a_2,\ldots,a_m]$ denotes a massive
particle with momentum equaling $\sum_{i=1}^{m}k_{a_i}$.

Applying this recurrence relation over the example in
(\ref{figure11}), one obtains (for presentation purpose here we omit
the superscript ${\rm PT^2}\oplus {\rm PT^2}$ and ${\rm PT^2}$)
{\footnotesize \bea
&&{\cal I}^{{\rm PT^2}\oplus {\rm PT^2}}_{7,3}(1,2,3,4,5,6,7\,|\, 8,9,10 )\label{I_7_3}\\
&=&{{\cal I}_{2,3}([3,4,5,6,7,1],2\,|\,8,9,10) \times {\cal
I}_{7}(1,[2,8,9,10 ],3,4,5,6,7) \over \widetilde{s}_{3,4,5,6,7,1}} +
{{\cal I}_{6,3}(1,[2,3],4,5,6,7\,|\, 8,9,10)\times{\cal I}_{3}([1,4,5,6,7,8,9,10],2,3)\over \widetilde{s}_{2,3}}\,  \nonumber\\
&&+ {{\cal I}_{5,3}(1,[2,3,4],5,6,7\,|\, 8,9,10)\times{\cal
I}_{4}([1,5,6,7,8,9,10],2,3,4)\over \widetilde{s}_{2,3,4}}+
{{\cal I}_{4,3}(1,[2,3,4,5],6,7\,|\, 8,9,10)\times{\cal I}_{5}([1,6,7,8,9,10],2,3,4,5)\over \widetilde{s}_{2,3,4,5}}\,  \nonumber\\
&&+ {{\cal I}_{3,3}(1,[2,3,4,5,6],7\,|\, 8,9,10)\times{\cal
I}_{6}([1,7,8,9,10],2,3,4,5,6)\over \widetilde{s}_{2,3,4,5,6}}+
{{\cal I}_{2,3}(1,[2,3,4,5,6,7]\,|\, 8,9,10)\times{\cal I}_{7}([1,8,9,10],2,3,4,5,6,7)\over \widetilde{s}_{2,3,4,5,6,7}}\,  \nonumber\\
&&+ {{\cal I}_{2,3 }([1,2,4,5,6,7],3\,|\, 8,9,10)\times{\cal
I}_{7}(1,2,[3, 8,9,10],4,5,6,7)\over \widetilde{s}_{3,8,9,10}}+
{{\cal I}_{3,3 }([1,2,5,6,7],3,4\,|\, 8,9,10)\times{\cal I}_{6}(1,2,[3,4, 8,9,10],5,6,7)\over \widetilde{s}_{3,4,8,9,10}}\,  \nonumber\\
&&+ {{\cal I}_{4,3 }([1,2,6,7],3,4,5\,|\, 8,9,10)\times{\cal
I}_{5}(1,2,[3,4,5, 8,9,10],6,7)\over \widetilde{s}_{3,4,5,8,9,10}}+
{{\cal I}_{5,3 }([1,2,7],3,4,5,6\,|\, 8,9,10)\times{\cal I}_{4}(1,2,[3,4,5,6, 8,9,10],7)\over \widetilde{s}_{3,4,5,6,8,9,10}}\,  \nonumber\\
&&+ {{\cal I}_{6,3 }([1,2],3,4,5,6,7\,|\, 8,9,10)\times{\cal
I}_{3}(1,2,[3,4,5,6,7, 8,9,10])\over
\widetilde{s}_{3,4,5,6,7,8,9,10}}~,~~~\nonumber\eea}

\noindent which is the right expression for the configurations given
in (\ref{figure11}). The terms ${\cal I}^{\rm PT^2}_{n},\, n=5,6,7,$
can easily be computed from the recurrence relation in
\eqref{recurrenceR}. Now, the terms ${\cal I}^{{\rm PT^2}\oplus {\rm
PT^2}}_{3,3},\,{\cal I}^{{\rm PT^2}\oplus {\rm PT^2}}_{4,3 },\,
{\cal I}^{{\rm PT^2}\oplus {\rm PT^2}}_{5,3 }$ and $\,{\cal I}^{{\rm
PT^2}\oplus {\rm PT^2}}_{6,3 }$ can be reduced using, iteratively,
the recurrence relation \eqref{recurrenceR2}. In addition, in
Appendix \ref{sec2PT2}  we have given the expressions for these
diagrams.

It is important to note that the relation in \eqref{recurrenceR2}
only works for CHY-integrands with $m>2$. In other words,
CHY-integrands such as ${\cal I}^{{\rm PT^2}\oplus {\rm
PT^2}}_{2,n-2}(1,2\,|\,3,...,n)$ with $n>4$ under the gauge-fixing
as in (\ref{figure11}) can not be solved just by the
$\Lambda$-algorithm, and therefore we should proceed to use the
cross-ratio identities. The main idea of the recurrence relation in
\eqref{recurrenceR2} is to use it to straightforwardly reduce the
original CHY-integrand ${\cal I}^{{\rm PT^2}\oplus {\rm
PT^2}}_{m,n-m}(1,2,\dots m,\,|\, m+1,\ldots, n )$ with $m>2~,~n-m>2$
to ${\cal I}^{{\rm PT^2}\oplus {\rm PT^2}}_{2,n'-2}$ with $n'$
smaller than $n$, and then apply the cross-ratio identities.

For our particular example in (\ref{figure11}), we were able to
reduce the whole expression in \eqref{I_7_3} as a linear combination
of  ${\cal I}^{{\rm PT^2}\oplus {\rm PT^2}}_{2,3}(a,b\,|c,d,e)$
diagram (see Appendix \ref{sec2PT2}). This diagram was solved
previously by using the cross-ratio identities, and its answer is
given in \eqref{I_2_3}.

For the most general case, we must solve the $n=(q+2)$-point
CHY-integrand
\begin{equation}\label{I_2_q}
\begin{aligned}
{\cal I}^{{\rm PT^2}\oplus {\rm PT^2}}_{2,q}(1,2\,|3,4,\ldots,q+2)~~~=
\raisebox{-15mm}{\includegraphics[keepaspectratio = true, scale = 0.4] {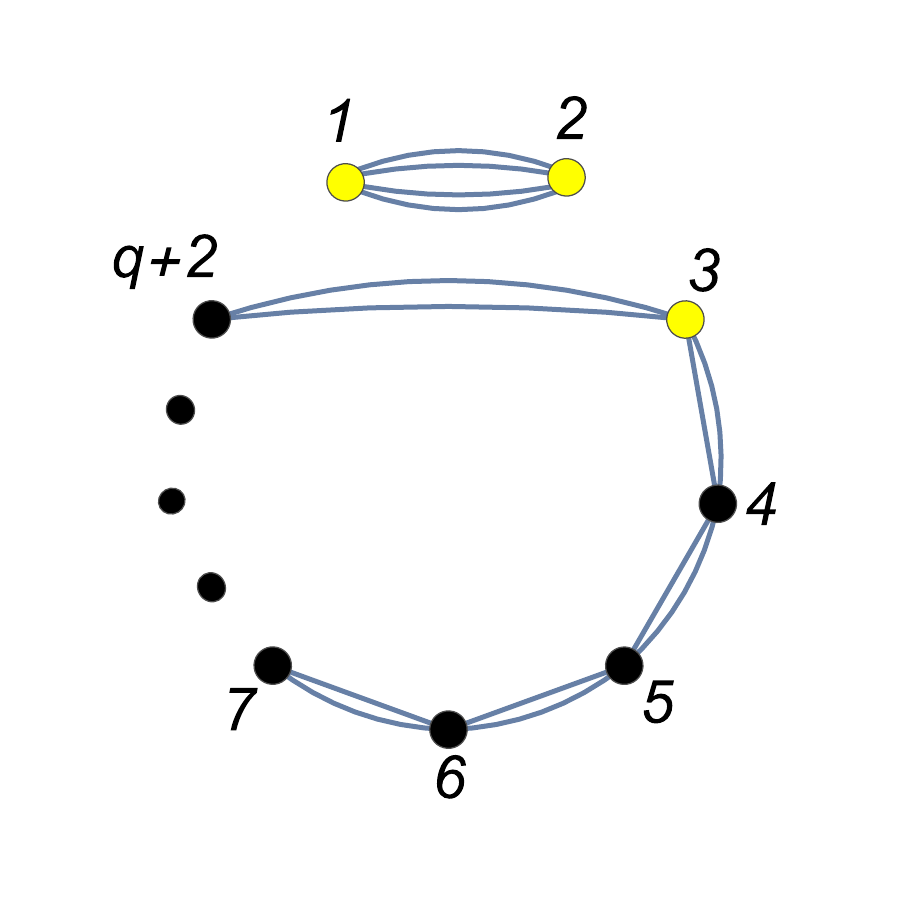}}\,\, ,
\end{aligned}
\end{equation}
where we use the cross-ratio identities for doing it. Similarly as
it was done in \S \ref{secSimpleexample} for the graph ${\cal
I}^{{\rm PT^2}\oplus {\rm PT^2}}_{2,3}(a,b\,|\,c,d,e)$, one notes
that this graph has a triple pole, $1/s^3_{34\ldots q+2}$. In order
to eliminate this pole, we can use the scattering equations, $\{
{\cal E}_3, {\cal E}_4,\ldots, {\cal E}_{q+2} \}$, to obtain the
cross-ratio identity
\begin{equation}
\widetilde{s}_{34\ldots q+2}= \sum_{a=4}^{q+2}\widetilde{s}_{1a} \left( \frac{z_{12} z_{3a} }{z_{23} z_{1a} }   \right)\, ,
\end{equation}
which has $q-1$ elements. Taking the square of this identity one
obtains
\begin{align}
\widetilde{s}_{34\ldots q+2}^2&= \sum_{a,b=4}^{q+2}\widetilde{s}_{1a}\,\widetilde{s}_{1b} \left( \frac{z_{12} z_{3a} }{z_{23} z_{1a} }   \right)\, \left( \frac{z_{12} z_{3b} }{z_{23} z_{1b} }   \right)\\
&= \sum_{a=4}^{q+2}\widetilde{s}_{1a}^2 \left( \frac{z_{12} z_{3a} }{z_{23} z_{1a} }   \right)
+2\sum_{a<b=4}^{q+2}\widetilde{s}_{1a}\,\widetilde{s}_{1b} \left( \frac{z_{12} z_{3a} }{z_{23} z_{1a} }   \right)\, \left( \frac{z_{12} z_{3b} }{z_{23} z_{1b} }   \right),
\end{align}
which has a total number of terms,
$q-1+\frac{(q-1)(q-2)}{2}=\frac{q\,(q-1)}{2}$.  Therefore, the
CHY-integrand
$${\cal I}^{{\rm PT^2}\oplus {\rm
PT^2}}_{2,q}(1,2\,|3,4,\ldots,q+2)$$ is solved in a straightforward
manner by computing a number of $\frac{q\,(q-1)}{2}$ trivial
CHY-integrands.

Finally, noticing that the recurrence relation (\ref{recurrenceR2})
is only a consequence of the iterative nature of the
$\Lambda$-algorithm, we can build many other relations by using the
same method. The only particularity that need to be satisfied in
order to have an ordered recurrence is that the graphs under
consideration should possess some symmetry such that the cutting
process preserves this "symmetry" for the smaller sub-graphs. For
instance, for the particular case in (\ref{figure11}), the original
 graph is build out from two disjoint pieces, and in
an appropriate gauge avoiding singular configurations,  all the
allowed cuts produce sub-graphs with the same topology as the
original graph. Examples where similar recurrence relations can be
also obtained are graphs of the type $\left({\rm
PT^2}\right)^n:=({\rm PT^2})\oplus ({\rm PT^2})\oplus\cdots
\oplus({\rm PT^2})$. Such CHY-integrands could have very large
$\Upsilon$, thus it is very necessary to apply $\Lambda$-algorithm
before decomposition algorithm with cross-ratio identity. By solely
using the $\Lambda$-algorithm we will reach sub-diagrams containing
singular configurations where the $\Lambda$-algorithm cease to work
and hence stop the recurrence. The remaining sub-diagrams should be
then rewritten by means of the cross-ratio identities in order to
either solving them directly by integration rules of simple poles or
by continuing the cutting process of $\Lambda$-algorithm.

\section{Conclusion}
\label{secConclusion}

Based on the extremely simple and efficient method of
integration rules for CHY-integrands with only simple poles
\cite{Baadsgaard:2015voa,Baadsgaard:2015ifa}, we propose a
systematic decomposition algorithm by use of cross-ratio identities,
which can be applied to decompose any CHY-integrand of higher-order
poles to those with only simple poles, suitable for evaluating by
the integration rules. The basic idea of the algorithm can be
described as follows. For any CHY-integrand with higher-order poles,
we multiply it with identities which are linear combination of terms
as
$${s_{A}\over s_{B}}{z_{a_1a_2}z_{a_3a_4}\over z_{a_1a_3}z_{a_2a_4}}~,$$
i.e., cross-ratios of $z_{ij}$. The cross-ratios of $z_{ij}$
reformulate the original CHY-integrands of higher-order poles as new
ones with only simple poles which can be evaluated trivially, while
the $s_{B}$ in denominator will compensate the extra degrees of
higher-order poles, such that the final result still possess
higher-order poles. This idea is exactly the same as the algorithm
proposed in \cite{Bjerrum-Bohr:2016juj}, but in
\cite{Bjerrum-Bohr:2016juj} identities from monodromy relations are
used instead of cross-ratio identities, and we have shown in the
Appendix \ref{secMonodromy} that a decomposition with the former is
far less efficient than that with the later for large number of
scattering particles and complicated higher-order pole structures,
due to the fact that the number of terms in an identity of monodromy
relation grows factorially while in cross-ratio identity
algebraically.

An eight-point CHY-integrand with one triple pole and six double
poles is computed in \S\ref{secAlgorithm}, resulting in around
10,000 terms of only simple poles, which summed up to produce the
amplitude by integration rules, with the evaluation time around a
few minutes. The increasing of scattering particles will not affect
the computation efficiency much, but the increasing of higher-order
poles will significantly increase the number of resulting terms,
since in order to compensate an extra $1\over s_{i_1\cdots i_k}$, we
need to multiply an identity with $(k-1)(n-k-1)$ terms, and
consequently the number of resulting terms will be about
$(k-1)(n-k-1)$ times larger. So for CHY-integrands with very large
$n$ and complicated higher-order pole structures, the number of
resulting CHY-integrands after decomposition will easily reach the
size of millions, which slows the computation. In need of these
situations, we seek help from $\Lambda$-algorithm
\cite{Gomez:2016bmv,Cardona:2016bpi}. The $\Lambda$-algorithm
rewrites a CHY-integrand as products of lower-point CHY-integrands,
until the resulting lower-point ones can not be computed by
$\Lambda$-algorithm. We show that for particular type of
CHY-integrands, recurrence relations can be deduced to iteratively
transform a specific CHY-integrand into lower-point ones which are
easy to compute by cross-ratio identities. The combination of
$\Lambda$-algorithm with decomposition algorithm makes a more
efficient method for evaluating large $n$-particle amplitude by use
of cross-ratio identities.

Some aspects are in the following. The CHY-integrand we consider in
this paper is quite general, with in fact any possible higher-order
poles. Thus the decomposition algorithm is suitable and ready for
evaluating Yang-Mills amplitudes and even gravity amplitudes in the
context of CHY-formulation, not only in principle but also in
practical. Since the cross-ratio identities deal with CHY-integrands
with higher-order poles, while in \cite{Huang:2016zzb} Feynman rules
for higher-order poles are conjectured, it would be interesting to
investigate if one can prove those Feynman rules by cross-ratio
identities and further derive rules for more higher-order poles.
Another interesting problem considers the recurrence relation by
$\Lambda$-algorithm. We have in this paper presented a recurrence
relation for CHY-integrands with $({\rm PT})^2\oplus ({\rm PT})^2$
geometry, to iteratively rewrite a CHY-integrand to other specific
ones which are easy to compute. It would be interesting to
generalize the study of recurrence relations to a broader range of
CHY-integrands with other complicated geometries where a direct
decomposition algorithm would take too much time.

\section*{Acknowledgments}

We would like to thank Christian Baadsgaard and Song He for helpful
conversation. C.C. thanks to the High Energy Group of the Institute
of Modern Physics and Center of Mathematical Science of Zhejiang
University in Hangzhou city, P.R.C. for the hospitality during the
first stages of this work. The work of B.F. and R.H. is supported by
Qiu-Shi Funding and the National Natural Science Foundation of China
(NSFC) with Grant No.11135006, No.11125523 and No.11575156. R.H.
would also like to acknowledge the supporting from Chinese
Postdoctoral Administrative Committee. The work of C.C. is supported
in part by the National Center for Theoretical Science (NCTS),
Taiwan, Republic of China. The work of H.G. is supported by CNPq
grant 403178/2014-2 and  USC grant DGI-COCEIN-No 935-621115-N22.

\appendix

\section{A Practical Algorithm to Determine the $\pm$ Sign in Integration Rules}
\label{secSignRule}

The systematic decomposition algorithm rewrites any CHY-integrand of
higher-order poles as terms of simple poles, and the last step
towards a final answer is to evaluate the resulting CHY-integrands
by integration rules of simple poles. This requires a practical
implementation of integration rules. In integration rules, the
evaluation in fact contains two parts. The first is the contributing
terms, which can be readily determined by working out the compatible
combinations. The second is to determine the $\pm$ sign of each
contributing term, which is the topic we want to discuss here.

Again let us start from a generic $n$-point CHY-integrand
(\ref{genericCHY}), as (assuming here the canonical ordering, i.e.,
always $i<j$),
\bea \mathcal{I}={1\over \prod_{1\leq i<j\leq n
}z_{ij}^{\beta_{ij}}}~,~~~\eea
respecting the M\"obius invariance. Assuming that it has $m$ simple
poles corresponding to $m$ subsets $A_i,i=1,\ldots,m$ of
$\chi(A_i)=0$. Assuming also that one can construct $m'$ compatible
combinations
$\{A_{\alpha_1},A_{\alpha_2},\ldots,A_{\alpha_{n-3}}\}$, etc., from
$m$ subsets. Then by integration rule, the result is given by the
summation of $m'$ terms, with each term from a compatible
combination as
\bea (-1)^{n+1+\mathcal{N}_{\inv}}{1\over
s_{\alpha_1}s_{\alpha_2}\cdots s_{\alpha_{n-3}}}~,~~~\eea
where $s_{\alpha_i}=(P_{A_{\alpha_i}})^2$, and $\mathcal{N}_{\inv}$
is the so-called {\sl inversion factor}\footnote{We thank Christian
Baadsgaard for explaining the inversion factor to
us.}\cite{Baadsgaard:2015abc}. In the original derivation, the
inversion factor is defined under a given gauge-fixing. One need to
pick up a specific gauge and then consider what the inversion factor
result. Briefly speaking, once a gauge is fixed, the inversion
factor $n_{\inv}$ of a subset $A_{\alpha}$ (avoiding the infinite
fixed point) is the number of factor $z_{ij}$ with $i\in A_{\alpha}$
and $j \notin A_{\alpha}$, and $\mathcal{N}_{\inv}$ is the sum of
all $n_{\inv}$ in a compatible combination. However, the choice of
gauge makes it difficult to automatically compute the inversion
factor for any given compatible combination.

From the practices , we found that it is not necessary to restrict
to a given gauge-fixing in order to compute the sign. Although the
inversion factor is not the same in different gauges, the parity of
$\mathcal{N}_{\inv}$ is invariant, while the $(\pm)$ sign depends
only on the parity. In fact, we can define the {\sl inversion
factor} without referring to any fixed points, as will be shown in
the following.

For a length-$n$ set $\Sigma=\{\sigma_1,\sigma_2,\ldots,
\sigma_n\}$, which is a permutation of $\{1,2,\ldots,n\}$, the {\sl
signature} of set $\Sigma$ is defined to be $(-1)^{\mathcal{N}}$,
where $\mathcal{N}$ is the number of times required to iteratively
permute two adjacent elements in order to arrive at the canonical
ordering $\{1,2,\ldots, n\}$. For example,
\bea \{1,3,2,5,4,6\}~~\xrightarrow[]{2\leftrightarrow
3}~~\{1,2,3,5,4,6\}~~\xrightarrow[]{4\leftrightarrow
5}~~\{1,2,3,4,5,6\}~,~~~\eea
so $\mathcal{N}=2$. Of course there are many different permutations
to do so, but the signature is invariant. Here we define the {\sl
weighted signature} $(-1)^{\mathcal{N}'}$, where $\mathcal{N}'$ is
the weighted number of times. For each permutation of two adjacent
elements $i\leftrightarrow j$, we count the number as $\beta_{ij}$
but not 1, where $\beta_{ij}$ is determined by a given
CHY-integrand. So for the above example, we have
$\mathcal{N}'=\beta_{23}+\beta_{45}$, and the {\sl weighted
signature} is $(-1)^{\beta_{23}+\beta_{45}}$.

Now we define the {\sl inversion factor}\footnote{Note again here
without referring to any gauge, which is different from the
definition in \cite{Baadsgaard:2015abc}.} for a compatible
combination $A=\{A_{\alpha_1},A_{\alpha_2},\ldots,
A_{\alpha_{n-3}}\}$ of a given CHY-integrand $\mathcal{I}(z_{ij})$.
Define the length-$n$ set $\Sigma=\{\sigma_1,\sigma_2,\ldots,
\sigma_n\}$ to be cyclically ordered, i.e., $\sigma_n$ and
$\sigma_1$ are also considered to be adjacent. An {\sl adjacent
subset} $A_{\alpha}$ of $\Sigma$ is defined to be a subset of
$\Sigma$ whose elements are adjacent in $\Sigma$ (but the ordering of
elements in $A_{\alpha}$ dose not need to respect the ordering in
$\Sigma$). For example, both $\{\sigma_2,\sigma_3,\sigma_4\}$ and
$\{\sigma_2,\sigma_4,\sigma_3\}$ are adjacent subsets of $\Sigma$,
while $\{\sigma_1,\sigma_2,\sigma_n\}$ is also an adjacent subset,
but $\{\sigma_2,\sigma_3,\sigma_5\}$ is not.

A cyclically ordered length-$n$ set
$\Sigma=\{\sigma_1,\sigma_2,\ldots, \sigma_n\}$ is said to be the
{\sl parent set} of a compatible combination
$\{A_{\alpha_1},A_{\alpha_2},\ldots, A_{\alpha_{n-3}}\}$ if all
subsets $A_{\alpha_i}$'s are {\sl adjacent subsets} of $\Sigma$.
Provided we have already found a {\sl parent set} $\Sigma$ of a
compatible combination $A$, then the {\sl inversion factor}
$\mathcal{N}_{\inv}$ of $A$ is defined to be the weighted number
$\mathcal{N}'$ for permuting $\Sigma=\{\sigma_1,\sigma_2,\ldots,
\sigma_n\}$ to canonical ordering $\{1,2,\ldots,n\}$, and the sign
of the term associated to the compatible combination $A$ is nothing
but $(-1)^{n+1+\mathcal{N}_{\inv}}=(-1)^{n+1+\mathcal{N}'}$,
proportional to the {\sl weighted signature} of {\sl parent set}
$\Sigma$. Of course, there will be more than one cyclically ordered
sets which could be the parent set of a given compatible
combination. Although the weighted number $\mathcal{N}'$ of them are
different, the parity of $\mathcal{N}'$ is invariant, so is the {\sl
weighted signature}. In this case, we only need to find one {\sl
parent set} for a compatible combination, and compute the $(\pm)$
sign with it.

For a given compatible combination
$A=\{A_{\alpha_1},A_{\alpha_2},\ldots, A_{\alpha_{n-3}}\}$ of
CHY-integrand $\mathcal{I}(z_{ij})$, now we shall find its {\sl
parent set}. Naively, one can generate all the permutation sets of
$\{1,2,\ldots,n\}$ in {\sc Mathematica}, and select one set which is
the {\sl parent set} of $A$. However, the number of permutation sets
grows factorially with $n$, so it is practically not efficient. We
propose the following strategy to construct a {\sl parent set} of
$A$. Remind that any two subsets in a compatible combination satisfy
{\sl compatible condition}, i.e., they should be {\sl nested}
($A_{\alpha_i}\bigcup A_{\alpha_j}=A_{\alpha_i}~\mbox{or}~
A_{\alpha_j}$) or {\sl disjoint} ($A_{\alpha_i}\bigcap
A_{\alpha_j}=\emptyset$). The strategy starts from
$B=\{A_{\alpha_1}\}$,
\begin{itemize}
\item If $A_{\alpha_2}$ is disjoint with $A_{\alpha_1}$, it will
be included to get $B=\{A_{\alpha_1},A_{\alpha_2}\}$,

\item If $A_{\alpha_2}$ is nested with $A_{\alpha_1}$, we will
replace $A_{\alpha_1}$ with a new subset $A'_{\alpha_1}$, and
consequently $B=\{A'_{\alpha_1}\}$.
\begin{itemize}
\item In the case that $A_{\alpha_1}\subset A_{\alpha_2}$, we
can define
$A'_{\alpha_1}=A_{\alpha_1}+\mathcal{C}[A_{\alpha_2},A_{\alpha_1}]$\footnote{$A_1+A_2$
stands for a new set with elements from $A_1$ followed by
elements from $A_2$, keeping the ordering.}, where
$\mathcal{C}[A_{\alpha_i},A_{\alpha_j}]$ denotes the
complement set of $A_{\alpha_j}$ with respective to
$A_{\alpha_i}$, i.e., a set whose elements are in
$A_{\alpha_i}$ but not $A_{\alpha_j}$. This construction is
to ensure that $A_{\alpha_1}$ as well as $A_{\alpha_2}$ are
the {\sl adjacent subsets} of a {\sl parent set}.
\item In
the case $A_{\alpha_2}\subset A_{\alpha_1}$, we can define
$A'_{\alpha_1}=A_{\alpha_2}+\mathcal{C}[A_{\alpha_1},A_{\alpha_2}]$.
\end{itemize}
\end{itemize}
After $A_{\alpha_2}$ is done, we continue to $A_{\alpha_3}$, where
now we should consider $A_{\alpha_3}$ to be nested or disjoint with
all subsets in $B$ from previous step. For example, if in previous
step we get $B=\{A_{\alpha_1},A_{\alpha_2}\}$, then
\begin{itemize}
  \item If $A_{\alpha_3}$ is disjoint with both $A_{\alpha_1},
  A_{\alpha_2}$ (any subsets in $B$), we renew the set as
  $B=\{A_{\alpha_1},A_{\alpha_2},A_{\alpha_3}\}$,
  \item If $A_{\alpha_3}\subset A_{\alpha_i}$ (here $i=1$ or 2. Because from the
  construction, the subsets included in $B$ are always disjoint
  with each other, so $A_{\alpha_3}$ could only be a subset of
  either $A_{\alpha_1}$ or $A_{\alpha_2}$), we define
  $A'_{\alpha_i}=A_{\alpha_3}+\mathcal{C}[A_{\alpha_i},A_{\alpha_3}]$,
  and renew the set as $B=\{A'_{\alpha_1},A_{\alpha_2}\}$ or
  $B=\{A_{\alpha_1},A'_{\alpha_2}\}$,
  \item If any subsets in $B$ are subsets of $A_{\alpha_3}$,
\begin{itemize}

  \item If only one subset in $B$ is a subset of the one under consideration, explicitly here $A_{\alpha_i}\subset A_{\alpha_3}$ ($i=1$ or 2), we
  define
  $A'_{\alpha_i}=A_{\alpha_i}+\mathcal{C}[A_{\alpha_3},A_{\alpha_i}]$,
  and renew the set as $B=\{A'_{\alpha_1},A_{\alpha_2}\}$ or
  $B=\{A_{\alpha_1},A'_{\alpha_2}\}$,
  \item If more than one subset in $B$ are subsets of the one under
  consideration, explicitly here $A_{\alpha_1}\subset
  A_{\alpha_3}$ and $A_{\alpha_2}\subset A_{\alpha_3}$,
  since $A_{\alpha_1}$ and $A_{\alpha_2}$ are disjoint, we
  can define a new subset
  $$A'_{\alpha_1}=A_{\alpha_1}+A_{\alpha_2}+\mathcal{C}[A_{\alpha_3},A_{\alpha_1}+A_{\alpha_2}]$$
  to replace $A_{\alpha_1},A_{\alpha_2}$, and renew the set
  as $B=\{A'_{\alpha_1}\}$.
\end{itemize}
\end{itemize}
With such strategy, we can enlarge set $B$ until the last
$A_{\alpha_{n-3}}$ in the compatible combination is considered. The
subsets in $B$ are by construction disjoint to each other. Assuming
$B=\{B_1,B_2,\ldots,B_{k}\}$, a {\sl parent set} of compatible
combination $A$ can then be given by
\bea
\Sigma=B_1+B_2+\cdots+B_k+\mathcal{C}[\{1,2,\ldots,n\},\bigcup_{i=1}^{k}B_k]~.~~~\eea
The construction of parent set is already done, but we can go a step
further. Because of the cyclic invariance, we can always fix $1$ in
the first position of $\Sigma$ as the convention. All above
operations can be easily implemented in {\sc Mathematica}.

It is better to understand  above algorithm with an example. Let us
consider the following six-point CHY-integrand with non-trivial
numerator
\bea \mathcal{I}=\frac{z_{14} z_{35}}{z_{12} z_{13}^2 z_{15} z_{16}
z_{23} z_{24} z_{26} z_{34}^2 z_{45}^2 z_{56}^2}~,~~~\eea
where explicitly, we have
\bea
&&\beta_{12}=1~~,~~\beta_{13}=2~~,~~\beta_{14}=-1~~,~~\beta_{15}=1~~,~~\beta_{16}=1~~,~~\beta_{23}=1~,~~~\nonumber\\
&&\beta_{24}=1~~,~~\beta_{26}=1~~,~~\beta_{34}=2~~,~~\beta_{35}=-1~~,~~\beta_{45}=2~~,~~\beta_{56}=2~,~~~\eea
and all others zero. It has six subsets of simple poles $\{1,3\}$,
$\{3,4\}$, $\{4,5\}$, $\{5,6\}$, $\{1,2,3\}$, $\{1,5,6\}$, and from
them we can construct three compatible combinations as
\bea {\cal B}_1\equiv \{\{1,3\},\{4,5\},\{1,2,3\}\}~~,~~{\cal
B}_2\equiv \{\{1,3\},\{5,6\},\{1,2,3\}\}~~,~~{\cal B}_3\equiv
\{\{3,4\},\{5,6\},\{1,5,6\}\}~.~~~\eea
Following the strategy, we have
\bea &&{\cal B}_1 :~~\{\{1,3\}\}~\xrightarrow[]{\mbox{consider~} \{4,5\}}~
\{\{1,3\},\{4,5\}\}~\xrightarrow[]{\mbox{consider~} \{1,2,3\}}~\{\{1,3,2\},\{4,5\}\}~,~~~\nonumber\\
&&{\cal B}_2:~~\{\{1,3\}\}~\xrightarrow[]{\mbox{consider~} \{5,6\}}~
\{\{1,3\},\{5,6\}\}~\xrightarrow[]{\mbox{consider~} \{1,2,3\}}~\{\{1,3,2\},\{5,6\}\}~,~~~\nonumber\\
&&{\cal B}_3:~~\{\{3,4\}\}~\xrightarrow[]{\mbox{consider~} \{5,6\}}~
\{\{3,4\},\{5,6\}\}~\xrightarrow[]{\mbox{consider~}
\{1,5,6\}}~\{\{3,4\},\{5,6,1\}\}~.~~~\nonumber\eea
So we can construct the {\sl parent set} $\{1,3,2,4,5,6\}$ for
$\{\{1,3\},\{4,5\},\{1,2,3\}\}$, {\sl parent} set $\{1,3,2,5,6,4\}$
for $\{\{1,3\},\{5,6\},\{1,2,3\}\}$, and
$\{3,4,5,6,1,2\}=\{1,2,3,4,5,6\}$ for
$\{\{3,4\},\{5,6\},\{1,5,6\}\}$. Since
\bea &&\{1,3,2,4,5,6\}~~\xrightarrow[]{2\leftrightarrow
3}~~\{1,2,3,4,5,6\}~,~~~\nonumber\\
&&\{1,3,2,5,6,4\}~~\xrightarrow[]{2\leftrightarrow
3}~~\{1,2,3,5,6,4\}~~\xrightarrow[]{4\leftrightarrow
6}~~\{1,2,3,5,4,6\}~~\xrightarrow[]{4\leftrightarrow
5}~~\{1,2,3,4,5,6\}~,~~~\nonumber\eea
and the last one is already in canonical order, we have
\bea
\mathcal{N}_{\inv}^{[1]}=\beta_{23}=1~~,~~\mathcal{N}_{\inv}^{[2]}=\beta_{23}+\beta_{46}+\beta_{45}=3~~,~~\mathcal{N}_{\inv}^{[3]}=0~.~~~\eea
The sign of each term is given by
$(-1)^{6+1+\mathcal{N}_{\inv}^{[i]}}$, so we get the correct result
\bea {1\over s_{13}s_{45}s_{123}}+{1\over
s_{13}s_{56}s_{123}}-{1\over s_{34}s_{56}s_{156}}~.~~~\eea

Before end, let us present an example of the parity invariance of
inversion factor. Consider again the afore-mentioned example and the
compatible combination $\{\{1,3\},\{4,5\},\{1,2,3\}\}$. Among the
$6!=720$ permutation sets of $\{1,2,3,4,5,6\}$, there are 96 sets
which could be the {\sl parent} set of $A$. Restricting to the
convention that $1$ is always placed in the first position, we still
get 16 {\sl parent} sets as
\bea
&&\{1,2,4,6,5,3\}~~,~~\{1,2,5,4,6,3\}~~,~~\{1,2,5,6,4,3\}~~,~~\{1,2,6,4,5,3\}~,~~~\nonumber\\
&&\{1,3,2,4,6,5\}~~,~~\{1,3,2,5,4,6\}~~,~~\{1,3,2,5,6,4\}~~,~~\{1,3,2,6,4,5\}~,~~~\nonumber\\
&&\{1,3,4,6,5,2\}~~,~~\{1,3,5,4,6,2\}~~,~~\{1,3,5,6,4,2\}~~,~~\{1,3,6,4,5,2\}~,~~~\nonumber\\
&&\{1,4,6,5,2,3\}~~,~~\{1,5,4,6,2,3\}~~,~~\{1,5,6,4,2,3\}~~,~~\{1,6,4,5,2,3\}~.~~~\nonumber\eea
The inversion factors $\mathcal{N}_{\inv}$ are consequently
\bea \mathcal{N}_{\inv}=1,~3,~ 3,~ 5,~ 1,~ 3,~ 3,~ 5,~ 3,~ 5,~ 5,~
7,~ 3,~ 5,~ 5,~7~,~~~\eea
which are all odd integers. Similarly, for compatible combination
$\{\{3,4\},\{5,6\},\{1,5,6\}\}$, there are in total also 16 {\sl
parent} sets requiring 1 in the first position, as
\bea
&&\{1,2,3,4,5,6\}~~,~~\{1,2,3,4,6,5\}~~,~~\{1,2,4,3,5,6\}~~,~~\{1,2,4,3,6,5\}~,~~~\nonumber\\
&&\{1,3,4,2,5,6\}~~,~~\{1,3,4,2,6,5\}~~,~~\{1,4,3,2,5,6\}~~,~~\{1,4,3,2,6,5\}~,~~~\nonumber\\
&&\{1,5,6,2,3,4\}~~,~~\{1,5,6,2,4,3\}~~,~~\{1,5,6,3,4,2\}~~,~~\{1,5,6,4,3,2\}~,~~~\nonumber\\
&&\{1,6,5,2,3,4\}~~,~~\{1,6,5,2,4,3\}~~,~~\{1,6,5,3,4,2\}~~,~~\{1,6,5,4,3,2\}~,~~~\nonumber\eea
and the inversion factors for them are respectively
\bea \mathcal{N}_{\inv}=0,~ 2,~ 2,~ 4,~ 2,~ 4,~ 4,~ 6,~ 2,~ 4,~ 4,~
6,~ 4,~ 6,~ 6,~ 8~,~~~\eea
which are all even integers. Thus the parity invariance of inversion
factor is clearly shown\footnote{Although this is confirmed by
numerous computations, we should remark that there is not yet a
proof on it.}.

\section{The Identities from Monodromy Relation and Cross-Ratio Identities}
\label{secMonodromy}

In \cite{Bjerrum-Bohr:2016juj}, an identity for generic pole
$s_{12\cdots k}$ is constructed from monodromy relations,
as\footnote{A derivation of the relation can be found in
\cite{Chen:2011jxa}.}
\bea \mbox{Id}_{\{1,\ldots,k\}}=-\sum_{\sigma\in
\widehat{{\mathcal{OP}}}(\{2,\ldots,k\}~,~\{k+1,\ldots,n-1\})}{PT(1,\sigma_1,\ldots,\sigma_{n-2},n)\over
PT(1,\ldots,n)s_{1\cdots k}}\Big(s_{1\cdots
k}+\sum_{\{i,j\}|\sigma_{i}>\sigma_j}s_{\sigma_i\sigma_j}\Big)~,~~~\label{monodromy}\eea
where $\widehat{\mathcal{OP}}(A,B)$ denotes the sets from {\sl
ordered permutation} of two sets $A,B$, i.e., all permutations among
$A,B$ while keeping the ordering of $A$ and $B$ respectively,
excluding the trivial one $\{2,3,\ldots,n-1\}$. $PT(1,2,\ldots,n)$
denotes the Parke-Taylor-like factor
\bea PT(1,2,\ldots,n)={1\over z_{12}z_{23}\cdots
z_{n-1,n}z_{n1}}~.~~~\eea

The fact that similar monodromy relations exist in CHY-integrands as
those in string and gauge theory amplitudes is itself very
interesting, while practically the identities from monodromy
relations can be applied to the decomposition of CHY-integrands with
higher-order poles. In fact, the systematic decomposition algorithm
proposed in \S\ref{secAlgorithm} can as well proceed with the
identities of monodromy relations as input, without any practical
modifications. Here we shall briefly compare the decomposition
algorithm with these two kinds of identities.

As mentioned, in a $n$-point scattering system, for a generic pole
$s_{i_1i_2\cdots i_k}$, there exists $k(n-k)$ cross-ratio
identities. The cross-ratio identity is {\sl almost} symmetric,
invariant under permutations on $A\setminus\{j\}$ and $\O A\setminus
\{p\}$. While for the identity of monodromy relation
(\ref{monodromy}), because of the {\sl ordered permutation}
$\widehat{\mathcal{OP}}(A,B)$, each ordering of $A,B$ would define
an identity for pole $s_{i_1i_2\cdots i_k}$. So naively, one would
expect $k!(n-k)!$ different identities of monodromy relations for a
pole $s_{i_1i_2\cdots i_k}$. However, as far as there are enough
identities to choose, this will not affect much the efficiency of
decomposition algorithm.

The most important point related to the efficiency of computation
via decomposition algorithm is the number of terms in an identity.
Especially for the CHY-integrands with large $n$ and large
$\Upsilon$ order of poles, the number of resulting terms of simples
poles is very sensitive to the number of terms in identities.
Furthermore, the more terms in an identity, the more troubles we
would meet, since the chance of producing CHY-integrands with other
higher-order poles will increase. For $s_{i_1i_2\cdots i_k}$ pole in
$n$-point scattering, the number of terms in the identity of
monodromy relation (MR) is
\bea \#[\mbox{MR}]={(n-2)!\over
(k-1)!(n-k-1)!}-1~,~~~\label{numMR}\eea
while the number of terms in the cross-ratio identity (CR) is
\bea \#[\mbox{CR}]=(n-k-1)(k-1)~.~~~\label{numCR}\eea
It can be seen that (\ref{numMR}) grows factorially while
(\ref{numCR}) grows algebraically. This matters a lot in the
decomposition algorithm. For example, below is a table listing the
number of terms in the identity of pole $s_{i_1i_2\cdots i_k}$ when
$n=16$,
\begin{center}
\begin{tabular}{|c|c|c|c|c|c|c|c|}
  \hline
  $k$ & ~~~~2~~~~ & ~~~~3~~~~ & ~~~~4~~~~ & ~~~~5~~~~ &~~~~6~~~~ &~~~~7~~~~ &~~~~8~~~~ \\
  \hline
 $\#[\mbox{MR}]$  & 13 & 90 & 363 & 1000 & 2001 & 3002 & 3431 \\
  \hline
  $\#[\mbox{CR}]$ & 13 & 24 & 33 & 40 & 45 & 48 & 49 \\
  \hline
\end{tabular}
\end{center}

\begin{figure}
  \centering
  \includegraphics[width=1.5in]{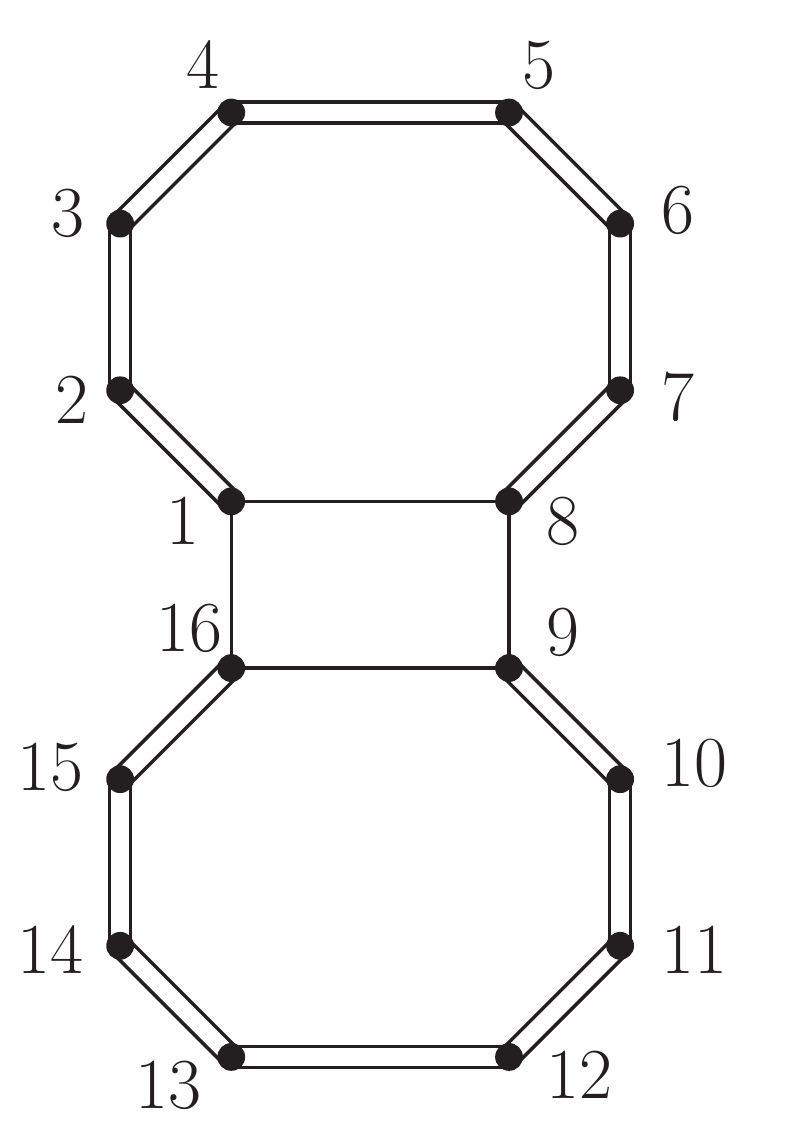}\\
  \caption{The {\sl 4-regular} graph of a 16-point CHY-integrand.}\label{FigA16}
\end{figure}

Although for pole $s_{i_1i_2}$, there is no difference, when
considering a CHY-integrand with higher-order pole $s_{i_1i_2\cdots
i_k}$ for large $k$, the difference becomes dramatic. For example,
let us consider a 16-point CHY-integrand
\bea \mathcal{I}_{16}=\frac{1}{z_{12}^2 z_{23}^2 z_{34}^2 z_{45}^2
z_{56}^2 z_{67}^2 z_{78}^2 z_{9,10}^2 z_{10,11}^2 z_{11,12}^2
z_{12,13}^2 z_{13,14}^2 z_{14,15}^2 z_{15,16}^2 z_{18} z_{89}
z_{9,16} z_{1,16}}~,~~~\eea
with its {\sl 4-regular} graph as shown in Figure \ref{FigA16}. It
has a double pole $s_{12345678}$, and a multiplication of
cross-ratio identity
\bea \mathcal{I}_{16}=\mathcal{I}_{16}\times
\mathbb{I}_{16}[\{1,2,3,4,5,6,7,8\},1,9]~~~~\eea
instantly produces 49 terms with simple poles. While for identity of
monodromy relations the decomposition is much more difficult,
leading to 3431 terms.

\section{$({\rm Parke- Taylor})\times ({\rm Parke- Taylor})$ Geometry and Its Recurrence Relation}
\label{secPT2}

In this section, let us consider the basic two-cycle CHY-integrands
given by $${\rm (Parke-Taylor)}\times {\rm (Parke-Taylor)} = {\rm
PT^2}~,$$ i.e., graphs such as the following
examples,\vspace{-0.2in}
\begin{equation}\label{appendixC1}
\begin{aligned}
\raisebox{-25mm}{\includegraphics[keepaspectratio = true, scale = 0.35] {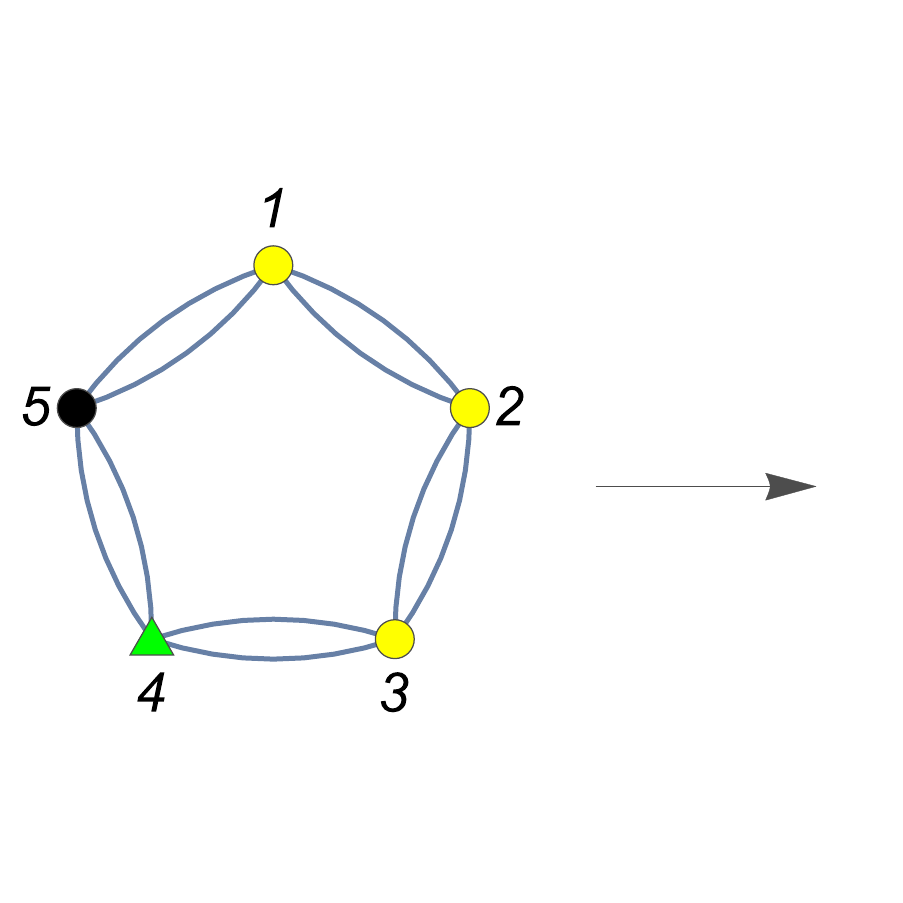}}
\raisebox{-25mm}{\includegraphics[keepaspectratio = true, scale = 0.35] {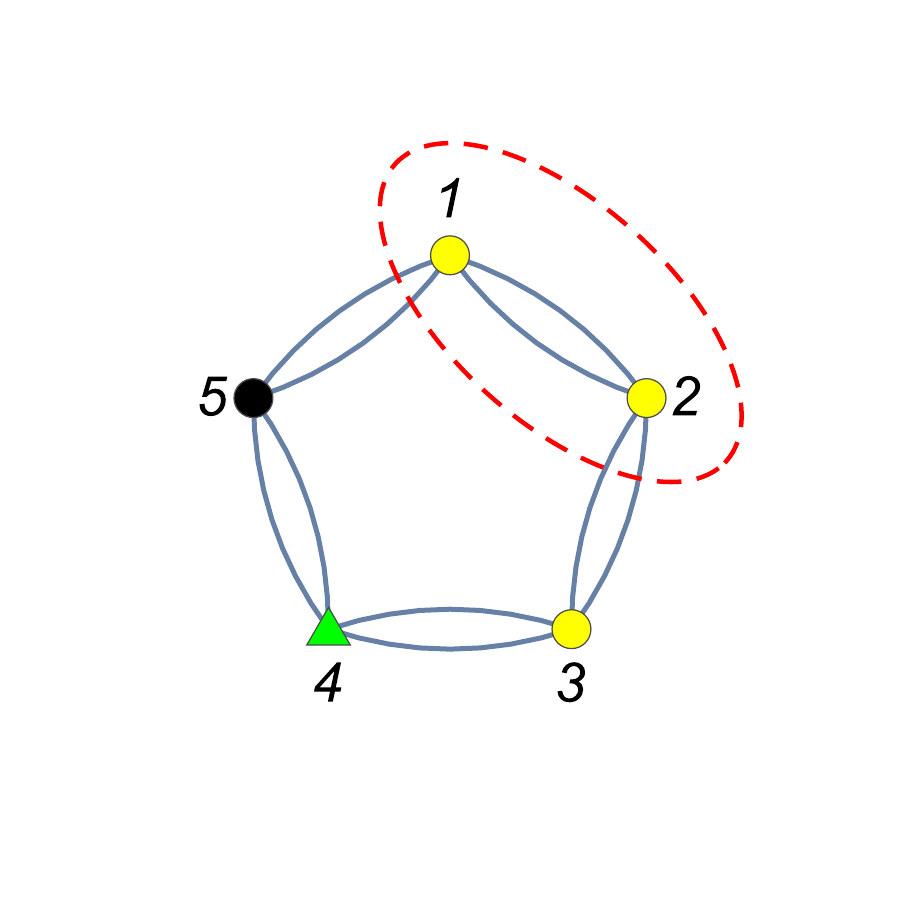}}
\raisebox{-25mm}{\includegraphics[keepaspectratio = true, scale = 0.35] {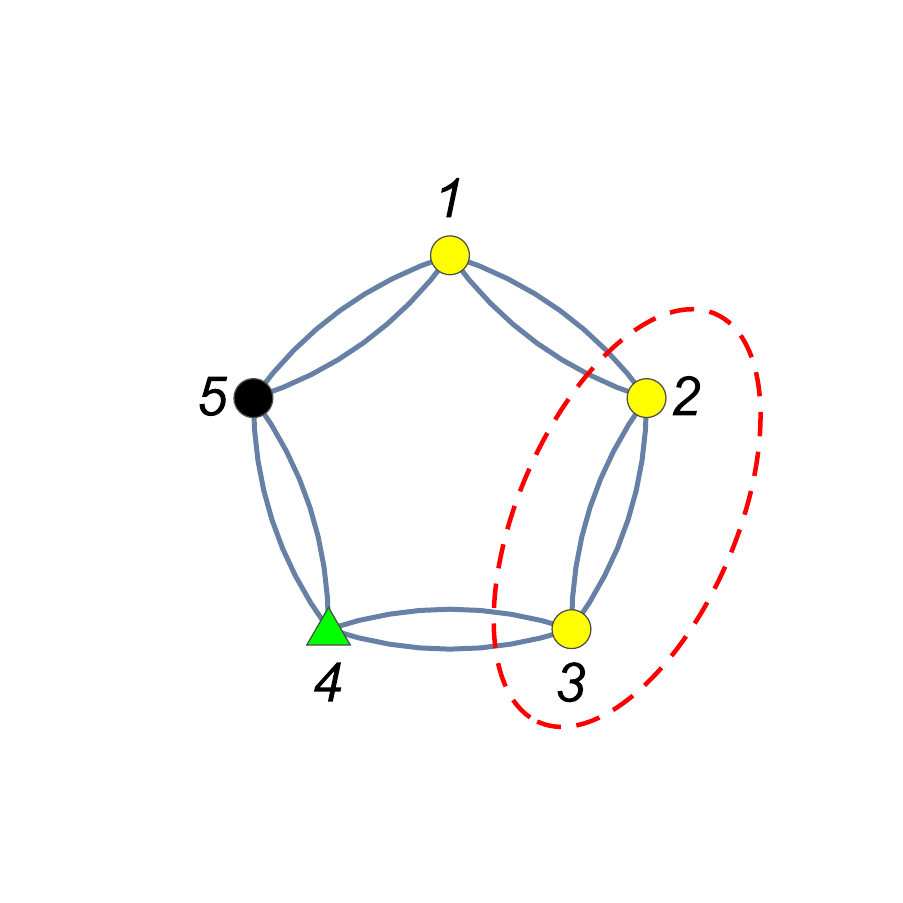}}
\raisebox{-25mm}{\includegraphics[keepaspectratio = true, scale = 0.35] {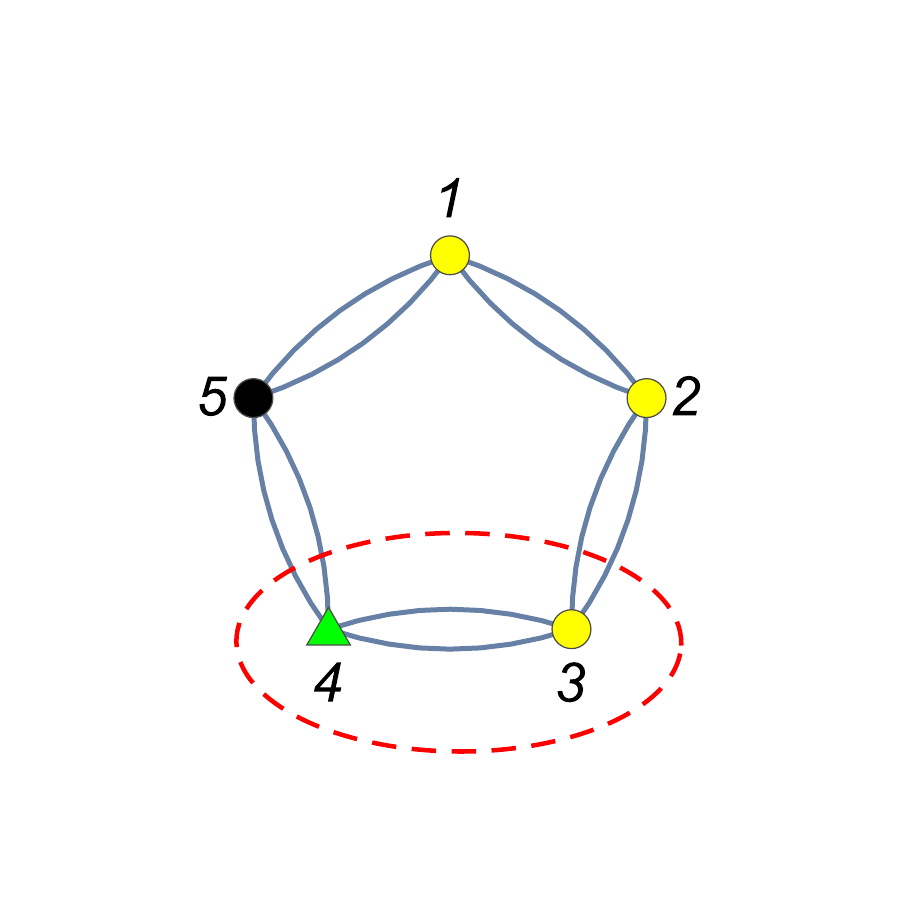}}
\end{aligned}~,~~~
\end{equation}
\vspace{-0.5in}

and

\vspace{-0.5in}
\begin{equation}\label{appendixC2}
\begin{aligned}
\raisebox{-25mm}{\includegraphics[keepaspectratio = true, scale = 0.34] {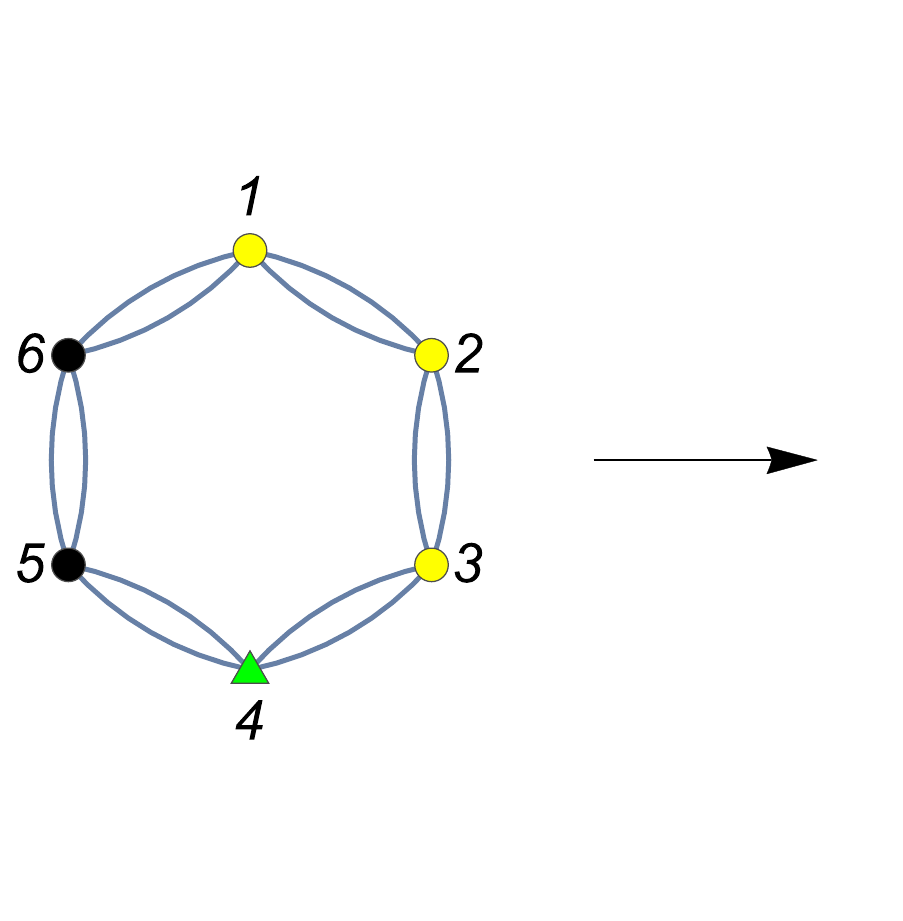}}
\raisebox{-25mm}{\includegraphics[keepaspectratio = true, scale = 0.34] {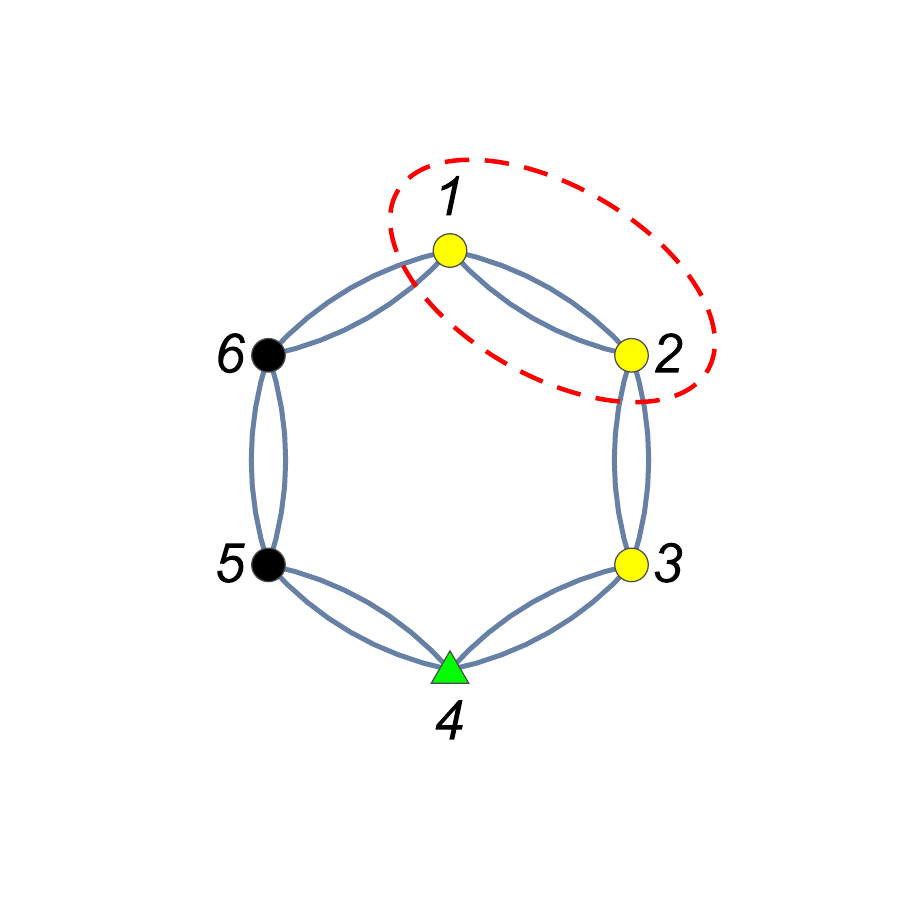}}
\raisebox{-25mm}{\includegraphics[keepaspectratio = true, scale = 0.34] {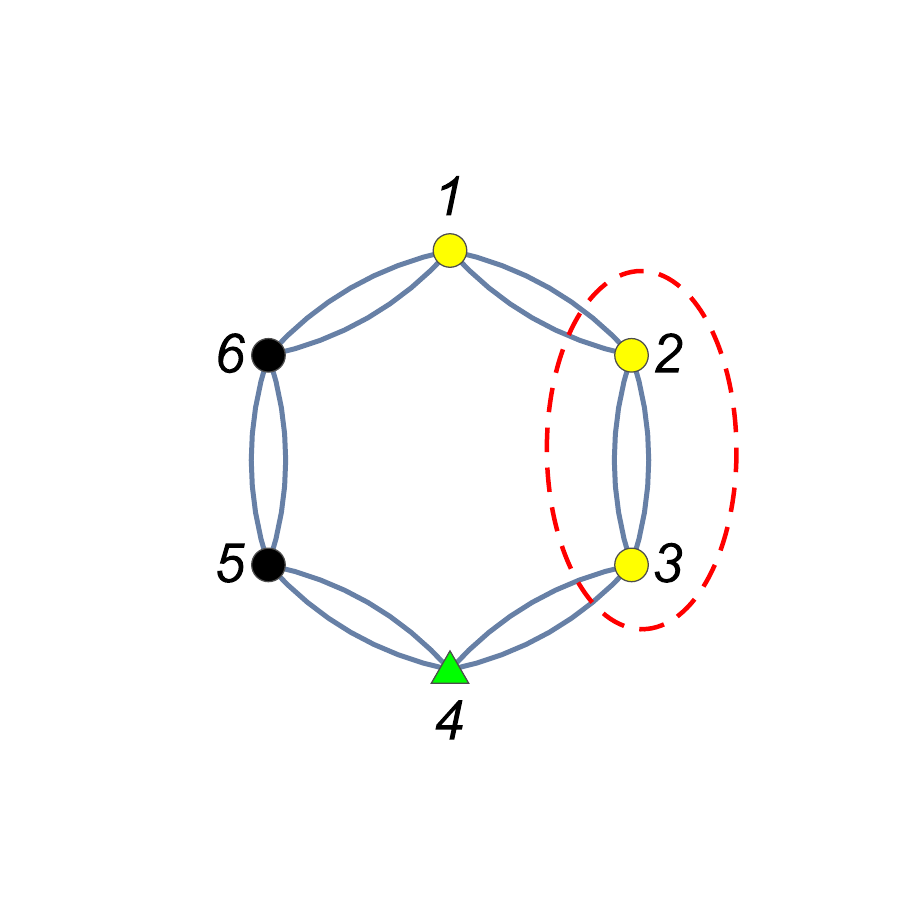}}
\raisebox{-25mm}{\includegraphics[keepaspectratio = true, scale = 0.34] {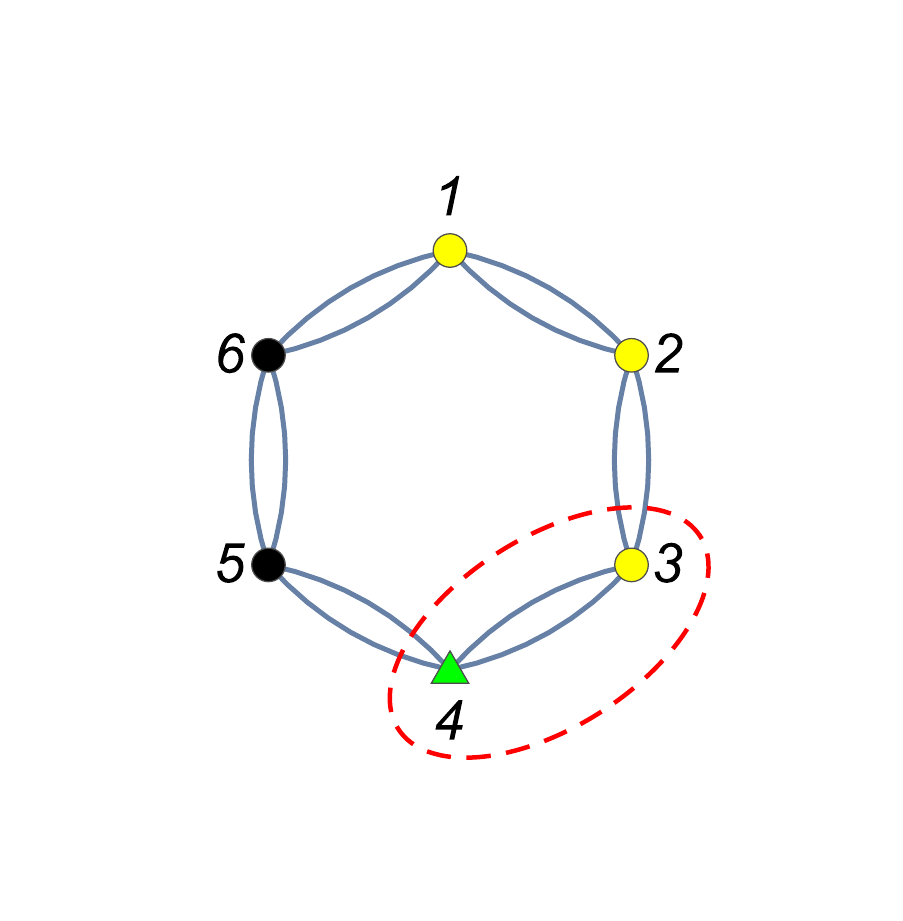}}
\raisebox{-25mm}{\includegraphics[keepaspectratio = true, scale = 0.34] {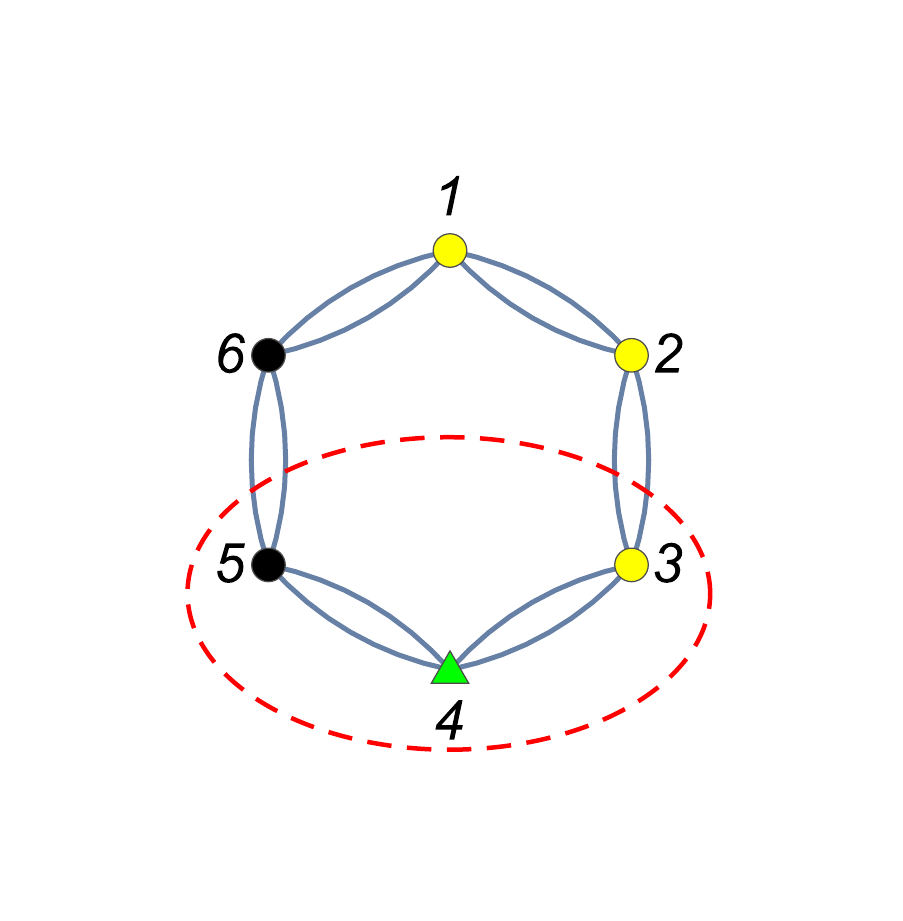}}
\end{aligned}~,~~~
\end{equation}
and
\begin{equation}\label{appendixC3}
\begin{aligned}
&\raisebox{-25mm}{\includegraphics[keepaspectratio = true, scale = 0.35] {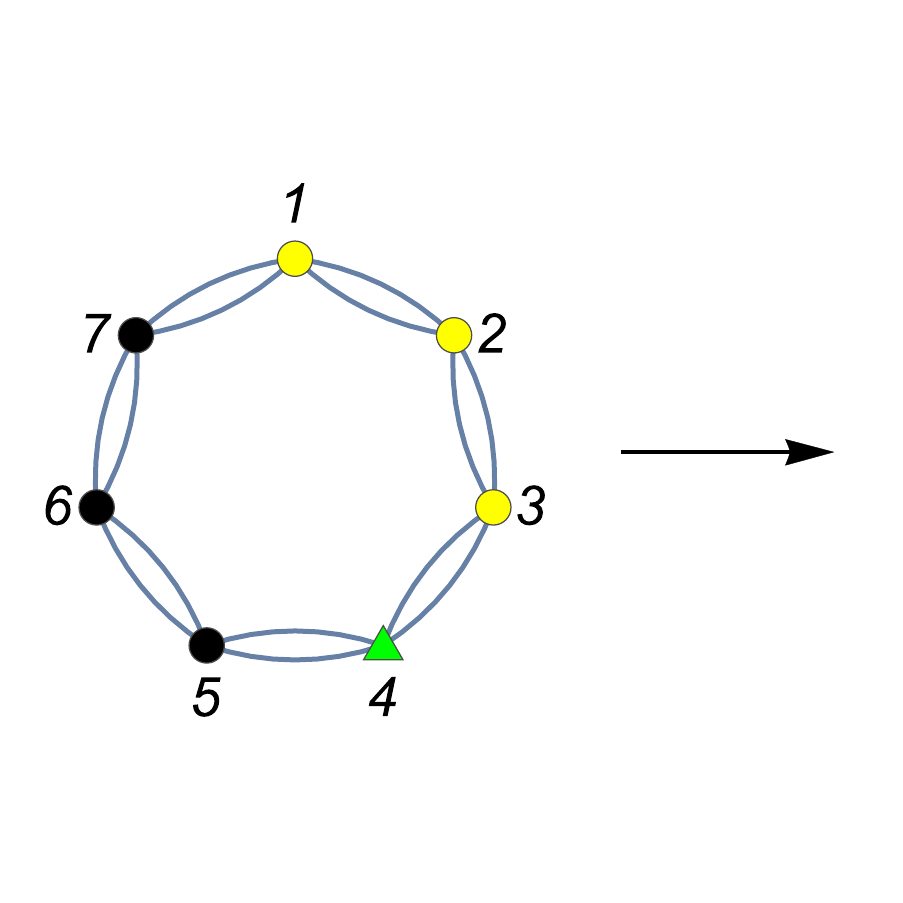}}
\raisebox{-25mm}{\includegraphics[keepaspectratio = true, scale = 0.35] {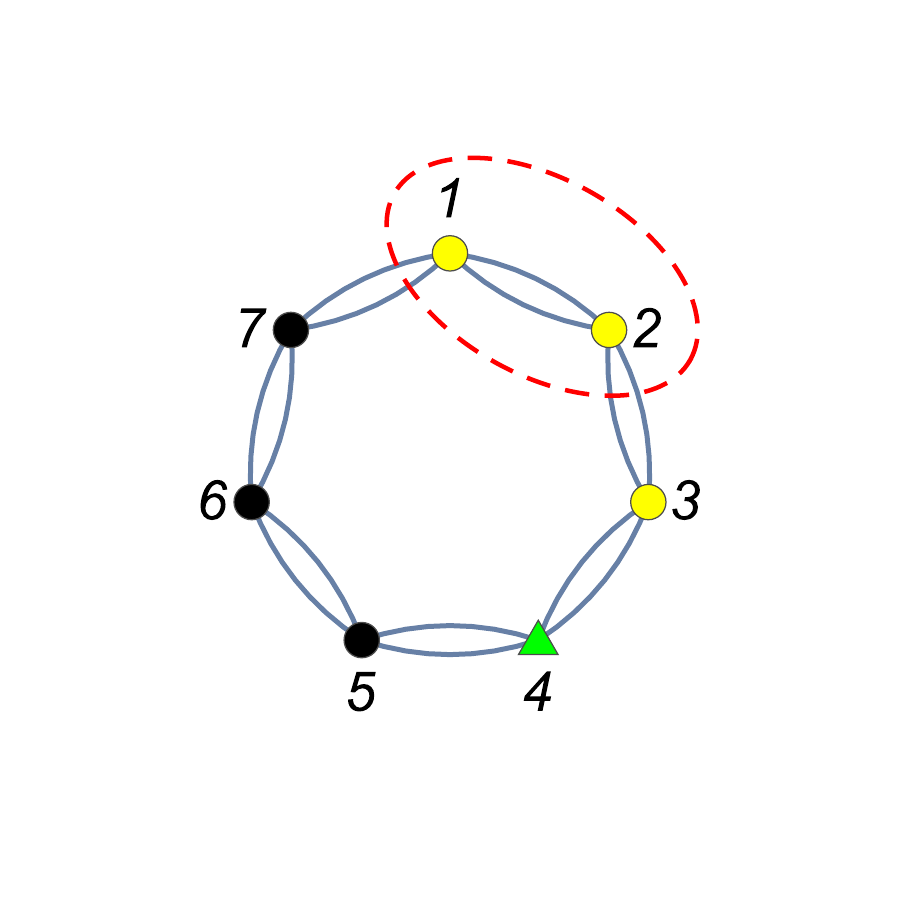}}
\raisebox{-25mm}{\includegraphics[keepaspectratio = true, scale = 0.35] {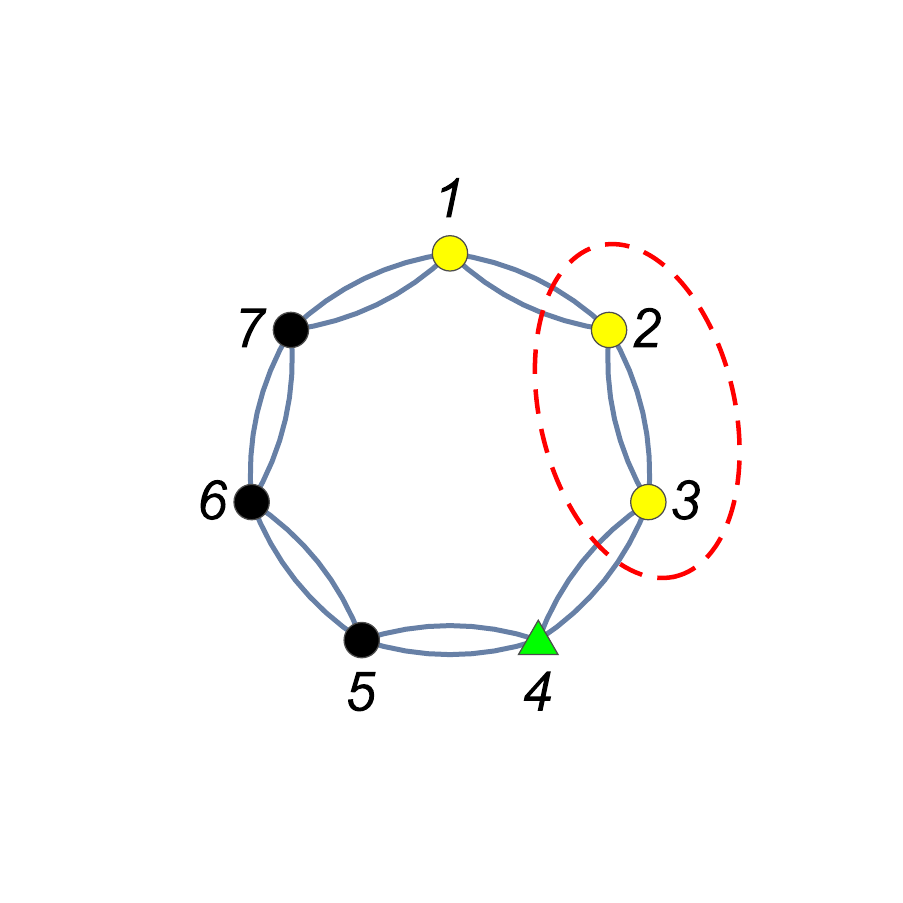}}\\
&~~~~~~~~~~~~~~~~~~~~~~~~~~~~~~~~\raisebox{-25mm}{\includegraphics[keepaspectratio = true, scale = 0.35] {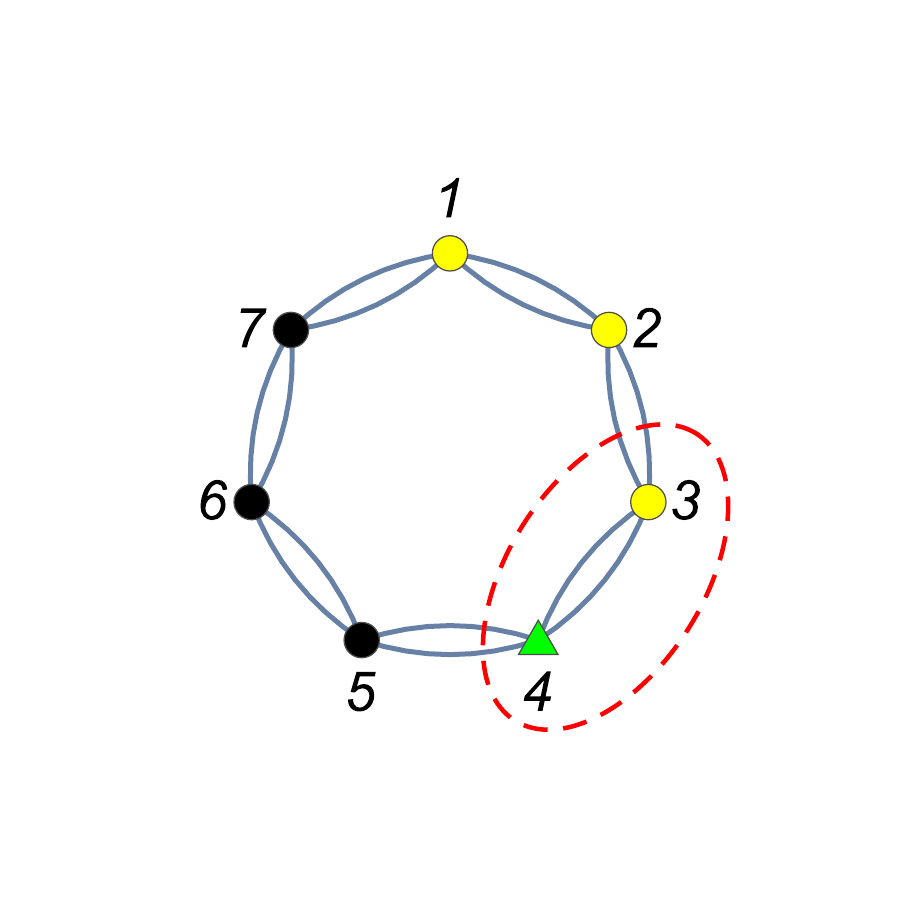}}
\raisebox{-25mm}{\includegraphics[keepaspectratio = true, scale = 0.35] {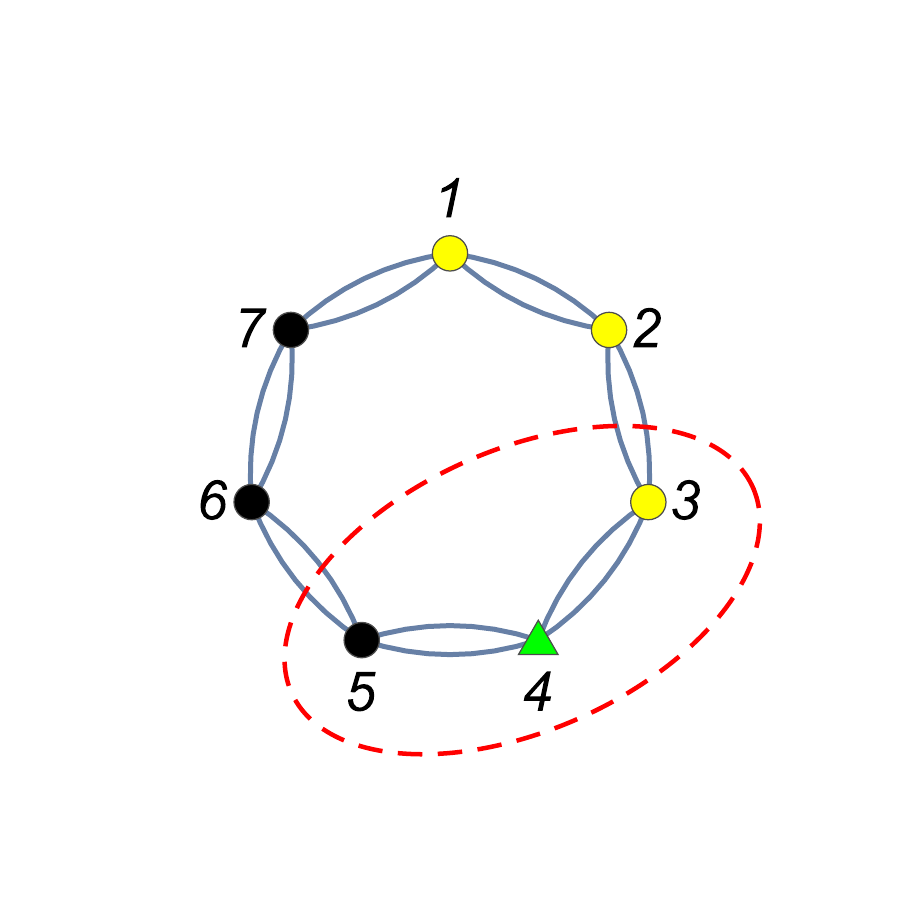}}
\raisebox{-25mm}{\includegraphics[keepaspectratio = true, scale = 0.35] {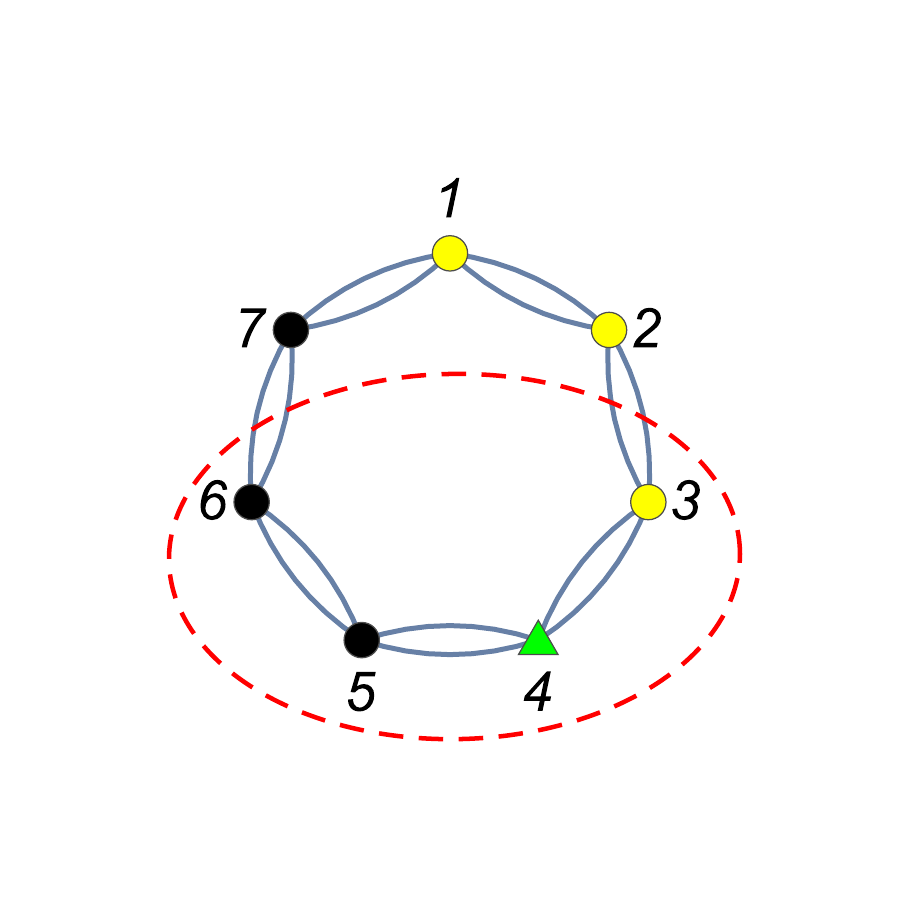}}
\end{aligned}~.~~~
\end{equation}
Although they can be computed trivially by integration rules of
simple poles, let us repeat here by the means of recurrence
relations of $\Lambda$-algorithm, with similar idea that can be
generalized to other recurrence relations of more complicated
geometry. Clearly, the number of all possible non-zero allowable
configurations is $(n-2)$. From the $\Lambda$-algorithm, it is very
well-known that each cut splits the original graph into two
sub-graphs, which are of the same type, i.e., $({\rm
Parke-Taylor})^2$.

Before writing this new recurrence relation, we give some
definitions which will be useful. Let $\mathbbm{o}_i$ be the set of
ordered elements given by
\begin{equation}
\mathbbm{o}_i:= \{5,6,7,\ldots i +4 \}~,~~ {\rm where}~~  i\in\{1,\ldots, n-5\} \, , ~~ n>5~,~~~
\end{equation}
and we defined $\mathbbm{o}_0=\emptyset$. Note that $
\mathbbm{o}_1=\{5\},\mathbbm{o}_2=\{5,6 \}$ and so on.  We also
denote $\O {\mathbbm{o}}_i$ as the ordered complement of
$\mathbbm{o}_i$
\begin{equation}
\O {\mathbbm{o}}_i:=\{5,6,7,\ldots, n-1\}  \setminus \mathbbm{o}_i~,~~~
\end{equation}
for example
\begin{equation}
\O {\mathbbm{o}}_0=\{5,\ldots , n-1  \}~~, ~~ \O {\mathbbm{o}}_1=\{6,\ldots , n-1  \}~~,~~\ldots~~,~~\O {\mathbbm{o}}_{n-5}=\emptyset~.~~~
\end{equation}

With these definitions, the recurrence relation\footnote{Let us
remember that the $\Lambda$-algorithm is an iterative process.}
takes the form,
\begin{align}\label{recurrenceR}
{\cal I}^{\rm PT^2}_{n}(1,2,\dots ,n )&=
{{\cal I}^{\rm PT^2}_{n-1}([1,2],3,\ldots,n)\over \widetilde{s}_{34\cdots n}}+{{\cal I}^{\rm PT^2}_{n-1}(1,[2,3],\ldots,n)\over \widetilde{s}_{4\cdots n 1}}\\
&~~~~+\sum_{i=0}^{n-5}
{{\cal I}^{\rm PT^2}_{n-i-1}(1,2,[3,4,\mathbbm{o}_i],\O {\mathbbm{o}}_i,n)\times{\cal I}^{\rm PT^2}_{i+3}([\O {\mathbbm{o}}_i,n,1,2],3,4,\mathbbm{o}_i)\over \widetilde{s}_{34\mathbbm{o}_i}}~,~~~\nonumber
\end{align}
where
$$
{\cal I}^{\rm PT^2}_3(a,b,c) =1~,
$$
which is the 3-point function. It is very  important to note that
\begin{equation}
{\cal I}_4^{\rm PT^2}(1,2,3,4)= B(4,2|1,3)~.~~~
\end{equation}

Finally, from the CHY-integrands given in (\ref{appendixC1}),
(\ref{appendixC2}), (\ref{appendixC3}) and the recurrence relation
in \eqref{recurrenceR}, it is simple to check the five, six and
seven-point examples
\bea &&{\cal I}_5^{\rm PT^2}(1,2,3,4,5)=\frac{{\cal I}_4^{\rm
PT^2}([1,2],3,4,5)}{\widetilde{s}_{345}}+\frac{{\cal I}_4^{\rm
PT^2}(1,[2,3],4,5)}{\widetilde{s}_{451}}+\frac{{\cal I}_4^{\rm
PT^2}(1,2,[3,4],5)}{\widetilde{s}_{34}}~,~~~\\
&&{\cal I}_6^{\rm PT^2}(1,2,3,4,5,6)=\frac{{\cal I}_5^{\rm PT^2}([1,2],3,4,5,6)}{\widetilde{s}_{3456}}+\frac{{\cal I}_5^{\rm PT^2}(1,[2,3],4,5,6)}{\widetilde{s}_{4561}}+\frac{{\cal I}_5^{\rm PT^2}(1,2,[3,4],5,6)}{\widetilde{s}_{34}}\nonumber \\
&&~~~~~~~~~~~~~~~~~~~~~~~~~~~~~~~ +\frac{{\cal I}_4^{\rm
PT^2}(1,2,[3,4,5],6)\times {\cal I}_4^{\rm
PT^2}([6,1,2],3,4,5)}{\widetilde{s}_{345}}~,~~~\eea
and
\bea
{\cal I}_7^{\rm PT^2}(1,2,3,4,5,6,7)&=&\,\frac{{\cal I}_6^{\rm PT^2}([1,2],3,4,5,6,7)}{\widetilde{s}_{34567}}+\frac{{\cal I}_6^{\rm PT^2}(1,[2,3],4,5,6,7)}{\widetilde{s}_{45671}}+\frac{{\cal I}_6^{\rm PT^2}(1,2,[3,4],5,6,7)}{\widetilde{s}_{34}}\nonumber\\
&&
+\frac{{\cal I}_5^{\rm PT^2}(1,2,[3,4,5],6,7)\times {\cal I}_4^{\rm PT^2}([6,7,1,2],3,4,5)}{\widetilde{s}_{345}} \nonumber\\
&&+\frac{{\cal I}_4^{\rm PT^2}(1,2,[3,4,5,6],7)\times {\cal
I}_5^{\rm PT^2}([7,1,2],3,4,5,6)}{\widetilde{s}_{3456}}~.~~~\eea

\section{Recurrence Relation for ${\cal I}^{{\rm PT^2}\oplus {\rm PT^2}}_{3,3}, {\cal I}^{{\rm PT^2}\oplus {\rm PT^2}}_{4,3}, {\cal I}^{{\rm PT^2}\oplus {\rm PT^2}}_{5,3}$ and ${\cal I}^{{\rm PT^2}\oplus {\rm PT^2}}_{6,3}$}
\label{sec2PT2}

In this section, we provide the results for the CHY-integrands
${\cal I}^{{\rm PT^2}\oplus {\rm PT^2}}_{3,3},\,{\cal I}^{{\rm
PT^2}\oplus {\rm PT^2}}_{4,3 },\, {\cal I}^{{\rm PT^2}\oplus {\rm
PT^2}}_{5,3 }$ and $\,{\cal I}^{{\rm PT^2}\oplus {\rm PT^2}}_{6,3
}$, which are needed in order to obtain the final expression of
\eqref{I_7_3}. Applying the recurrence relation given in
\eqref{recurrenceR2} one obtains
{\small \bea {\cal I}^{{\rm PT^2}\oplus {\rm
PT^2}}_{3,3}(\sigma_1,\sigma_2,\sigma_3|\sigma_4,\sigma_5,\sigma_6)&=&
\frac{{\cal I}^{{\rm PT^2}\oplus {\rm
PT^2}}_{2,3}([\sigma_3,\sigma_1],\sigma_2|\sigma_4,\sigma_5,\sigma_6)}{\widetilde{s}_{\sigma_3\sigma_1}}+
\frac{{\cal I}^{{\rm PT^2}\oplus {\rm
PT^2}}_{2,3}(\sigma_1,[\sigma_2,\sigma_3]|\sigma_4,\sigma_5,\sigma_6)}{\widetilde{s}_{\sigma_2\sigma_3}}\nonumber\\
&&+\frac{{\cal I}^{{\rm PT^2}\oplus {\rm
PT^2}}_{2,3}([\sigma_1,\sigma_2],\sigma_3|\sigma_4,\sigma_5,\sigma_6)}{\widetilde{s}_{\sigma_3\sigma_4\sigma_5\sigma_6}}~,~~~\eea}
and
{\footnotesize \bea {\cal I}^{{\rm PT^2}\oplus {\rm
PT^2}}_{4,3}(\sigma_1,\sigma_2,\sigma_3,\sigma_4|\sigma_5,\sigma_6,\sigma_7)&=&\frac{{\cal
I}^{{\rm PT^2}\oplus {\rm
PT^2}}_{2,3}([\sigma_3,\sigma_4,\sigma_1],\sigma_2|\sigma_5,\sigma_6,\sigma_7)
\times {\cal I}^{{\rm
PT^2}}_{4}(\sigma_1,[\sigma_2,\sigma_5,\sigma_6,\sigma_7],\sigma_3,\sigma_4)}
{\widetilde{s}_{\sigma_3\sigma_4\sigma_1}}
\nonumber\\
&& + \frac{{\cal I}^{{\rm PT^2}\oplus {\rm
PT^2}}_{3,3}(\sigma_1,[\sigma_2,\sigma_3]\sigma_4|\sigma_5,\sigma_6,\sigma_7)
\times {\cal I}^{{\rm
PT^2}}_{3}([\sigma_1,\sigma_4,\sigma_5,\sigma_6,\sigma_7],\sigma_2,\sigma_3)
}{\widetilde{s}_{\sigma_2\sigma_3}}
\nonumber\\
&& + \frac{{\cal I}^{{\rm PT^2}\oplus {\rm
PT^2}}_{2,3}(\sigma_1,[\sigma_2,\sigma_3,\sigma_4]|\sigma_5,\sigma_6,\sigma_7)\times
{\cal I}^{{\rm PT^2}}_{4}([\sigma_1,\sigma_5,\sigma_6,\sigma_7],
\sigma_2,\sigma_3,\sigma_4)}{\widetilde{s}_{\sigma_2\sigma_3\sigma_4}}
\nonumber\\
&& + \frac{{\cal I}^{{\rm PT^2}\oplus {\rm
PT^2}}_{2,3}([\sigma_1,\sigma_2,\sigma_4],\sigma_3|\sigma_5,\sigma_6,\sigma_7)\times
{\cal I}^{{\rm
PT^2}}_{4}(\sigma_1,\sigma_2,[\sigma_3,\sigma_5,\sigma_6,\sigma_7],\sigma_4)}{\widetilde{s}_{\sigma_3\sigma_5\sigma_6\sigma_7}}
\nonumber\\
&& + \frac{{\cal I}^{{\rm PT^2}\oplus {\rm
PT^2}}_{3,3}([\sigma_1,\sigma_2]\sigma_3,\sigma_4|\sigma_5,\sigma_6,\sigma_7)\times
{\cal I}^{{\rm
PT^2}}_{3}(\sigma_1,\sigma_2,[\sigma_3,\sigma_4,\sigma_5,\sigma_6,\sigma_7])}{\widetilde{s}_{\sigma_3\sigma_4\sigma_5\sigma_6\sigma_7}}~,~~~\eea}
as well as
{\footnotesize \bea &&{\cal I}^{{\rm PT^2}\oplus {\rm
PT^2}}_{5,3}(\sigma_1,\sigma_2,\sigma_3,\sigma_4,\sigma_5|\sigma_6,\sigma_7,\sigma_8)\\
&=& \frac{{\cal I}^{{\rm PT^2}\oplus {\rm
PT^2}}_{2,3}([\sigma_3,\sigma_4,\sigma_5,\sigma_1],\sigma_2|\sigma_6,\sigma_7,\sigma_8)
\times {\cal I}^{{\rm
PT^2}}_{5}(\sigma_1,[\sigma_2,\sigma_6,\sigma_7,\sigma_8],\sigma_3,\sigma_4,\sigma_5)}
{\widetilde{s}_{\sigma_3\sigma_4\sigma_5\sigma_1}}
\nonumber\\
&& + \frac{{\cal I}^{{\rm PT^2}\oplus {\rm
PT^2}}_{4,3}(\sigma_1,[\sigma_2,\sigma_3]\sigma_4,\sigma_5|\sigma_6,\sigma_7,\sigma_8)
\times {\cal I}^{{\rm
PT^2}}_{3}([\sigma_1,\sigma_4,\sigma_5,\sigma_6,\sigma_7,\sigma_8],\sigma_2,\sigma_3)
}{\widetilde{s}_{\sigma_2\sigma_3}}
\nonumber\\
&& + \frac{{\cal I}^{{\rm PT^2}\oplus {\rm
PT^2}}_{3,3}(\sigma_1,[\sigma_2,\sigma_3,\sigma_4],\sigma_5|\sigma_6,\sigma_7,\sigma_8)\times
{\cal I}^{{\rm
PT^2}}_{4}([\sigma_1,\sigma_5,\sigma_6,\sigma_7,\sigma_8],
\sigma_2,\sigma_3,\sigma_4)}{\widetilde{s}_{\sigma_2\sigma_3\sigma_4}}
\nonumber\\
&& + \frac{{\cal I}^{{\rm PT^2}\oplus {\rm
PT^2}}_{2,3}(\sigma_1,[\sigma_2,\sigma_3,\sigma_4,\sigma_5]|\sigma_6,\sigma_7,\sigma_8)\times
{\cal I}^{{\rm PT^2}}_{5}([\sigma_1,\sigma_6,\sigma_7,\sigma_8],
\sigma_2,\sigma_3,\sigma_4,\sigma_5)}{\widetilde{s}_{\sigma_2\sigma_3\sigma_4\sigma_5}}
\nonumber\\
&& + \frac{{\cal I}^{{\rm PT^2}\oplus {\rm
PT^2}}_{2,3}([\sigma_1,\sigma_2,\sigma_4,\sigma_5],\sigma_3|\sigma_6,\sigma_7,\sigma_8)\times
{\cal I}^{{\rm
PT^2}}_{5}(\sigma_1,\sigma_2,[\sigma_3,\sigma_6,\sigma_7,\sigma_8],\sigma_4,\sigma_5)}{\widetilde{s}_{\sigma_3\sigma_6\sigma_7\sigma_8}}
\nonumber\\
&& + \frac{{\cal I}^{{\rm PT^2}\oplus {\rm
PT^2}}_{3,3}([\sigma_1,\sigma_2,\sigma_5],\sigma_3,\sigma_4|\sigma_6,\sigma_7,\sigma_8)\times
{\cal I}^{{\rm
PT^2}}_{4}(\sigma_1,\sigma_2,[\sigma_3,\sigma_4,\sigma_6,\sigma_7,\sigma_8],\sigma_5)}{\widetilde{s}_{\sigma_3\sigma_4\sigma_6\sigma_7\sigma_8}}
\nonumber\\
&& + \frac{{\cal I}^{{\rm PT^2}\oplus {\rm
PT^2}}_{4,3}([\sigma_1,\sigma_2],\sigma_3,\sigma_4,\sigma_5|\sigma_6,\sigma_7,\sigma_8)\times
{\cal I}^{{\rm
PT^2}}_{3}(\sigma_1,\sigma_2,[\sigma_3,\sigma_4,\sigma_5,\sigma_6,\sigma_7,\sigma_8])}{\widetilde{s}_{\sigma_3\sigma_4\sigma_5\sigma_6\sigma_7\sigma_8}}~,~~~
\nonumber \eea}
and
{\footnotesize \bea
&&{\cal I}^{{\rm PT^2}\oplus {\rm PT^2}}_{6,3}(\sigma_1,\sigma_2,\sigma_3,\sigma_4,\sigma_5,\sigma_6\,|\, \sigma_7,\sigma_8,\sigma_9 )\\
&=&{{\cal I}^{{\rm PT^2}\oplus {\rm
PT^2}}_{2,3}([\sigma_3,\sigma_4,\sigma_5,\sigma_6,\sigma_1],2\,|\,\sigma_7,\sigma_8,\sigma_9)
\times {\cal I}^{{\rm
PT^2}}_{6}(\sigma_1,[\sigma_2,\sigma_7,\sigma_8,\sigma_9
],\sigma_3,\sigma_4,\sigma_5,\sigma_6)
\over \widetilde{s}_{\sigma_3\sigma_4\sigma_5\sigma_6\sigma_1}} \nonumber  \\
&&+
{{\cal I}^{{\rm PT^2}\oplus {\rm PT^2}}_{5,3}(\sigma_1,[\sigma_2,\sigma_3],\sigma_4,\sigma_5,\sigma_6\,|\, \sigma_7,\sigma_8,\sigma_9)\times{\cal I}^{\rm PT^2}_{3}([\sigma_1,\sigma_4,\sigma_5,\sigma_6,\sigma_7,\sigma_8,\sigma_9],\sigma_2,\sigma_3)\over \widetilde{s}_{\sigma_2\sigma_3}}\,  \nonumber\\
&&+
{{\cal I}^{{\rm PT^2}\oplus {\rm PT^2}}_{4,3}(\sigma_1,[\sigma_2,\sigma_3,\sigma_4],\sigma_5,\sigma_6\,|\, \sigma_7,\sigma_8,\sigma_9)\times{\cal I}^{\rm PT^2}_{4}([\sigma_1,\sigma_5,\sigma_6,\sigma_7,\sigma_8,\sigma_9],\sigma_2,\sigma_3,\sigma_4)\over \widetilde{s}_{\sigma_2\sigma_3\sigma_4}}\,  \nonumber\\
&&+
{{\cal I}^{{\rm PT^2}\oplus {\rm PT^2}}_{3,3}(\sigma_1,[\sigma_2,\sigma_3,\sigma_4,\sigma_5],\sigma_6\,|\, \sigma_7,\sigma_8,\sigma_9)\times{\cal I}^{\rm PT^2}_{5}([\sigma_1,\sigma_6,\sigma_7,\sigma_8,\sigma_9],\sigma_2,\sigma_3,\sigma_4,\sigma_5)\over \widetilde{s}_{\sigma_2\sigma_3\sigma_4\sigma_5}}\,  \nonumber\\
&&+
{{\cal I}^{{\rm PT^2}\oplus {\rm PT^2}}_{2,3}(\sigma_1,[\sigma_2,\sigma_3,\sigma_4,\sigma_5,\sigma_6]\,|\, \sigma_7,\sigma_8,\sigma_9)\times{\cal I}^{\rm PT^2}_{6}([\sigma_1,\sigma_7,\sigma_8,\sigma_9],\sigma_2,\sigma_3,\sigma_4,\sigma_5,\sigma_6)\over \widetilde{s}_{\sigma_2\sigma_3\sigma_4\sigma_5\sigma_6}}\,  \nonumber\\
&&+
{{\cal I}^{{\rm PT^2}\oplus {\rm PT^2}}_{2,3 }([\sigma_1,\sigma_2,\sigma_4,\sigma_5,\sigma_6],\sigma_3\,|\,\sigma_7, \sigma_8,\sigma_9)\times{\cal I}^{\rm PT^2}_{6}(\sigma_1,\sigma_2,[\sigma_3,\sigma_7, \sigma_8,\sigma_9],\sigma_4,\sigma_5,\sigma_6)\over \widetilde{s}_{\sigma_3\sigma_7\sigma_8\sigma_9}}\,  \nonumber\\
&&+
{{\cal I}^{{\rm PT^2}\oplus {\rm PT^2}}_{3,3 }([\sigma_1,\sigma_2,\sigma_5,\sigma_6],\sigma_3,\sigma_4\,|\,\sigma_7, \sigma_8,\sigma_9)\times{\cal I}^{\rm PT^2}_{5}(\sigma_1,\sigma_2,[\sigma_3,\sigma_4,\sigma_7, \sigma_8,\sigma_9],\sigma_5,\sigma_6)\over \widetilde{s}_{\sigma_3\sigma_4\sigma_7\sigma_8\sigma_9}}\,  \nonumber\\
&&+
{{\cal I}^{{\rm PT^2}\oplus {\rm PT^2}}_{4,3 }([\sigma_1,\sigma_2,\sigma_6],\sigma_3,\sigma_4,\sigma_5\,|\,\sigma_7, \sigma_8,\sigma_9)\times{\cal I}^{\rm PT^2}_{4}(\sigma_1,\sigma_2,[\sigma_3,\sigma_4,\sigma_5,\sigma_7, \sigma_8,\sigma_9],\sigma_6)\over \widetilde{s}_{\sigma_3\sigma_4\sigma_5\sigma_7\sigma_8\sigma_9}}\,  \nonumber\\
&&+ {{\cal I}^{{\rm PT^2}\oplus {\rm PT^2}}_{5,3
}([\sigma_1,\sigma_2],\sigma_3,\sigma_4,\sigma_5,\sigma_6\,|\,\sigma_7,
\sigma_8,\sigma_9)\times{\cal I}^{\rm
PT^2}_{3}(\sigma_1,\sigma_2,[\sigma_3,\sigma_4,\sigma_5,\sigma_6,\sigma_7,\sigma_8,\sigma_9])\over
\widetilde{s}_{\sigma_3\sigma_4\sigma_5\sigma_6\sigma_7\sigma_8\sigma_9}}~.~~~
\nonumber\eea}
These results are checked numerically.


\bibliographystyle{JHEP}
\bibliography{cross}

\providecommand{\href}[2]{#2}\begingroup\raggedright\begin{thebibliography}{10}

\bibitem{Cachazo:2013gna}
F.~Cachazo, S.~He, and E.~Y. Yuan, {\it {Scattering equations and
  Kawai-Lewellen-Tye orthogonality}},  {\em Phys. Rev.} {\bf D90} (2014), no.~6
  065001, [\href{http://arxiv.org/abs/1306.6575}{{\tt arXiv:1306.6575}}].

\bibitem{Cachazo:2013hca}
F.~Cachazo, S.~He, and E.~Y. Yuan, {\it {Scattering of Massless Particles in
  Arbitrary Dimensions}},  {\em Phys. Rev. Lett.} {\bf 113} (2014), no.~17
  171601, [\href{http://arxiv.org/abs/1307.2199}{{\tt arXiv:1307.2199}}].

\bibitem{Cachazo:2013iea}
F.~Cachazo, S.~He, and E.~Y. Yuan, {\it {Scattering of Massless Particles:
  Scalars, Gluons and Gravitons}},  {\em JHEP} {\bf 07} (2014) 033,
  [\href{http://arxiv.org/abs/1309.0885}{{\tt arXiv:1309.0885}}].

\bibitem{Cachazo:2014nsa}
F.~Cachazo, S.~He, and E.~Y. Yuan, {\it {Einstein-Yang-Mills Scattering
  Amplitudes From Scattering Equations}},  {\em JHEP} {\bf 01} (2015) 121,
  [\href{http://arxiv.org/abs/1409.8256}{{\tt arXiv:1409.8256}}].

\bibitem{Cachazo:2014xea}
F.~Cachazo, S.~He, and E.~Y. Yuan, {\it {Scattering Equations and Matrices:
  From Einstein To Yang-Mills, DBI and NLSM}},  {\em JHEP} {\bf 07} (2015) 149,
  [\href{http://arxiv.org/abs/1412.3479}{{\tt arXiv:1412.3479}}].

\bibitem{Dolan:2014ega}
L.~Dolan and P.~Goddard, {\it {The Polynomial Form of the Scattering
  Equations}},  {\em JHEP} {\bf 07} (2014) 029,
  [\href{http://arxiv.org/abs/1402.7374}{{\tt arXiv:1402.7374}}].

\bibitem{He:2014wua}
Y.-H. He, C.~Matti, and C.~Sun, {\it {The Scattering Variety}},  {\em JHEP}
  {\bf 10} (2014) 135, [\href{http://arxiv.org/abs/1403.6833}{{\tt
  arXiv:1403.6833}}].

\bibitem{Kalousios:2013eca}
C.~Kalousios, {\it {Massless scattering at special kinematics as Jacobi
  polynomials}},  {\em J. Phys.} {\bf A47} (2014) 215402,
  [\href{http://arxiv.org/abs/1312.7743}{{\tt arXiv:1312.7743}}].

\bibitem{Weinzierl:2014vwa}
S.~Weinzierl, {\it {On the solutions of the scattering equations}},  {\em JHEP}
  {\bf 04} (2014) 092, [\href{http://arxiv.org/abs/1402.2516}{{\tt
  arXiv:1402.2516}}].

\bibitem{Lam:2014tga}
C.~S. Lam, {\it {Permutation Symmetry of the Scattering Equations}},  {\em
  Phys. Rev.} {\bf D91} (2015), no.~4 045019,
  [\href{http://arxiv.org/abs/1410.8184}{{\tt arXiv:1410.8184}}].

\bibitem{Du:2016blz}
Y.-j. Du, F.~Teng, and Y.-s. Wu, {\it {CHY formula and MHV amplitudes}},  {\em
  JHEP} {\bf 05} (2016) 086, [\href{http://arxiv.org/abs/1603.08158}{{\tt
  arXiv:1603.08158}}].

\bibitem{Kalousios:2015fya}
C.~Kalousios, {\it {Scattering equations, generating functions and all massless
  five point tree amplitudes}},  {\em JHEP} {\bf 05} (2015) 054,
  [\href{http://arxiv.org/abs/1502.07711}{{\tt arXiv:1502.07711}}].

\bibitem{Cardona:2015eba}
C.~Cardona and C.~Kalousios, {\it {Comments on the evaluation of massless
  scattering}},  {\em JHEP} {\bf 01} (2016) 178,
  [\href{http://arxiv.org/abs/1509.08908}{{\tt arXiv:1509.08908}}].

\bibitem{Cardona:2015ouc}
C.~Cardona and C.~Kalousios, {\it {Elimination and recursions in the scattering
  equations}},  {\em Phys. Lett.} {\bf B756} (2016) 180--187,
  [\href{http://arxiv.org/abs/1511.05915}{{\tt arXiv:1511.05915}}].

\bibitem{Dolan:2015iln}
L.~Dolan and P.~Goddard, {\it {General Solution of the Scattering Equations}},
  \href{http://arxiv.org/abs/1511.09441}{{\tt arXiv:1511.09441}}.

\bibitem{Huang:2015yka}
R.~Huang, J.~Rao, B.~Feng, and Y.-H. He, {\it {An Algebraic Approach to the
  Scattering Equations}},  {\em JHEP} {\bf 12} (2015) 056,
  [\href{http://arxiv.org/abs/1509.04483}{{\tt arXiv:1509.04483}}].

\bibitem{Sogaard:2015dba}
M.~Sogaard and Y.~Zhang, {\it {Scattering Equations and Global Duality of
  Residues}},  {\em Phys. Rev.} {\bf D93} (2016), no.~10 105009,
  [\href{http://arxiv.org/abs/1509.08897}{{\tt arXiv:1509.08897}}].

\bibitem{Bosma:2016ttj}
J.~Bosma, M.~Sogaard, and Y.~Zhang, {\it {The Polynomial Form of the Scattering
  Equations is an H-Basis}},  \href{http://arxiv.org/abs/1605.08431}{{\tt
  arXiv:1605.08431}}.

\bibitem{Zlotnikov:2016wtk}
M.~Zlotnikov, {\it {Polynomial reduction and evaluation of tree- and loop-level
  CHY amplitudes}},  \href{http://arxiv.org/abs/1605.08758}{{\tt
  arXiv:1605.08758}}.

\bibitem{Cachazo:2015nwa}
F.~Cachazo and H.~Gomez, {\it {Computation of Contour Integrals on ${\cal
  M}_{0,n}$}},  {\em JHEP} {\bf 04} (2016) 108,
  [\href{http://arxiv.org/abs/1505.03571}{{\tt arXiv:1505.03571}}].

\bibitem{Gomez:2016bmv}
H.~Gomez, {\it {$\Lambda$ Scattering Equations}},
  \href{http://arxiv.org/abs/1604.05373}{{\tt arXiv:1604.05373}}.

\bibitem{Cardona:2016bpi}
C.~Cardona and H.~Gomez, {\it {Elliptic scattering equations}},
  \href{http://arxiv.org/abs/1605.01446}{{\tt arXiv:1605.01446}}.

\bibitem{Baadsgaard:2015voa}
C.~Baadsgaard, N.~E.~J. Bjerrum-Bohr, J.~L. Bourjaily, and P.~H. Damgaard, {\it
  {Integration Rules for Scattering Equations}},  {\em JHEP} {\bf 09} (2015)
  129, [\href{http://arxiv.org/abs/1506.06137}{{\tt arXiv:1506.06137}}].

\bibitem{Baadsgaard:2015ifa}
C.~Baadsgaard, N.~E.~J. Bjerrum-Bohr, J.~L. Bourjaily, and P.~H. Damgaard, {\it
  {Scattering Equations and Feynman Diagrams}},  {\em JHEP} {\bf 09} (2015)
  136, [\href{http://arxiv.org/abs/1507.00997}{{\tt arXiv:1507.00997}}].

\bibitem{Lam:2015sqb}
C.~S. Lam and Y.-P. Yao, {\it {Role of M\"bius constants and scattering
  functions in Cachazo-He-Yuan scalar amplitudes}},  {\em Phys. Rev.} {\bf D93}
  (2016), no.~10 105004, [\href{http://arxiv.org/abs/1512.05387}{{\tt
  arXiv:1512.05387}}].

\bibitem{Lam:2016tlk}
C.~S. Lam and Y.-P. Yao, {\it {Evaluation of the Cachazo-He-Yuan gauge
  amplitude}},  {\em Phys. Rev.} {\bf D93} (2016), no.~10 105008,
  [\href{http://arxiv.org/abs/1602.06419}{{\tt arXiv:1602.06419}}].

\bibitem{Baadsgaard:2015hia}
C.~Baadsgaard, N.~E.~J. Bjerrum-Bohr, J.~L. Bourjaily, P.~H. Damgaard, and
  B.~Feng, {\it {Integration Rules for Loop Scattering Equations}},  {\em JHEP}
  {\bf 11} (2015) 080, [\href{http://arxiv.org/abs/1508.03627}{{\tt
  arXiv:1508.03627}}].

\bibitem{Mafra:2016ltu}
C.~R. Mafra, {\it {Berends-Giele recursion for double-color-ordered
  amplitudes}},  \href{http://arxiv.org/abs/1603.09731}{{\tt
  arXiv:1603.09731}}.

\bibitem{Huang:2016zzb}
R.~Huang, B.~Feng, M.-x. Luo, and C.-J. Zhu, {\it {Feynman Rules of
  Higher-order Poles in CHY Construction}},
  \href{http://arxiv.org/abs/1604.07314}{{\tt arXiv:1604.07314}}.

\bibitem{Bjerrum-Bohr:2016juj}
N.~E.~J. Bjerrum-Bohr, J.~L. Bourjaily, P.~H. Damgaard, and B.~Feng, {\it
  {Analytic Representations of Yang-Mills Amplitudes}},
  \href{http://arxiv.org/abs/1605.06501}{{\tt arXiv:1605.06501}}.

\bibitem{Baadsgaard:2015abc}
C.~Baadsgaard, {\it {Amplitudes from sting theory and CHY formalism}},
  Master's thesis, Copenhagen University, 2015.
\newblock Available at
  \url{http://discoverycenter.nbi.ku.dk/teaching/thesis_page/}.

\bibitem{Chen:2011jxa}
Y.-X. Chen, Y.-J. Du, and B.~Feng, {\it {A Proof of the Explicit Minimal-basis
  Expansion of Tree Amplitudes in Gauge Field Theory}},  {\em JHEP} {\bf 02}
  (2011) 112, [\href{http://arxiv.org/abs/1101.0009}{{\tt arXiv:1101.0009}}].

\end{thebibliography}\endgroup

\end{document}